\documentclass[sigconf,nonacm]{acmart}

\copyrightyear{2023}
\acmYear{2023}
\setcopyright{rightsretained}
\acmConference[WSDM '23]{Proceedings of the Sixteenth ACM International Conference on Web Search and Data Mining}{February 27-March 3, 2023}{Singapore, Singapore}
\acmBooktitle{Proceedings of the Sixteenth ACM International Conference on Web Search and Data Mining (WSDM '23), February 27-March 3, 2023, Singapore, Singapore}\acmDOI{10.1145/3539597.3570441}
\acmISBN{978-1-4503-9407-9/23/02}

\usepackage{etoolbox}
\makeatletter
\patchcmd{\maketitle}{\@copyrightpermission}{
  \begin{minipage}{0.3\columnwidth}
    \href{https://creativecommons.org/licenses/by-nc/4.0/}{\includegraphics[width=0.90\textwidth]{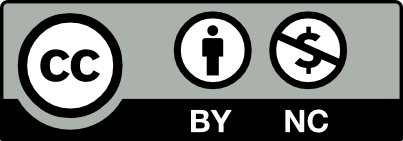}}
  \end{minipage}\hfill
  \begin{minipage}{0.7\columnwidth}
    \href{https://creativecommons.org/licenses/by-nc/4.0/}{This work is licensed under a Creative Commons Attribution-NonCommercial International 4.0 License.}
  \end{minipage}

  \vspace{5pt}
}{}{}

\makeatother
\AtBeginDocument{%
  \providecommand\BibTeX{{%
    \normalfont B\kern-0.5em{\scshape i\kern-0.25em b}\kern-0.8em\TeX}}}

\usepackage[lined, boxed, ruled, commentsnumbered, noend]{algorithm2e}
\usepackage{algpseudocode}
\usepackage{bbding}
\usepackage{bm}
\usepackage{cancel}
\usepackage{colortbl}
\usepackage{enumitem}
\usepackage{multirow}
\usepackage{savesym}
\usepackage{subcaption}
\usepackage{thmtools,thm-restate}
\usepackage{tikz}
\usepackage{float}
\usepackage{caption}
\usepackage{booktabs}

\usetikzlibrary{arrows,shapes,shadows,decorations.text,positioning,automata,shapes.geometric}

\savesymbol{u}
\savesymbol{v}
\savesymbol{x}
\savesymbol{z}
\savesymbol{y}
\savesymbol{w}
\savesymbol{U}
\savesymbol{a}
\savesymbol{d}
\savesymbol{t}
\savesymbol{L}
\savesymbol{do}
\savesymbol{S}
\savesymbol{P}
\savesymbol{G}
\savesymbol{E}
\savesymbol{A}
\savesymbol{H}

\DeclareMathOperator*{\argmax}{arg\,max}
\newtheorem{problem}{Problem}

\newcommand{\R}{\mathbb{R}}
\newcommand{\E}{\mathbb{E}}
\newcommand{\identitymat}{\mathbb{I}}

\newcommand{\bigObound}{\ensuremath{\mathcal{O}}\xspace}

\newcommand{\norm}[1]{\|#1\|}

\newcommand{\round}[1]{\left(#1\right)}
\newcommand{\braces}[1]{\left\{#1\right\}}
\newcommand{\squares}[1]{\left[#1\right]}

\newcommand{\spara}[1]{\subsubsection*{\textnormal{\textbf{#1}}}}

\definecolor{forestgreen}{rgb}{0.13, 0.5, 0.13}
\newcommand{\revision}[1]{\textcolor{black}{{#1}}}

\newcommand{\specsumm}{\textsc{SpecSumm}\xspace}
\newcommand{\random}{\textsc{Random}\xspace}
\newcommand{\deepwalk}{\textsc{DeepWalk}\xspace}
\newcommand{\sls}{\textsc{S2L}\xspace}
\newcommand{\grass}{\textsc{GraSS}\xspace}

\newcommand{\ssumm}{\textsc{SSumM}\xspace}
\newcommand{\opt}{\textsc{Ocsa}\xspace}
\newcommand{\optstiefelgbb}{\textsc{OptStiefelGBB}\xspace}

\newcommand{\lmeigvecs}{\textsc{LM-EigVecs}\xspace}
\newcommand{\reassignment}{\textsc{Reassignment}\xspace}
\newcommand{\reassign}{\textsc{R}\xspace}

\newcommand{\kmeans}{\summarysize-\textsc{Means}\xspace}
\newcommand{\km}{\textsc{KM}\xspace}

\newcommand{\sbm}{\textsc{SBM}\xspace}
\newcommand{\cora}{\textsc{Cora}\xspace}
\newcommand{\PPI}{\textsc{PPI}\xspace}
\newcommand{\caGrQc}{\textsc{ca-GrQc}\xspace}
\newcommand{\lastfm}{\textsc{LastFM-Asia}\xspace}
\newcommand{\blogcatalog}{\textsc{Blogcatalog}\xspace}
\newcommand{\facebook}{\textsc{Facebook}\xspace}
\newcommand{\enron}{\textsc{Email-Enron}\xspace}
\newcommand{\Amazon}{\textsc{Amazon}\xspace}
\newcommand{\Youtube}{\textsc{Youtube}\xspace}
\newcommand{\Wikitalk}{\textsc{Wikitalk}\xspace}

\newcommand{\graph}{\ensuremath{\mathcal{G}}\xspace}
\newcommand{\vertexset}{\ensuremath{\mathcal{V}}\xspace}
\newcommand{\edgeset}{\ensuremath{\mathcal{E}}\xspace}
\newcommand{\numnodes}{\ensuremath{n}\xspace}
\newcommand{\numedges}{\ensuremath{m}\xspace}
\newcommand{\Ag}{\ensuremath{A}\xspace}

\newcommand{\summarysize}{\ensuremath{k}\xspace}
\newcommand{\neigs}{\ensuremath{d}\xspace}
\newcommand{\dimension}{\ensuremath{k}\xspace}
\newcommand{\S}{\ensuremath{S}\xspace}
\newcommand{\ksummary}{\ensuremath{\summarysize\text{-summary}}\xspace}
\newcommand{\partition}{\ensuremath{V}\xspace}
\newcommand{\normloss}{\ensuremath{L}\xspace}

\newcommand{\Aslift}{\ensuremath{A_{\S}^{\uparrow}}\xspace}
\newcommand{\FZ}[1]{\ensuremath{\mathcal{F}_{#1}}\xspace}

\newcommand{\X}[1]{\ensuremath{X_{#1}}\xspace}
\newcommand{\Z}[1]{\ensuremath{Z_{#1}}\xspace}
\newcommand{\P}[1]{\ensuremath{P_{#1}}\xspace}
\newcommand{\PAP}{\ensuremath{\P{\S} \Ag \P{\S}}\xspace}
\newcommand{\ZAZ}{\ensuremath{\Z{\S}^{\top} \Ag \Z{\S}}\xspace}
\newcommand{\tr}[1]{\ensuremath{\bm{tr}[#1]}}

\newcommand{\v}{\ensuremath{z}\xspace}
\newcommand{\e}{\ensuremath{\bm{e}}\xspace}
\newcommand{\B}{\ensuremath{B}\xspace}
\newcommand{\D}{\ensuremath{\Lambda}\xspace}
\newcommand{\u}{\ensuremath{\bm{u}}\xspace}
\newcommand{\x}{\ensuremath{\bm{x}}\xspace}
\newcommand{\y}{\ensuremath{\bm{y}}\xspace}
\newcommand{\L}{\ensuremath{\mathcal{L}}\xspace}

\newcommand{\Q}[1]{\ensuremath{Q_{#1}}\xspace}
\newcommand{\G}[1]{\ensuremath{G_{#1}}\xspace}
\newcommand{\Y}[1]{\ensuremath{Y_{#1}}\xspace}
\newcommand{\U}[1]{\ensuremath{U_{#1}}\xspace}
\newcommand{\Zinit}{\ensuremath{\Z{}^{\round{0}}}\xspace}
\newcommand{\J}[1]{\ensuremath{\bm{J}^{#1}}\xspace}
\newcommand{\t}{\ensuremath{\round{t}}\xspace}
\newcommand{\tone}{\ensuremath{\round{t+1}}\xspace}
\newcommand{\Zt}{\ensuremath{\Z{}^{\t}}\xspace}
\newcommand{\Pt}{\ensuremath{\P{}^{\t}}\xspace}
\newcommand{\Gt}{\ensuremath{\G{}^{\t}}\xspace}
\newcommand{\Yt}{\ensuremath{\Y{}^{\t}\round{\tau}}\xspace}
\newcommand{\Qt}{\ensuremath{\Q{}^{\t}}\xspace}
\newcommand{\Ztone}{\ensuremath{\Z{}^{\tone}}\xspace}
\newcommand{\g}[1]{\ensuremath{\bm{g}\round{#1}}\xspace}

\newcommand{\ZT}{\ensuremath{\Z{}^{\round{T}}}\xspace}
\newcommand{\rounds}{\ensuremath{T}\xspace}
\newcommand{\samples}{\ensuremath{D}\xspace}
\newcommand{\bestcost}{\ensuremath{C_{\text{best}}}\xspace}

\newcommand{\newcost}{\ensuremath{C_{\text{new}}}\xspace}

\begin{document}

\title{Graph Summarization via Node Grouping: A Spectral Algorithm}

%%%%%%%%%%%%%%%%%%%%%%%%%%%%%%%%%%%%%%%%%%%%%%%%%%%%%%%%%%%%%%%%%%%%%%%%%%%%%%%%
% WSDM
%%%%%%%%%%%%%%%%%%%%%%%%%%%%%%%%%%%%%%%%%%%%%%%%%%%%%%%%%%%%%%%%%%%%%%%%%%%%%%%%

\author{Arpit Merchant}
\email{arpit.merchant@helsinki.fi}
\orcid{0000-0001-8143-1539}
\affiliation{%
  \institution{University of Helsinki}
  \city{Helsinki}
  \country{Finland}
  \postcode{00014}
}

\author{Michael Mathioudakis}
\email{michael.mathioudakis@helsinki.fi}
\orcid{0000-0003-0074-3966}
\affiliation{%
  \institution{University of Helsinki}
  \city{Helsinki}
  \country{Finland}
  \postcode{00014}
}

\author{Yanhao Wang}
\email{yhwang@dase.ecnu.edu.cn}
\orcid{0000-0002-7661-3917}
\affiliation{%
  \institution{East China Normal University}
  \city{Shanghai}
  \country{China}
  \postcode{200062}
}

\renewcommand{\shortauthors}{Merchant, et al.}

\begin{abstract}
Graph summarization via node grouping is a popular method to build concise graph representations by grouping nodes from the original graph into supernodes and encoding edges into superedges such that the loss of adjacency information is minimized. 
Such summaries have immense applications in large-scale graph analytics due to their small size and high query processing efficiency.
In this paper, we reformulate the loss minimization problem for summarization into an equivalent integer maximization problem.
By initially allowing relaxed (fractional) solutions for integer maximization, we analytically expose the underlying connections to the spectral properties of the adjacency matrix.
Consequently, we design an algorithm called \specsumm that consists of two phases.
In the first phase, motivated by spectral graph theory, we apply \summarysize-means clustering on the \summarysize largest (in magnitude) eigenvectors of the adjacency matrix to assign nodes to supernodes.
In the second phase, we propose a greedy heuristic that updates the initial assignment to further improve summary quality.
Finally, via extensive experiments on 11 datasets, we show that \specsumm efficiently produces high-quality summaries compared to state-of-the-art summarization algorithms and scales to graphs with millions of nodes.
\end{abstract}

	

\maketitle

\section{Introduction}
\label{sec:introduction}

Graphs have become ubiquitous in diverse fields such as sociology, bioinformatics, and computer science to model different types of relations among objects~\cite{leskovec2009community,sen2008collective}.
Understanding their structure, querying their properties, and designing meaningful visualizations of such graphs can lead to deeper insights about various phenomena.
With increasing graph sizes, a necessary first step for graph analytics is to build an accurate yet small representation of the original graph that is more efficient to process~\cite{tang2016graph}.
To this end, we study the graph summarization problem wherein the goal is to concisely preserve overall graph structure while reducing its size.

Graph summarization has been extensively studied in literature (see~\cite{liu2018graph} for a comprehensive survey).
In general, algorithms for this task can be broadly classified into three categories based on different objectives, namely, \emph{(a) query efficiency}, \emph{(b) space reduction}, and \emph{(c) reconstruction error}.
Respectively, these categories include \textit{(i) application-based} methods tailored for efficiently processing specific types of queries such as reachability~\cite{fan2012query}, distances~\cite{toivonen2011compression}, neighborhoods~\cite{maserrat2010neighbor}, etc., \textit{(ii) compression-based} methods that encode graph structure using fewer bits~\cite{boldi2004webgraph,chierichetti2009compressing,navlakha2008graph}, and \textit{(iii) aggregation-based} methods that combine adjacent nodes and edges into supernodes and superedges to best preserve topology information~\cite{lefevre2010grass,riondato2017graph,lee2020ssumm}.

One popular approach among the above, as well as the focus of this work, is to create aggregation-based supergraph summaries~\cite{lefevre2010grass,riondato2017graph,lee2020ssumm,beg2018scalable} or \summarysize-summaries, for short.
Informally, a $k$-summary is constructed as follows: given size \summarysize as input, each node in the original graph is assigned to one of \summarysize supernodes.
Then, a superedge is added between each pair of supernodes.
Each superedge is assigned a weight equal to the number of edges in the original graph between the nodes within the corresponding supernode pair.
The quality of a \ksummary is measured by the reconstruction error, typically $l_2$-error, defined as the entry-wise difference between the original and recovered adjacency matrices.
Thus, the summarization objective is to minimize the $l_2$-error. 
Aggregation-based algorithms for this task in literature exhibit two primary limitations. First, most algorithms including \grass~\cite{lefevre2010grass} and \sls~\cite{riondato2017graph} cannot scale to large graphs because of high time complexity, dimensionality, or memory footprint. Second, algorithms that can scale such as \textsc{SSumM}~\cite{lee2020ssumm} produce summaries with higher reconstruction errors and poorly preserved graph topologies (eg. number of triangles).

\begin{figure*}[t]
  \captionsetup{skip=3pt}
  \centering
  \scalebox{0.59}{\input{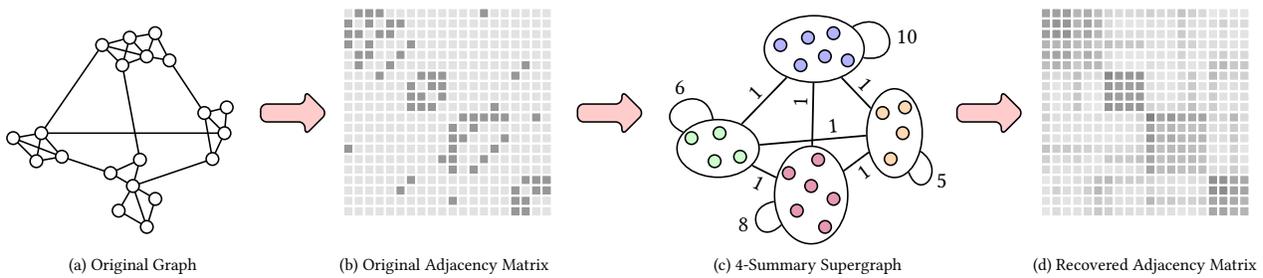}}
  \caption{Illustration of a 4-summary created by \specsumm and adjacency matrix recovered from the summary on a toy graph.}
  \label{fig:toy_graph_illustration}
\end{figure*}

\spara{Our Contributions} 
To address these limitations, we design a scalable algorithm to build a \ksummary that best preserves adjacency information.
We reformulate the $l_2$-error minimization problem into an equivalent integer trace maximization problem.
An integral solution indicates the supernode that each node of the original graph belongs to.
We start by relaxing the integer problem to allow fractional memberships.
We theoretically prove that the \summarysize largest \textit{in magnitude} eigenvectors of the adjacency matrix provide a non-trivial lower bound for the relaxed problem.
We also propose an orthonormality-constrained steepest ascent algorithm (called \opt) adapted from \cite{wen2013feasible} to show that the eigenvectors represent at least a locally optimal solution.
Our approach to building the summary, which we call \specsumm, comprises of two phases.
In the first phase, motivated by spectral graph theory, we apply \summarysize-means clustering on the eigenvector solution to obtain an initial membership matrix.
The second phase comprises of a heuristic that samples nodes uniformly at random and greedily updates their membership to a different supernode if the reassignment improves the objective.
The \ksummary is constructed from the final membership matrix after reassignments. Figure~\ref{fig:toy_graph_illustration} illustrates a 4-summary of a toy graph obtained via \specsumm and the recovered adjacency matrix.

In addition, we provide extensive empirical evidence for the efficacy of our approach.
We implement three variants of \opt using the eigenvectors, a QR-decomposition of a random matrix, and a DeepWalk embedding~\cite{perozzi2014deepwalk} as initial feasible solutions for the relaxed problem.
We show that \opt converges to the eigenvector solution after sufficiently many iterations.
We compare \specsumm, with and without the reassignment heuristic, with \opt and two state-of-the-art baselines over 11 real graphs ranging from 1,000 to 2.3 million nodes.
Across different datasets and summary sizes, results show that \specsumm consistently and efficiently builds summaries with low reconstruction errors.
Lastly, we analyze the scalability of \specsumm on three large graphs via an ablation study for construction time and summary quality as a function of the number of eigenvectors and summary size.
We observe that smaller summaries based on more eigenvectors can be built up to 17\texttimes~faster than larger summaries based on fewer eigenvectors while maintaining comparable quality, thereby offering useful trade-offs for real-world applications.
Our main contributions include:
\begin{itemize}[leftmargin=*]
  \item We introduce a novel reformulation for the \ksummary problem and analytically motivate the design of our algorithms.
  \item We propose \specsumm that clusters the eigenvectors of the original adjacency matrix to create an initial high-quality \ksummary and further refines the summary using a greedy heuristic.
  \item We show via extensive experiments that \specsumm constructs summaries of upto 22.5\% and 76.1\% higher quality on small to medium sized graphs compared to state-of-the-art baselines \sls and \ssumm while running upto 200\texttimes~faster than \sls. Further, \specsumm scales well to massive graphs with millions of nodes and produces concise, meaningful summaries within 3 hours.
\end{itemize} 

\section{Related Work}
\label{sec:related_work}

% The goal of a summary is to build concise representations of input graphs. 
We categorize previous studies into three broad classes based on their summarization objectives. We refer interested readers to \citet{liu2018graph} for a more extensive survey.

\spara{Query Efficiency}
Methods in this class construct summaries tailored for processing specific types of graph queries.
\citet{maserrat2010neighbor} and \citet{DBLP:journals/isci/NejadJT21} designed summaries that efficiently search for neighbors of a query vertex.
\citet{toivonen2011compression} and \citet{DBLP:journals/is/SadriSRZCS17} summarize weighted graphs to preserve the distances between vertices.
\citet{fan2012query} and \citet{DBLP:journals/isci/LiangCWLYL20} devised graph summaries for efficient reachability queries. %while the summary by \citet{DBLP:conf/sdm/ZhouLLSSC21} preserves degrees.
% Several methods~\cite{DBLP:journals/snam/KhanA17,DBLP:conf/bigdataconf/NelsonRCS17,DBLP:journals/tkde/TsalouchidouBMB20,DBLP:conf/icde/Gou0Z019} were proposed to summarize graph streams for dynamic graph queries.
A separate but related set of methods in this class construct summaries for user-specified utilities~\cite{DBLP:journals/pvldb/KumarE18,DBLP:conf/kdd/HajiabadiS0T21}, modularity~\cite{gorke2010modularity}, and motifs~\cite{dunne2013motif}.
However, these summaries do not include adjacency recovery procedures and further, our goal is to build a general-purpose summary for different types of queries. This makes a direct comparison infeasible.

\spara{Space Reduction}
Methods in this class store a (lossless or lossy) representation of a graph using minimum possible space.
For instance, VoG~\cite{koutra2014vog} uses Minimum Description Length for compression to encode a vocabulary of subgraphs such as stars and cliques.
Subsequent studies proposed different graph reordering and encoding schemes to improve compression ratios~\cite{boldi2004webgraph,DBLP:conf/wsdm/BuehrerC08,chierichetti2009compressing,DBLP:conf/www/BoldiRSV11,DBLP:conf/kdd/DhulipalaKKOPS16,NEURIPS2021_9a4d6e86}.
Aggregation-based schemes for compression proposed by \citet{navlakha2008graph} among others~\cite{DBLP:conf/sigmod/TianHP08,DBLP:journals/computing/KhanNL15,DBLP:conf/www/ShinG0R19,DBLP:conf/kdd/KoKS20,DBLP:conf/sigmod/YongH0T21,DBLP:conf/sigmod/FanLLL21} maintain extra edge corrections to recover the missing information due to node/edge grouping.
However, unlike our paradigm, these methods either do not create hypergraphs or they do not minimize reconstruction loss or both.
Within this class, \ssumm~\cite{lee2020ssumm} presents the closest summary specification to ours and thus we include it as a baseline for comparison. 
Note, \ssumm has a different objective: it minimizes the number of bits required for storage jointly with the reconstruction error, which is achieved by coarsening supernodes and pruning superedges.
As a result, \ssumm cannot guarantee that the summary size is exactly equal to the user-specified input $k$ and, as shown in the experiments, it exhibits higher reconstruction errors than our algorithms while having higher or comparable efficiencies.

\spara{Reconstruction Error}
Methods in this class build supergraph summaries such that the error in reconstructing adjacency matrices is minimized and are thus closely related to our work.
\grass~\cite{lefevre2010grass} constructs a \ksummary by repeatedly merging a pair of supernodes that maximally decreases the reconstruction error until only $k$ supernodes remain.
\textsc{ScalableSumm}~\cite{beg2018scalable} adopts a similar merging-based scheme as \grass.
Additionally, it utilizes a sampling method for candidate pair selection and a count-min sketch~\cite{cormode2005improved} for reconstruction error estimation.
However, merging-based schemes suffer from low summary quality when the summary size \summarysize is small.
\citet{riondato2017graph} proposed \sls which employs $\summarysize$-means clustering on the rows of the adjacency matrix to create supernodes.
\sls provides a theoretical guarantee on the $l_p$-reconstruction error of the output summary.
% \ssumm and \sls have both been shown to outperform \grass.
Nevertheless, \sls incurs costly distance computations given the high dimensionality of the adjacency matrix and thus is not scalable to massive graphs.
We compare with \sls in the experiments and the results confirm that \specsumm outperforms \sls in terms of both summary quality and efficiency.

\section{Preliminaries}
\label{sec:background}

Consider an unweighted, undirected graph $\graph = \round{\vertexset, \edgeset}$ where \vertexset is a set of $\numnodes$ nodes and \edgeset is a set of $\numedges$ edges.
We denote its adjacency matrix by $\Ag \in \braces{0,1}^{\numnodes \times \numnodes}$.
A \summarysize-partition of \vertexset is defined as $\partition = \braces{V_1, \ldots, V_{\summarysize}}$ such that $\forall i \neq j \in \squares{\summarysize}, V_i \cap V_j = \emptyset$ and $\bigcup_{i=1}^{\summarysize} V_{i} = \vertexset$.
Let $\X{\partition} \in \braces{0, 1}^{\numnodes \times \summarysize}$ represent a \emph{membership matrix} corresponding to partition $\partition$, where the $\round{i, j}$-th entry is 1 if node $i$ belongs to set $V_j$ and 0 otherwise. \revision{Each node is assigned to exactly one partition and thus $\X{\partition}$ is orthogonal.}
Let $\Z{\partition} = \X{\partition} (\X{\partition}^\top \X{\partition})^{-1/2}$ be the associated normalized membership matrix where $\Z{\partition}^\top \Z{\partition} = \identitymat$. 
We denote $\P{\partition} = \Z{\partition} \Z{\partition}^\top$ as a smoothing operator, i.e., the orthogonal projection onto the subspace spanned by the columns of \Z{\partition}. 

Given a \summarysize-partition \partition of \vertexset, let $\partition \times \partition$ denote the set of all superedges between every pair of subsets in \partition.
Then, a \ksummary of \graph is defined as a weighted, supergraph $\S_{\graph,\partition} = \braces{\partition, \partition \times \partition}$ of $|\partition| = \summarysize$ supernodes and $\summarysize \round{\summarysize-1}/ 2$ superedges. For $i, j \in \squares{\summarysize}$, the weight of a superedge between supernodes $V_i$ and $V_j$ is given by:
\begin{equation} \label{eq:summary_edge_weight}
    A_{\S}(V_i, V_j) := \frac{\sum_{u \in V_i, v \in V_j} \Ag(u, v)}{|V_i| \cdot |V_j|},
\end{equation}
where $A_{\S}$ is called the density matrix of \S.\footnote{We omit \graph and \partition from the subscript for notational convenience.}
This weight denotes the fraction of actual edges in \graph between the nodes in $V_i$ and $V_j$ divided by the maximum possible number of edges.
We use $A_{\S}$ to approximate the original adjacency matrix. This approximation recovered from a summary is referred to as a \emph{lifted adjacency matrix}~\cite{riondato2017graph}.
Its $\round{u,v}$-th entry captures the probability of the existence of an edge between $u$ and $v$ in \graph. Specifically, $\Aslift(u, v) = A_{\S}(\S(u), \S(v))$, where $\S(u)$ represents the supernode that $u$ belongs to.
In matrix notation, the lifted adjacency matrix is written as
$\Aslift = \P{\partition} \Ag \P{\partition}$~\cite{riondato2017graph}.
The quality of a \dimension-summary \S is measured by the $l_2$-norm of the entry-wise difference between \Ag and \Aslift~\cite{riondato2017graph,lee2020ssumm}. Formally:
\begin{equation}\label{eq:normloss}
    \normloss\round{\Ag, \Aslift} = \norm{\Ag - \Aslift}_2^2 = \sum\limits_{u \in \vertexset}\sum\limits_{v \in \vertexset} \round{\Ag\round{u,v} - \Aslift\round{u,v}}^2
\end{equation}
\revision{This $l_2$-norm error is exactly twice that of the $l_1$-norm error, thereby making these errors equivalent~\cite{riondato2017graph}.} Thus, we focus on finding a summary \S that minimizes $\normloss(\Ag, \Aslift)$.
We rewrite $l_2$-error as follows:
\begin{lemma}\label{lemma:traceloss}
    $\normloss\round{\Ag, \Aslift} = \tr{\Ag^2} - \underbrace{\tr{\round{\ZAZ}^2}}_{\FZ{\Z{\S}}}$
\end{lemma}

\noindent
\revision{We defer all proofs to Appendix A.}
Since the first term, $\tr{\Ag^2} = 2\cdot |\edgeset|$ is a constant, the matrix \Z{\S} that maximizes the second term, \FZ{\Z{\S}}, \revision{also minimizes $\normloss\round{\Ag, \cdot}$}. Formally, we recast the graph summarization problem given graph \graph and size \summarysize as the following integer trace maximization problem:
\begin{problem}\label{prob:trace_maximization_integer}[Graph \summarysize-Summarization]
    \begin{equation*}
        \begin{aligned}
            \argmax_{Z}       \quad & \tr{\round{Z^{\top} \Ag Z}^2}  \\
            \textnormal{s.t.} \quad & \revision{Z^{\top}Z = \identitymat \text{ where } Z = X \round{X^\top X}^{-1/2}} \\
                                    & X \in \braces{0,1}^{\numnodes \times \summarysize}\\
        \end{aligned}
    \end{equation*}
\end{problem}

\section{Algorithms}
\label{sec:algorithms}

In this section, we present our approach for graph $k$-summarization (i.e., Problem~\ref{prob:trace_maximization_integer}) along with the underlying analytical motivations.
Our approach consists of three steps. First, we relax the membership matrix \X{} to accept real entries with all other conditions remaining intact.
Formally, this gives us the following relaxed problem:
\begin{problem}\label{prob:trace_maximization_relaxed}[Relaxed Graph \summarysize-Summarization]
    \begin{equation*}
        \begin{aligned}
            \argmax_{Z}       \quad & \tr{\round{Z^{\top} \Ag Z}^2} \\
            \textnormal{s.t.} \quad & \revision{Z^{\top} Z = \identitymat \text{ where } Z = X \round{X^\top X}^{-1/2}}\\
                                    & X \in \R^{\numnodes \times \summarysize}\\
        \end{aligned}
    \end{equation*}
\end{problem}
Second, in Section~\ref{subsec:relaxed_problem}, we design two solutions for Problem~\ref{prob:trace_maximization_relaxed}. And third, in Section~\ref{subsec:integer_problem}, we define a heuristic rounding algorithm to convert the relaxed solution to an integral solution for Problem~\ref{prob:trace_maximization_integer}.

\subsection{Relaxed Graph \summarysize-Summarization}
\label{subsec:relaxed_problem}

Consider the trivial solution when $\summarysize = \numnodes$.
The following result is obtained immediately via substitution:
\begin{restatable}[]{lemma}{kequalsn}\label{lemma:k_equals_n}
    Given an adjacency matrix \Ag and $\summarysize = \numnodes$, $\Z{} = [\e_1, \dots,$ $\e_{\summarysize}]$ optimally solves Problem~\ref{prob:trace_maximization_relaxed} where $\e_i$ are the eigenvectors of \Ag.
\end{restatable}

For general values of \summarysize, we write the objective in vector form. Let $\Z{} = \squares{\v_1, \dots, \v_{\summarysize}}$ where $\v_i$ represents the $i$-th column of \Z{}. Then:
\begin{equation}
    \begin{aligned}
        \tr{\round{\Z{}^{\top} \Ag \Z{}}^2}
        & = \tr{\round{\squares{\v_1, \ldots, \v_{\summarysize}}^\top \Ag \squares{\v_1, \ldots, \v_{\summarysize}}}^2} \\
        & = \underbrace{\sum\limits_{j=1}^{\summarysize} \round{\v^{\top}_j \Ag \v_j}^2}_{T_1} + \underbrace{\sum\limits_{j=1}^{\summarysize} \sum\limits_{i \in \squares{\summarysize} \backslash \braces{j}} (\v^\top_j \Ag \v_i)^2}_{T_2}
    \end{aligned}
    \label{eq:vectorform}
\end{equation}
A trivial lower bound for \FZ{\Z{}} is $0$ since the individual terms in Equation~\ref{eq:vectorform} are squares of scalar numbers. Below, for $\summarysize = \braces{1,\ldots,\numnodes}$, we analyze the two terms, $T_1$ and $T_2$, to obtain non-trivial solutions.

\spara{Largest-Magnitude Eigenvectors}
Our main result proves that the \summarysize largest (\emph{in magnitude}) eigenvectors of \Ag represent a non-trivial lower bound on the value of the relaxed objective function and thus a non-trivial feasible solution to Problem~\ref{prob:trace_maximization_relaxed}.

\begin{restatable}[]{theorem}{lmeigvecslowerbound}\label{thm:lmeigvecs_lower_bound}
    A constructive lower bound for the maximization objective (Problem~\ref{prob:trace_maximization_relaxed}) is given as follows:
    \begin{equation}\label{eq:lmeigvecs_lower_bound}
        \tr{\round{\Z{}^{\top} \Ag \Z{}}^2} \geq \sum\limits_{j=1}^{\summarysize} \lambda_{j}^2,
    \end{equation}
    where, for $j \in \squares{\summarysize}$, $\lambda_j$ is the $j$-th largest (in magnitude) eigenvalue of \Ag. Further, this lower bound is achieved when $\Z{} = \squares{\e_1, \ldots, \e_{\summarysize}}$ where each $\e_j$ is the eigenvector corresponding to $\lambda_j$.
\end{restatable}

\spara{Proof Sketch}
We prove the above result by induction over \summarysize.
In the base case when $\summarysize=1$, there are no cross-terms (i.e., $T_2$) and $T_1$ consists of just one term. This yields the following result:
\begin{restatable}[]{lemma}{kequalsone}\label{lemma:k_equals_one}
    Given an adjacency matrix \Ag and $\summarysize = 1$, the maximum value of the relaxed objective in Problem~\ref{prob:trace_maximization_relaxed} is achieved by the largest-magnitude eigenvector $\e_1$ of \Ag, i.e.,
    \begin{equation}\label{eq:k_equals_one}
        \argmax_{\Z{}} T_1 = \argmax_{\v} \tr{\round{\v^{\top} \Ag \v}^2} = \e_1
    \end{equation}
\end{restatable}

\begin{algorithm}[t]
\small

\nl {\bf Input}: Adjacency matrix \Ag of \graph; summary size \summarysize. \\
\nl {\bf Output}: Feasible solution \Z{} for Problem~\ref{prob:trace_maximization_relaxed}. \\

\tcp*[h]{\textcolor{blue}{Compute the \summarysize largest (in magnitude) eigenvectors}}\\
\nl $\Z{} \gets \text{\textsc{ComputeEigVecs}}\round{\Ag, \summarysize}$ \\

\nl \textbf{return} \Z{}

\caption{\lmeigvecs (Relaxed Problem)}
\label{algo:lm_eigvecs}

\end{algorithm}

In the induction step, we show that for higher values of \summarysize, $T_1$ is maximized by the \summarysize largest (in magnitude) eigenvectors of \Ag.
\begin{restatable}[]{lemma}{kbetweenonen} \label{lemma:higher_k}
    Given an adjacency matrix \Ag and $\summarysize \in \braces{2, \ldots, \numnodes}$, the set of self-terms, $T_1$, in the relaxed objective (Equation~\ref{eq:vectorform}) is maximized by the \summarysize largest (magnitude) eigenvectors $\e_1,\ldots,\e_{\summarysize}$ of \Ag.
    \begin{equation}\label{eq:higher_k}
        \argmax_{\Z{}} T_1 
            = \argmax_{\Z{}} \sum\limits_{j=1}^{\summarysize} \round{\v^{\top}_j \Ag \v_j}^2
            = \squares{\e_1, \ldots, \e_{\summarysize}}
    \end{equation}
\end{restatable}
Here, the cross-terms ($T_2$) always reduce to 0. It follows from the definition of eigenvectors and their mutual orthogonality whereby, for any $i,j \in \squares{\summarysize}, \e_i \neq \e_j, \round{\e_i^\top \Ag \e_j}^2 = \round{\e_i^\top \lambda_j \e_j}^2  = 0$. Putting Lemmas~\ref{lemma:k_equals_one} and~\ref{lemma:higher_k} together proves Theorem~\ref{thm:lmeigvecs_lower_bound}. Algorithm~\ref{algo:lm_eigvecs} codifies it into a subroutine we refer to as \lmeigvecs.

We conjecture that these eigenvectors represent an optimal solution for the entire relaxed objective. That is,
$\tr{\round{\Z{}^{\top} \Ag \Z{}}^2} \leq \sum_{j=1}^{\summarysize} \lambda_{j}^2$ (\textit{Conjecture 1}).
However, this upper bound is not straightforward to determine for general \summarysize values and arbitrary graphs. Lemma~\ref{lemma:nonzero_cross_terms} proves a non-constructive result identifying some cases when Conjecture 1 holds true.
\begin{restatable}[]{lemma}{nonzerocrossterms} \label{lemma:nonzero_cross_terms}
    Given $\summarysize \geq 2$ and a fixed adjacency matrix \Ag such that the largest magnitude eigenvalue has multiplicity $m < \summarysize$, there exist feasible non-eigenvector solutions $\Z{} = \squares{\v_1, \ldots, \v_{\summarysize}}$ such that $T_2$ in the relaxed objective (Equation~\ref{eq:vectorform}) is non-zero. Otherwise, if $m \geq \summarysize$, then $T_2 = 0$ and the eigenvector solution is optimal for Problem~\ref{prob:trace_maximization_relaxed}.
    \begin{equation}\label{eq:nonzero_cross_terms}
        \exists~ \Z{}, \text{ s.t. } \Z{}^\top \Z{} = \identitymat, \text{ and } T_2 = \sum\limits_{j=1}^{\summarysize} \sum\limits_{i \in \squares{\summarysize} \backslash \braces{j}} (\v^\top_j \Ag \v_i)^2 > 0
    \end{equation}
\end{restatable}

In other words, Lemma~\ref{lemma:nonzero_cross_terms} implies that there exist feasible orthonormal solutions \Z{} for Problem~\ref{prob:trace_maximization_relaxed} that are different from the eigenvector solution and the value of $T_2$ for these solutions is larger than the corresponding value of $T_2$ for the eigenvector solution (which is 0). Due to the non-constructive nature of the result, it is an open problem to obtain exact upper bounds for $T_1$ and $T_2$ in arbitrary graphs. So we design a heuristic algorithm called \opt to construct alternative candidates for such \Z{}. In Section~\ref{sec:experiments}, we provide empirical evidence supporting our conjecture. We show that the solution returned by \opt converges to the eigenvector solution.

\begin{algorithm}[tb]
\small

\nl {\bf Input}: Adjacency matrix \Ag of graph \graph; summary size \summarysize; error tolerance $\epsilon$; number of iterations \rounds.\\
\nl {\bf Output}: Feasible solution \Z{} for Problem~\ref{prob:trace_maximization_relaxed}. \\

\tcp*[h]{\textcolor{blue}{Initial feasible solution}}\\
\nl Draw a random matrix from $\R^{\numnodes \times \summarysize}$ as $R$ \\
\nl $\Z{}^{\round{0}} \gets \text{\textsc{QR-Decomposition}}\round{R}$ \\

\For {$t \gets 1 ~ \text{\textbf{to}} ~\rounds$}
{
    \tcp*[h]{\textcolor{blue}{Preparation for gradient ascent}}\\
    \nl Compute gradients \Gt (cf.~Equation~\ref{eq:trace_square_gradient})

    \nl Compute $\tau \gets$ \textsc{Newton-Line-Search}~\cite{wright1999numerical} \\
    \nl Compute skew-symmetric matrix \Pt (cf.~Equation~\ref{eq:skew_symmetric_matix}) \\
    \nl Compute new iterate \Yt (cf.~Equation~\ref{eq:cayley_transform}) \\

    \tcp*[h]{\textcolor{blue}{Update the current solution}}\\
    \nl $\Ztone \gets \Zt + \frac{\tau}{2} \Pt \round{\Zt + \Yt}$ (cf.~Equation~\ref{eq:crank_nicolson_update_scheme})

    \nl \textbf{if} $\frac{\FZ{\Ztone}-\FZ{\Zt}}{\FZ{\Zt}} \leq \epsilon$, \textbf{then break}.
}

\nl \textbf{return} $\Z{}^{\round{\rounds}}$

\caption{\opt (Relaxed Problem)}
\label{algo:optstiefelgbb}

\end{algorithm}

\spara{Orthogonality-Constrained Optimization Heuristic}
\revision{Our algorithm, \opt, is directly adapted from \citet{wen2013feasible}.}

The set of feasible solutions $\mathcal{M} = \{\Z{} \in \R^{\numnodes \times \summarysize} \,:\, \Z{}^\top \Z{} = \identitymat\}$ is called a Stiefel Manifold.
In problems involving such manifolds, there are usually no guarantees for obtaining the global maximizer~\cite{wen2013feasible}.
Our iterative heuristic solution then relies on constraint-preserving steepest ascent.
It proceeds as follows: As a first step, we construct a feasible initial solution denoted as \Zinit.
For instance, \Zinit may be the eigenvector solution obtained previously, or the \Q{} matrix from the QR decomposition of a random $\numnodes \times \summarysize$ matrix.
The second step comprises of $T$ iterations.
At each iteration $t \in \squares{T}$, we first compute the gradient of the objective function with respect to the current solution \Zt and then update \Zt.
\begin{restatable}[]{lemma}{gradienttrace} \label{lemma:gradienttrace}
    Given an adjacency matrix \Ag and a solution \Z{}, denote \G{} as the $\round{\numnodes \times \summarysize}$-dimensional gradient matrix of the trace objective with respect to \Z{}. Then, the $\round{i,j}$-th entry of the gradient is:
    \begin{equation}\label{eq:trace_square_gradient}
        \G{ij} = \frac{\partial~\tr{\round{\Z{}^{\top} \Ag \Z{}}^2}}{\partial~\Z{ij}} = \tr{2 (\Z{}^\top \Ag \Z{}) \times (\Z{}^\top\Ag \J{ij} + \J{ji}\Ag\Z{})}
    \end{equation}
    where \J{ij} is the single-entry matrix of appropriate dimensions whose $\round{i,j}$-th entry is 1 and all other entries are 0.
\end{restatable}

Given \Z{} and the gradient matrix \G{}, we define \P{} as:
\begin{equation}\label{eq:skew_symmetric_matix}
    \P{} = \G{} \Z{}^\top + \Z{} \G{}^\top
\end{equation}
Using steepest ascent, we find the best gradient direction and set the new solution as $\Ztone = \Zt + \tau \Pt \Zt$ where $\tau$ is the best step size computed using Newton's Line Search method (cf.~Algorithm 3.2~\cite{wright1999numerical}).
However, \Ztone may not necessarily be orthonormal.
Thus, we use the Cayley transformation as defined in \optstiefelgbb~\cite{wen2013feasible} to create the next constraint-preserving iterate, i.e.,
\begin{equation}\label{eq:crank_nicolson_update_scheme}
    \Ztone = \Zt + \frac{\tau}{2} \Pt \round{\Zt + \Yt}
\end{equation}
where \Yt is given by:
\begin{equation}\label{eq:cayley_transform}
    \Yt = \Zt \Qt \, \text{ and } \,
    \Qt = \round{\identitymat + \frac{\tau}{2}\Pt}^{-1}\round{\identitymat - \frac{\tau}{2}\Pt}
\end{equation}

\noindent
\citet{wen2013feasible} show that the update scheme in Equation~\ref{eq:crank_nicolson_update_scheme} preserves orthonormality, maintains a smooth curve for \Yt over $\tau$, and converges to a stationary point given sufficient iterations (cf.~Lemma 3~\cite{wen2013feasible}). Algorithm~\ref{algo:optstiefelgbb} presents the pseudocode for \opt.

\begin{algorithm}[tb]
\small

\nl {\bf Input}: Adjacency matrix \Ag of graph $\graph = \round{\vertexset, \edgeset}$; summary size \summarysize; 
% $\text{\textsc{Reassign}} \in \braces{\textsc{True}, \text{\textsc{False}}}$; number of rounds \rounds; 
number of samples per round \samples.\\
\nl {\bf Output}: Membership and density matrices \X{\S{}}, $A_{\S}$ of summary \S.\\

\tcp*[h]{\textcolor{blue}{Phase 1: Create initial node membership assignment}}\\
\nl $\Z{} \gets \lmeigvecs(\Ag, \summarysize) \text{ or } \opt(\Ag, \summarysize)$ \\

% \tcp*[h]{\textcolor{blue}{Assign nodes to supernodes}}\\
\nl $\X{}^{\round{0}} \gets \text{\kmeans}(\Z{}, \summarysize)$\\
\nl Compute the current best cost $\bestcost \gets \FZ{\X{}^{\round{0}}}$

\tcp*[h]{\textcolor{blue}{Phase 2: Update node memberships (optional)}}\\
\For{$r \gets 1 ~ \text{\textbf{to}} ~\rounds$}{
    \nl Sample \samples nodes from \vertexset without replacement\\
    \For{$v \in \braces{v_{1}, \ldots, v_{\samples}}$}{
        \nl Get the current supernode of $v$ as $\S(v)$\\
        \For{$j \in \squares{\summarysize} \backslash \braces{\S(v)}$}{
            \nl Reassign node $v$ to supernode $j$ \\
            \nl Build a temporary membership matrix $\tilde{\X{v}}$\\
            \nl Compute the new cost $\newcost \gets \FZ{\tilde{\X{v}}}$\\
            \If{$\newcost > \bestcost$}{
                \nl $\bestcost \gets \newcost$\\
                \nl Update the membership $\X{}^{\round{r}} \gets \tilde{\X{v}}$
            }
        }
    }
}
\nl $\X{\text{final}} \gets \X{}^{\round{\rounds}}$

% \tcp*[h]{\textcolor{blue}{Create \summarysize-summary}}\\
\nl Compute densities $A_{\S}$ (cf.~Equation~\ref{eq:summary_edge_weight})\\

\nl \textbf{return} \X{\text{final}}, $A_{\S}$

\caption{\specsumm}
\label{algo:specsumm}

\end{algorithm}

\subsection{The \specsumm Algorithm}
\label{subsec:integer_problem}

We now propose our algorithm called \specsumm which consists of two phases, namely \kmeans and \reassignment. In the first phase, \specsumm converts the relaxed solution (obtained previously) into an integral solution using \summarysize-means clustering. In the second (optional) phase, \specsumm improves the \summarysize-means solution using a greedy heuristic. We discuss each of these in further detail below.

\spara{$k$-Means Clustering}
A good-quality summary \S{}, as per Problem~\ref{prob:trace_maximization_integer}, implies placing nearby nodes in the same supernode and distant nodes in different supernodes. The final relaxed solution \ZT represents an embedding of nodes in \summarysize-dimensional Euclidean space such that the summarization objective is optimized. Let $a_1, \ldots, a_{\numnodes} \in \R^{\summarysize}$ denote this embedding of \numnodes points where $a_i$ is the $i$-th row of \ZT. To create supernodes, we use the continuous \kmeans algorithm which constructs a set of \summarysize centroids $c_{1}, \ldots, c_{\summarysize} \in \R^{\summarysize}$ such that the following cost function is minimized:
\begin{equation} \label{eq:kmeans_cost_function}
    \min_{c_1, \ldots, c_{\summarysize}} \quad \sum\limits_{i=1}^{\numnodes} \norm{a_i - c_{l(i)}}_2^{2}
\end{equation}
where $l(i)$ is the centroid closest to $a_i$.
Then, the $\round{i,j}$-th entry of the membership matrix \X{} is $1$ if node $l(i)=j$ and $0$ otherwise. Thus, each node is assigned to exactly one supernode.
% \footnote{As per our notation, \sls~\cite{riondato2017graph} sets $\ZT = \Ag$ and applies \kmeans on it.}
% That is, they create clusters based on the rows of the adjacency matrix of the original graph used as vector representations of the nodes. They analytically show that the loss of the summary obtained from \kmeans on \Ag is at most $4$ times as large as the loss of the best possible \summarysize-summary (cf.~Theorem 1). This depicts the connection between the two distinct problems of clustering and summarization.

\spara{Reassignment}
One limitation of using \kmeans alone is that it does not directly optimize the objective, \FZ{\Z{}}, in Problem~\ref{prob:trace_maximization_integer}.
To improve the quality of the summary returned by \kmeans, we propose \reassignment as a secondary heuristic.
Let \rounds denote the number of rounds.
In each round $r \in \squares{\rounds}$, we proceed as follows: Let $\X{}^{\round{r}}$ denote the current membership matrix. We randomly sample \samples nodes from \vertexset without replacement. For each sampled node $v$, we check if moving $v$ from its current supernode, say $\S(v)$, to another supernode improves the objective value ($\FZ{\X{}^{\round{r}}}$).
If yes, then we reassign $v$ to that supernode.
If there are more than one such candidate supernodes, we reassign $v$ to that supernode which results in the maximum increase in the current $\FZ{\X{}^{\round{r}}}$.
Otherwise, we do not reassign $v$.
At each step, and thus after \rounds rounds, this ensures that \reassignment returns a feasible solution that is at least as good as the solution obtained from \kmeans in the context of Problem~\ref{prob:trace_maximization_integer}.
Finally, we use the final membership matrix $\X{}^{\round{T}}$ to create the \summarysize-summary by computing edge densities according to Equation~\ref{eq:summary_edge_weight}. Algorithm~\ref{algo:specsumm} presents the pseudocode of \specsumm.

\spara{Time Complexity.}
The complexity of computing the top-\summarysize eigenvectors of a sparse symmetric matrix is $\bigObound\round{\numedges \summarysize t_1}$~\cite{baglama2005lanczos} where $t_1$ is the number of Arnoldi iterations. The complexity of computing a clustering using mini-batch \kmeans is $\bigObound\round{\numnodes \summarysize t_2}$ where $t_2$ is the number of clustering iterations~\cite{sculley2010web,yan2009fast}. Finally, computing the densities requires $\bigObound\round{\numedges}$ time~\cite{riondato2017graph}. Thus, the total computation complexity of our algorithm is $\bigObound\round{\numedges \summarysize t_1 + \numnodes \summarysize t_2}$. However, the widespread use and study of each of the components involved in \specsumm indicates that scaling summarization to massive graphs is feasible.

\begin{figure*}[ht]
	\captionsetup{skip=3pt}
	\captionsetup[sub]{skip=0pt}
	\centering
	\includegraphics[width=0.89\textwidth]{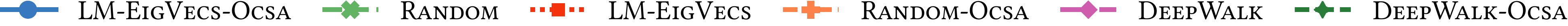}
	\\
	\begin{subfigure}{0.20\textwidth}
	  \centering
	  \includegraphics[width=\textwidth]{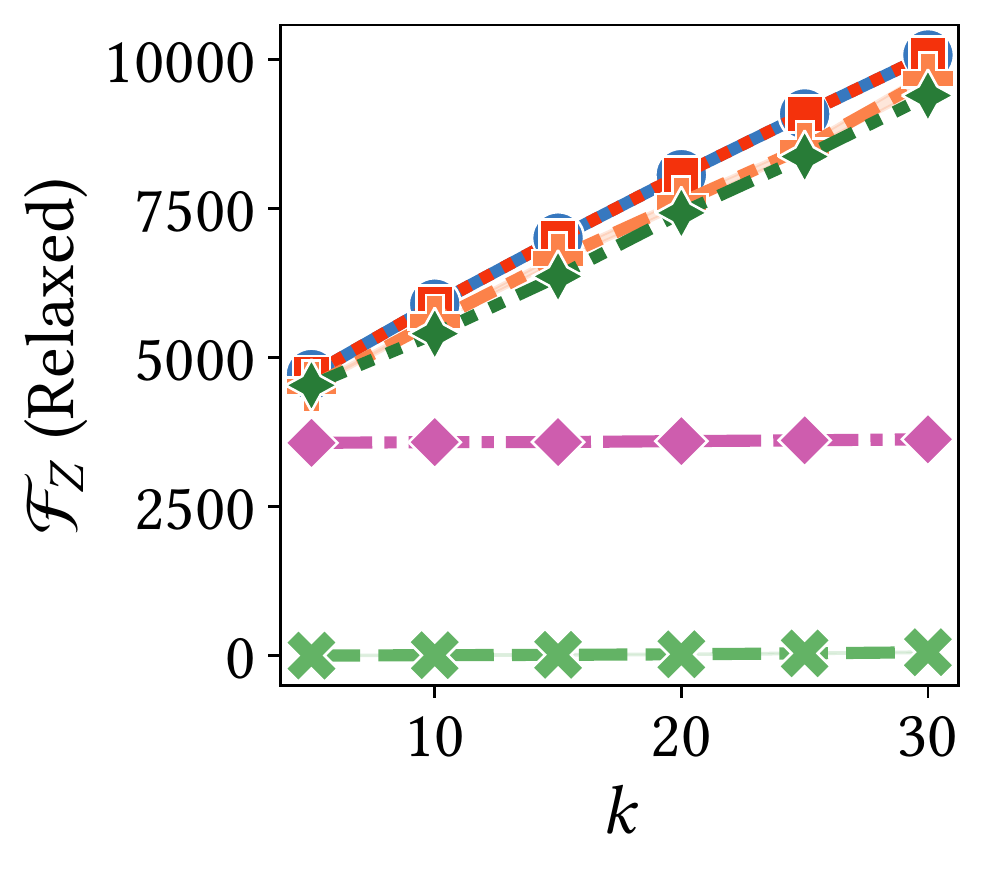}
	  \caption{\sbm}
	  \label{fig:SBM_relaxed_trace}
	\end{subfigure}
	\hspace{5mm}
	\begin{subfigure}{0.20\textwidth}
	  \centering
	  \includegraphics[width=\textwidth]{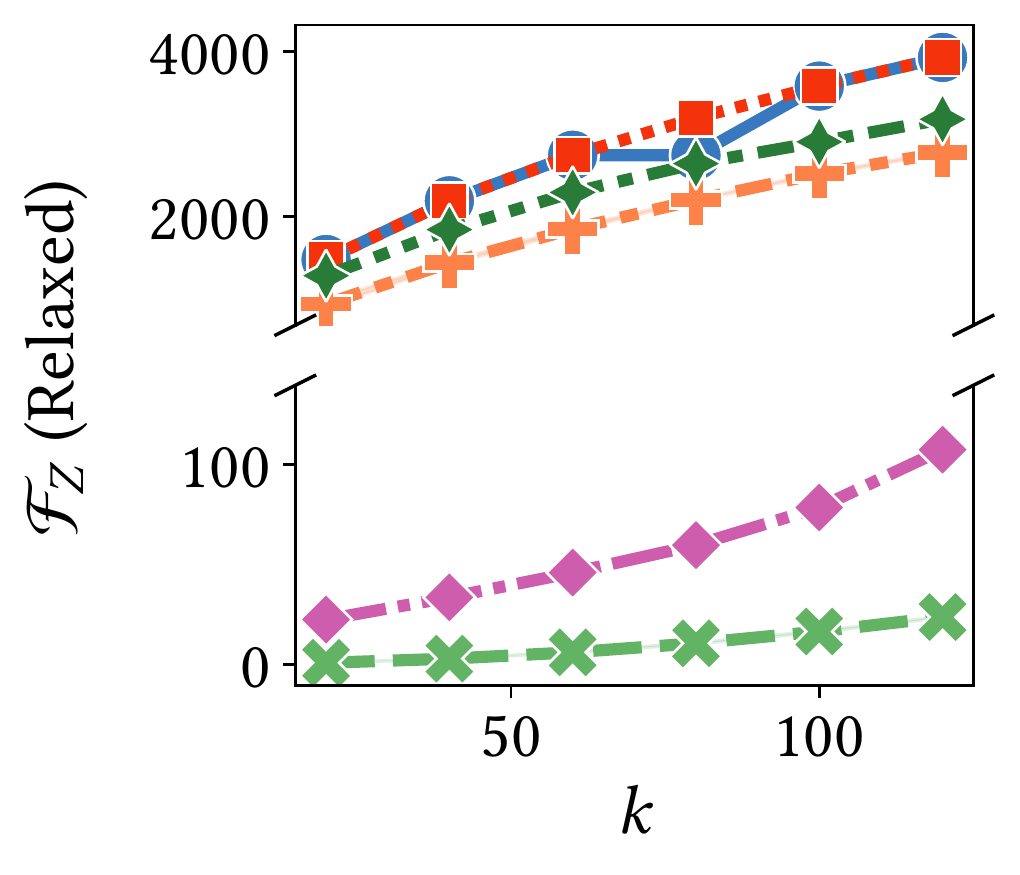}
	  \caption{\cora}
	  \label{fig:Cora_relaxed_trace}
	\end{subfigure}
	\hspace{5mm}
	\begin{subfigure}{0.20\textwidth}
	  \centering
	  \includegraphics[width=\textwidth]{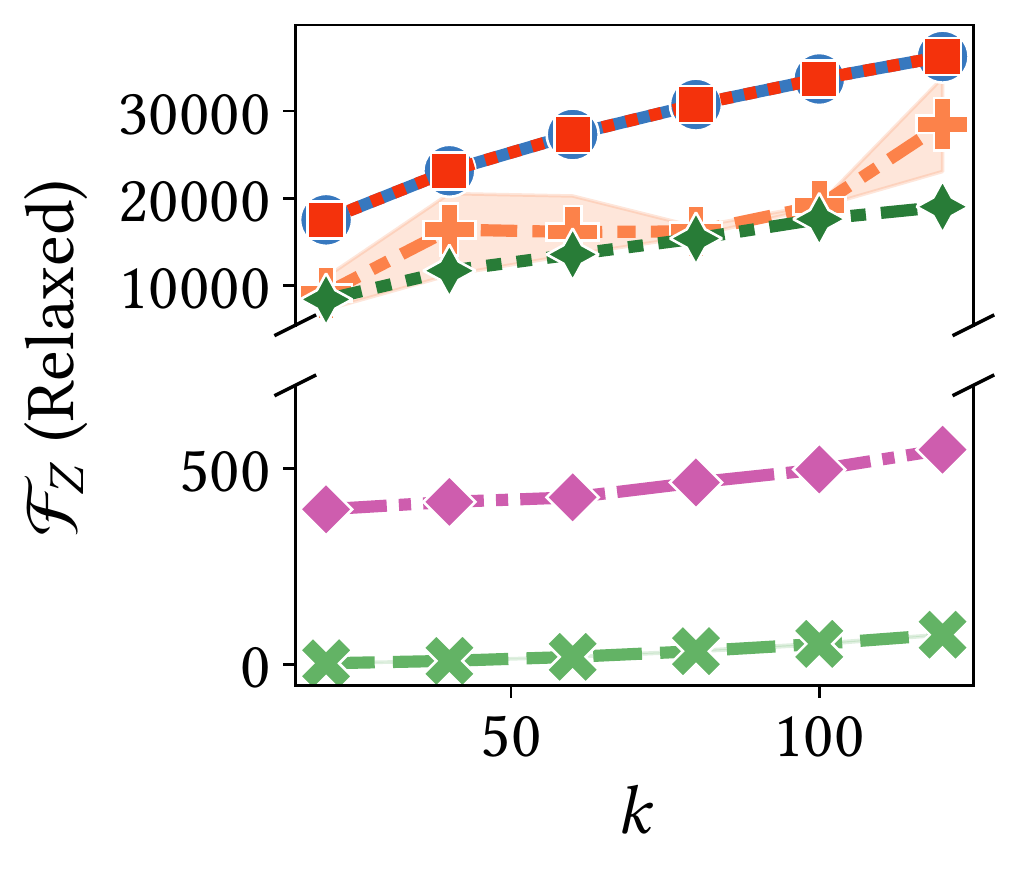}
	  \caption{\PPI}
	  \label{fig:PPI_relaxed_trace}
	\end{subfigure}
	\hspace{5mm}
	\begin{subfigure}{0.20\textwidth}
		\centering
		\includegraphics[width=\textwidth]{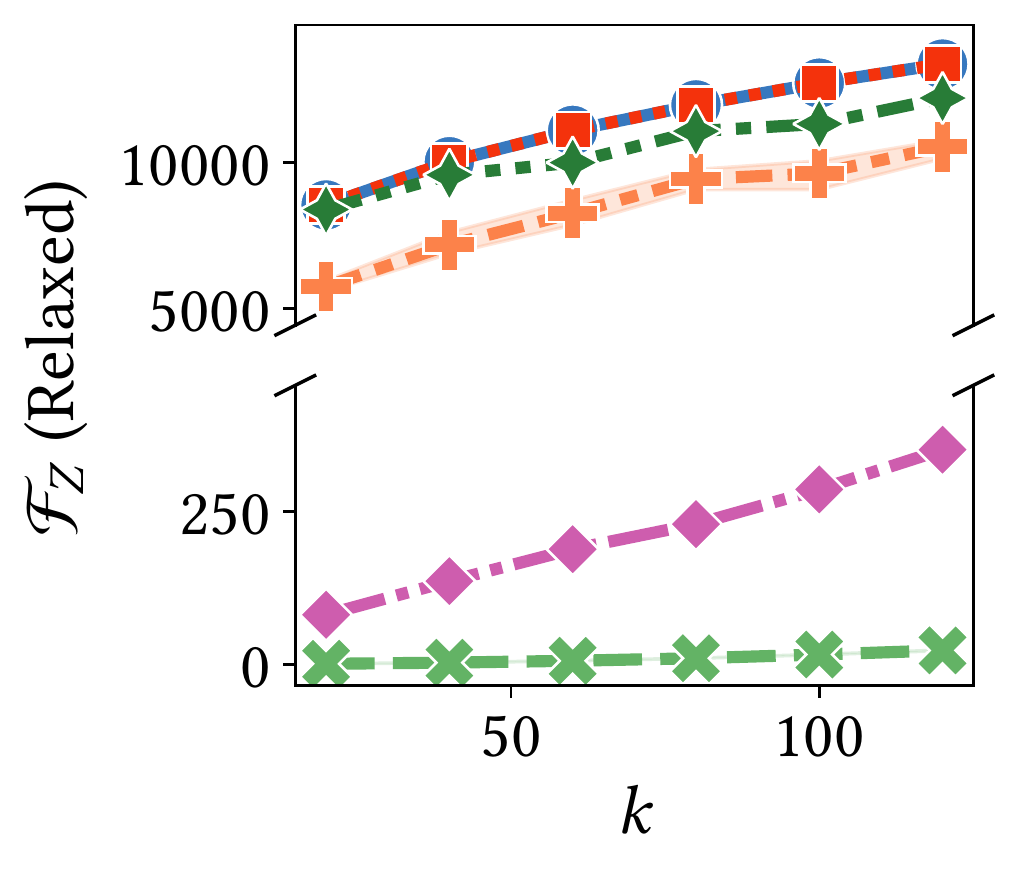}
		\caption{\caGrQc}
	    \label{fig:caGrQc_relaxed_trace}
    \end{subfigure}
	\\
	\begin{subfigure}{0.20\textwidth}
		\centering
		\includegraphics[width=\textwidth]{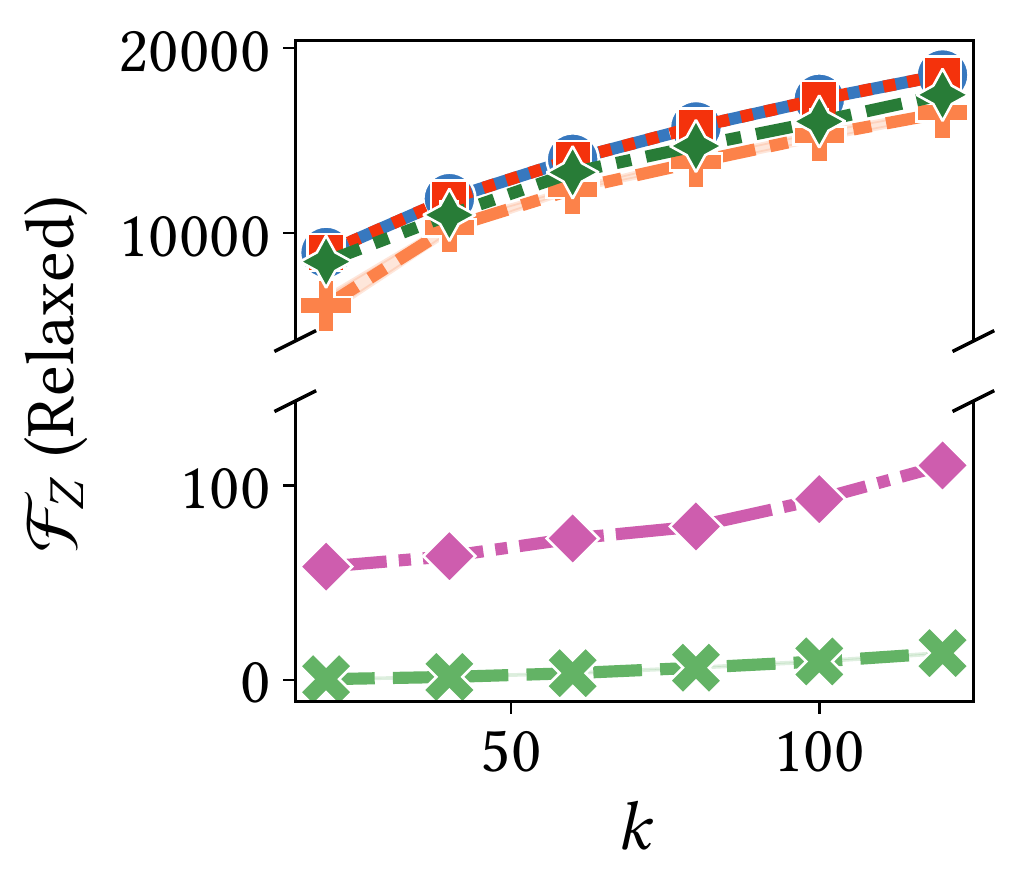}
		\caption{\lastfm}
		\label{fig:LastFM_Asia_relaxed_trace}
	\end{subfigure}
	\hspace{5mm}
	\begin{subfigure}{0.20\textwidth}
		\centering
		\includegraphics[width=\textwidth]{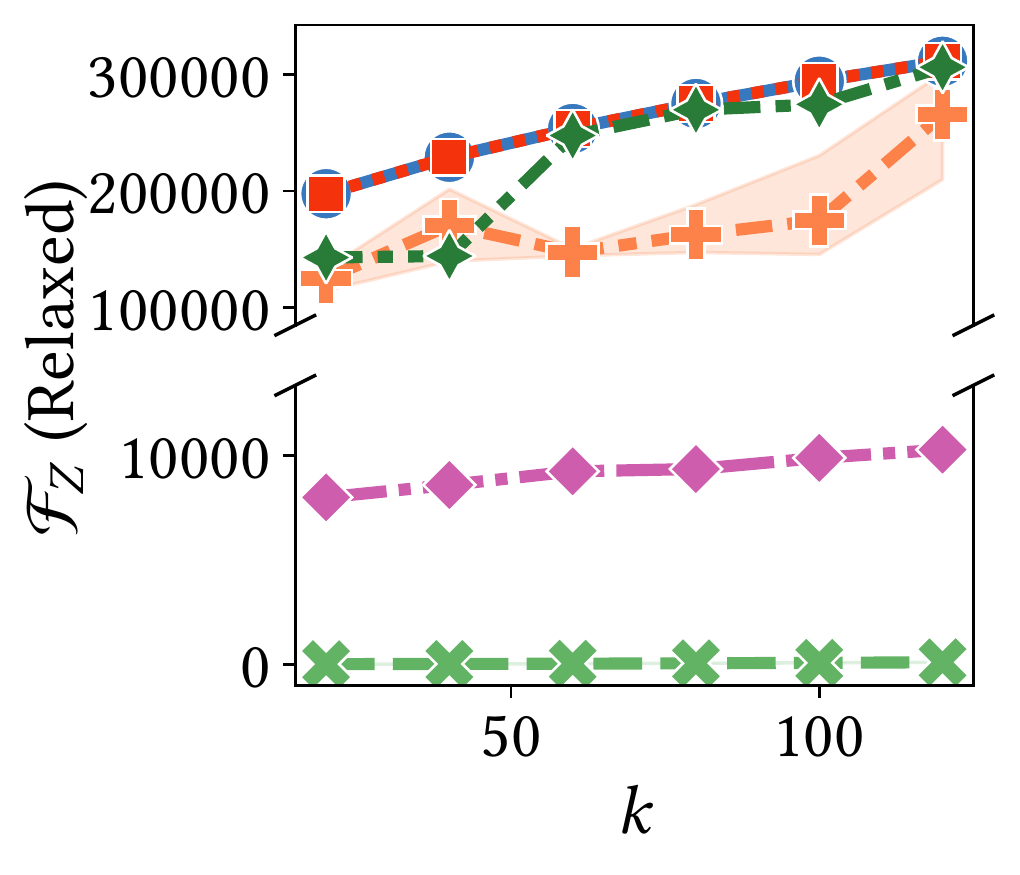}
		\caption{\blogcatalog}
	    \label{fig:Blogcatalog_relaxed_trace}
	\end{subfigure}
	\hspace{5mm}
	\begin{subfigure}{0.20\textwidth}
		\centering
		\includegraphics[width=\textwidth]{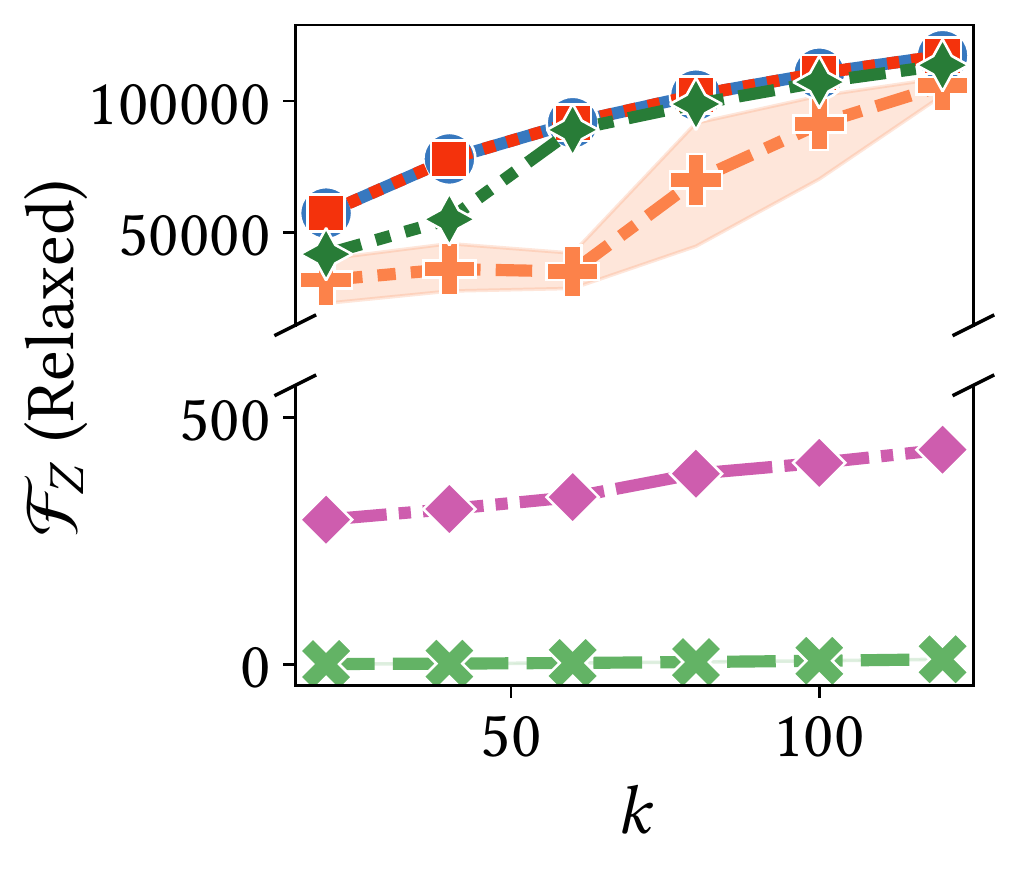}
		\caption{\facebook}
	    \label{fig:Facebook_relaxed_trace}
	\end{subfigure}
	\hspace{5mm}
	\begin{subfigure}{0.20\textwidth}
		\centering
		\includegraphics[width=\textwidth]{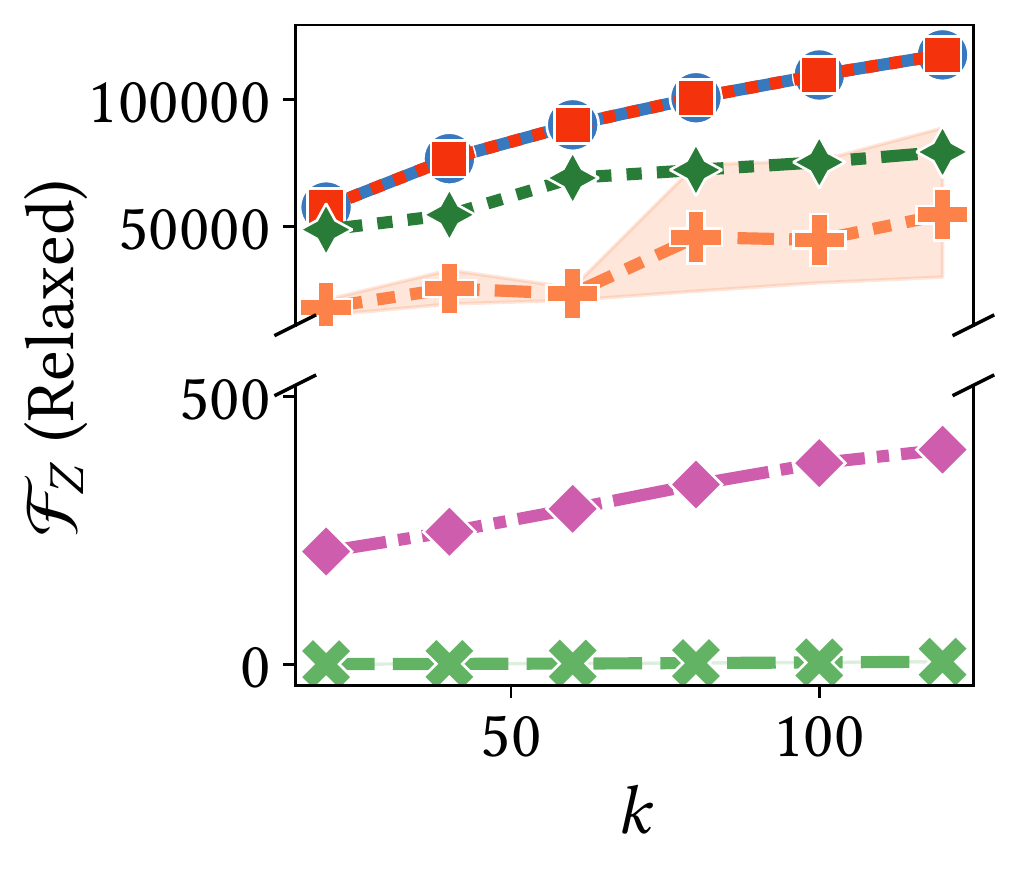}
		\caption{\enron}
	    \label{fig:Email_Enron_relaxed_trace}
	\end{subfigure}
	\caption{Objective value ($\FZ{\Z{}}$) of Problem~\ref{prob:trace_maximization_relaxed} with respect to summary size \summarysize for different variants of \lmeigvecs and \opt.}
	\label{fig:relaxed_baselines_objval}
	\Description{Experiment 1}
  \end{figure*}

\section{Experiments} \label{sec:experiments}

We perform extensive experiments to evaluate the efficacy of our algorithms.
Section~\ref{subsec:setup} describes our setup.
Section~\ref{subsec:results} presents our main results.
\revision{Extended results are deferred to Appendix B.}

\subsection{Setup}
\label{subsec:setup}

\begin{table}[t]
  \footnotesize
  \centering
  \caption{Dataset Statistics: number of nodes ($|\vertexset|$), number of edges ($|\edgeset|$), average degree ($d_{avg}$), density ($\rho$), diameter ($D$), clustering coefficient ($C$). 
  % $\dagger$ denotes graphs that are originally disconnected for which we use its largest connected component.}
  \revision{$\dagger$ denotes originally disconnected graphs for which we use their largest connected component.}}
  \label{tab:dataset_statistics}
  \vspace{-1em}
  \begin{tabular}{lcccccc}
    \toprule
    \multirow{2}[3]{*}{\textbf{Dataset}} & \multicolumn{2}{c}{\textbf{Size}} & \multicolumn{4}{c}{\textbf{Graph Properties}} \\
    \cmidrule(lr){2-3} \cmidrule(lr){4-7} & $|\vertexset|$ & $|\edgeset|$ & $ d_{avg} $ & $\rho$ & $D$ & $C$ \\
    \midrule
    \textsc{SBM}~\cite{abbe2015exact} & 1,000 & 29,872 & 59.74 & 5.98$\times 10^{-2}$ & 3 & 0.06 \\
    \revision{\textsc{Cora}$\dagger$~\cite{sen2008collective}} & 2,485 & 5,069 & 4.08 & 1.64$\times 10^{-3}$ & 19 & 0.24 \\
    \revision{\textsc{PPI}~$\dagger$\cite{qiu2018network}} & 3,852 & 37,841 & 19.65 & 5.10$\times 10^{-3}$ & 8 & 0.15 \\
    \revision{\textsc{ca-GrQc$\dagger$}~\cite{leskovec2007graph}} & 4,158 & 13,428 & 6.46 & 1.55$\times 10^{-3}$ & 17 & 0.56 \\
    \textsc{LastFM-Asia}~\cite{feather} & 7,624 & 27,806 & 7.29 & 9.57$\times 10^{-4}$ & 15 & 0.22 \\
    \revision{\textsc{BlogCatalog}$\dagger$~\cite{qiu2018network}} & 10,312 & 333,983 & 64.78 & 6.28$\times 10^{-3}$ & 5 & 0.46 \\
    \textsc{Facebook}~\cite{rozemberczki2019multiscale} & 22,470 & 171,002 & 15.22 & 6.77$\times 10^{-4}$ & 15 & 0.36 \\
    \revision{\textsc{email-Enron}$\dagger$~\cite{leskovec2009community}} & 33,696 & 180,811 & 10.73 & 3.19$\times 10^{-4}$ & 13 & 0.51 \\
    \textsc{Amazon}~\cite{leskovec2012define} & 334,863 & 925,872 & 5.52 & 1.65$\times 10^{-5}$ & 44 & 0.40 \\
    \textsc{Youtube}~\cite{leskovec2012define} & 1,134,890 & 2,987,624 & 5.26 & 4.63$\times 10^{-6}$ & 20 & 0.08 \\
    \textsc{Wikitalk}~\cite{leskovec2010predict} & 2,394,385 & 5,021,410 & 4.19 & 1.75$\times 10^{-6}$ & 9 & 0.05 \\
    \bottomrule
  \end{tabular}
\end{table}

\spara{Datasets}
We evaluate our algorithms on 11 publicly available datasets spanning various domains and with sizes ranging from 1K to 2.39M nodes.
\sbm~\cite{abbe2015exact} is a stochastic block model graph comprising of 20 clusters of 50 nodes each, with intra-cluster and inter-cluster probabilities set to $0.25$ and $0.05$, respectively.
\cora~\cite{sen2008collective} and \caGrQc~\cite{leskovec2007graph} are academic citation and collaboration networks.
\PPI~\cite{qiu2018network} is a protein-protein interaction network.
\lastfm~\cite{feather}, \blogcatalog~\cite{qiu2018network}, and \Youtube~\cite{leskovec2012define} are social networks.
\Amazon~\cite{leskovec2012define} is a product co-purchasing network.
\facebook~\cite{rozemberczki2019multiscale} is a web-graph of Facebook sites.
\enron~\cite{leskovec2009community} and \Wikitalk~\cite{leskovec2010predict} are communication networks.
\revision{If a graph is disconnected, we extract its largest connected component for our experiments.
Table~\ref{tab:dataset_statistics} summarizes the statistics of the processed datasets.}

\begin{figure}[t]
    \captionsetup{skip=3pt}
    \captionsetup[sub]{skip=0pt}
    \centering
    \includegraphics[width=0.475\textwidth]{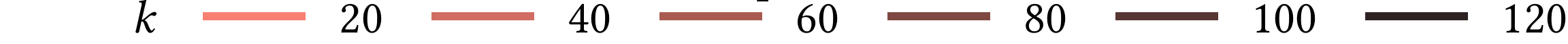}
    \\
    \begin{subfigure}{0.21\textwidth}
        \centering
        \includegraphics[width=\textwidth]{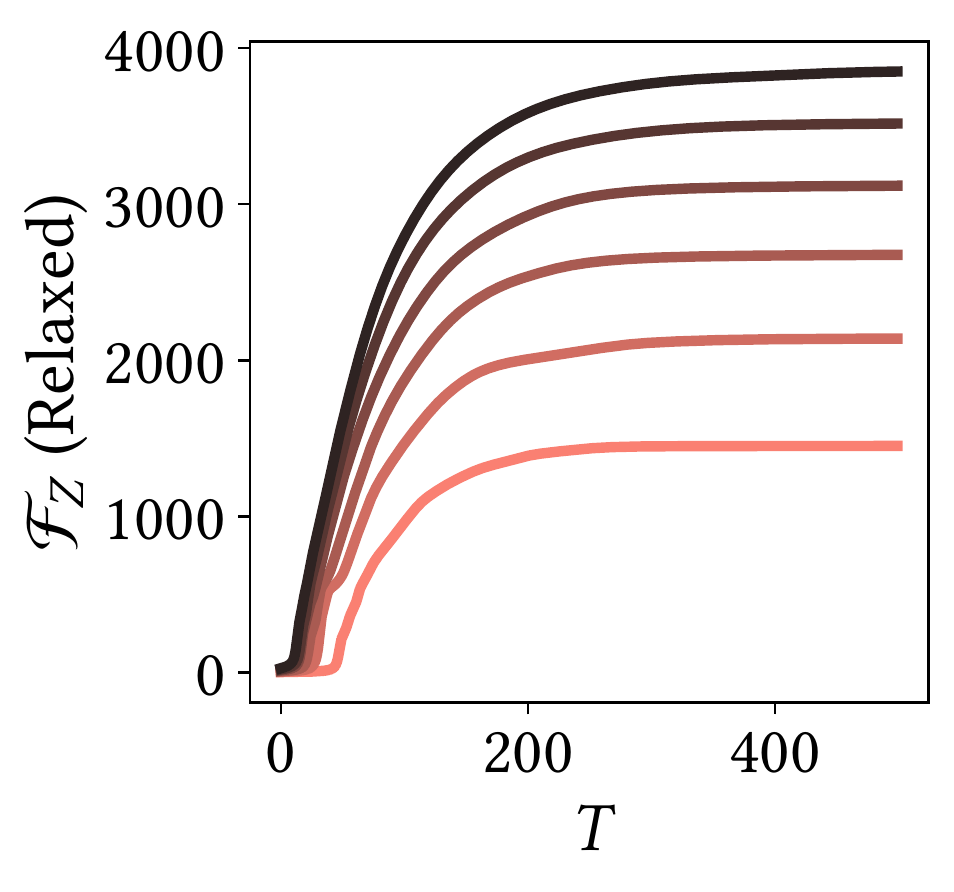}
        \caption{\cora}
        \label{fig:Cora_opt_iterations}
    \end{subfigure}
    \begin{subfigure}{0.21\textwidth}
        \centering
        \includegraphics[width=\textwidth]{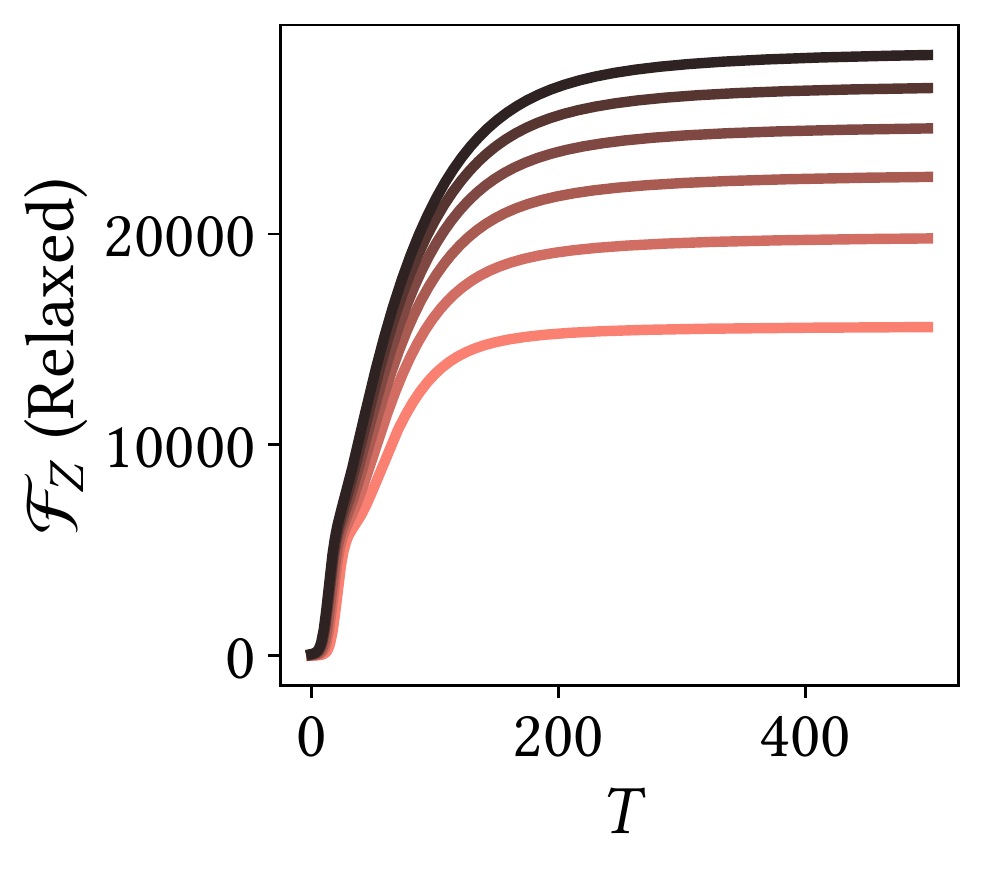}
        \caption{\PPI}
        \label{fig:PPI_opt_iterations}
    \end{subfigure}
    \caption{Objective value ($\FZ{\Z{}}$) of Problem~\ref{prob:trace_maximization_relaxed} as a function of the number of iterations (\rounds) for \random-\opt.}
    \label{fig:opt_iterations}
    \Description{Experiment 2}
\end{figure}

\spara{Algorithms}
We evaluate the following algorithms for the relaxed problem: (i) \lmeigvecs (Algorithm~\ref{algo:lm_eigvecs}), and three variants of \opt (Algorithm~\ref{algo:optstiefelgbb}) depending on the choice of the initial feasible solution, namely (ii) \lmeigvecs-\opt (largest-magnitude eigenvectors) (iii) \random-\opt (random QR matrix), and (iv) \deepwalk-\opt (QR decomposition of a DeepWalk~\cite{perozzi2014deepwalk} node embedding).

For the integer problem, we consider two variants of our algorithm: (i) \specsumm-\reassign and (ii) \specsumm that apply \kmeans on the eigenvectors with and without the \reassignment heuristic, respectively.
We compare against (iii) \deepwalk-\opt-\km (\kmeans on the relaxed solution returned by \deepwalk-\opt) and two state-of-the-art competitors (iv) \sls~\cite{riondato2017graph} and (v) \ssumm~\cite{lee2020ssumm}.

\spara{Parameter Setting}
We construct graph summaries of size $\summarysize \in \braces{5, 10, \dots, 30}$ for \sbm, $\summarysize \in \braces{20, 40, \dots, 120}$ for small graphs, and $\summarysize \in \braces{100, 250, 500, 1000, 2000, 5000}$ for large graphs.
Unless otherwise specified, the number of eigenvectors is set to \summarysize.
\opt is executed for $\rounds=100$ iterations with initial step size $\tau = 0.001$ and tolerance $\epsilon = 0.001$.
For fair comparison, all algorithms use the same Mini-Batch \kmeans algorithm by Sculley~\cite{sculley2010web} with \textsc{kmeans++} initialization.
% \footnote{The original C++ implementation of \sls~\cite{riondato2017graph} uses Lloyd's algorithm~\cite{lloyd1982least} for \kmeans with \textsc{kmeans++} initialization.}
For \reassignment, we set $\rounds=4$ (number of rounds) and $\samples=500$ (number of samples per round) for each dataset and \summarysize.
% While this does not result in the best possible \summarysize-summary, we find that it is sufficient for demonstrating its efficacy in improving the \kmeans solution. Further, we show that increasing \samples and \rounds leads to higher quality summaries.

\spara{Implementation}
We implement our algorithms in Python 3.
For \ssumm, we use the Java version by \citet{lee2020ssumm}.
All experiments were conducted on a Linux machine with 32 cores and 50GB RAM.
\revision{Our code is available at \url{https://version.helsinki.fi/ads/specsumm}.}

\begin{table*}[t]
    \captionsetup{skip=3pt}
    \captionsetup[sub]{skip=0pt}
    \centering
    \footnotesize
	\caption{Objective value, $\mathcal{F}_{Z}$ ($ \times 10^3$), of Problem~\ref{prob:trace_maximization_integer} for the summaries computed by each algorithm across different datasets. The values highlighted in blue denote the best quality and the underlined values denote the second-best quality.}
	\label{tab:fz:small}
	\begin{subtable}{.485\linewidth}
	    \centering
		\begin{tabular}{lcccccc}
		\toprule
		\multirow{2}[2]{*}{\textbf{Algorithm}} & \multicolumn{6}{c}{\textbf{$k$}} \\
		\cmidrule(lr){2-7}
		& $5$ & $10$ & $15$ & $20$ & $25$ & $30$ \\
		\midrule
        \textsc{SSumM} & 3.32 & 3.21 & 3.39 & 3.26 & 3.55 & 3.56 \\
        \textsc{S2L} & 3.57 & 3.59 & 3.61 & 3.62 & 3.63 & 3.64 \\
        \textsc{DeepWalk}-\textsc{Ocsa}-\textsc{KM} & 3.74 & 3.79 & 3.86 &  3.9 & 3.93 & 4.12 \\
        \textsc{SpecSumm} & \underline{3.88} & \underline{4.23} & \underline{4.56} & \underline{4.89} & \underline{5.08} & \underline{5.11} \\
        \textsc{SpecSumm}-\textsc{R} & \cellcolor{blue!25}3.97 & \cellcolor{blue!25}4.39 & \cellcolor{blue!25}4.86 & \cellcolor{blue!25}5.22 & \cellcolor{blue!25}5.42 & \cellcolor{blue!25}5.48 \\
        \bottomrule 
    \end{tabular} 
    \caption{\textsc{SBM}}
    \end{subtable}%
    \begin{subtable}{.485\linewidth}% 
		\centering% 
		\begin{tabular}{lcccccc} 
		\toprule 
		\multirow{2}[2]{*}{\textbf{Algorithm}} & \multicolumn{6}{c}{\textbf{$k$}} \\
		\cmidrule(lr){2-7} 
		& $20$ & $40$ & $60$ & $80$ & $100$ & $120$ \\ 
		\midrule 
	    \textsc{SSumM} & 0.33 & 0.51 & 0.74 & 0.88 & 0.86 & 0.96 \\
        \textsc{S2L} & 0.22 & 0.42 & 0.7 & 0.89 & 0.94 & 1.05 \\
        \textsc{DeepWalk}-\textsc{Ocsa}-\textsc{KM} & 0.32 & 0.85 & 1.09 & 1.3 & 1.37 & \underline{1.6} \\
        \textsc{SpecSumm} & \underline{0.49} & \underline{0.86} & \underline{1.25} & \underline{1.48} & \underline{1.43} & 1.59 \\
        \textsc{SpecSumm}-\textsc{R} & \cellcolor{blue!25}0.58 & \cellcolor{blue!25}1.03 & \cellcolor{blue!25}1.4 & \cellcolor{blue!25}1.7 & \cellcolor{blue!25}1.72 & \cellcolor{blue!25}1.9 \\
        \bottomrule 
    \end{tabular} 
    \caption{\textsc{Cora}}
    \end{subtable}% 
    \\\vspace{1mm}
    \begin{subtable}{.485\linewidth}% 
	    \centering% 
		\begin{tabular}{lcccccc} 
		\toprule 
		\multirow{2}[2]{*}{\textbf{Algorithm}} & \multicolumn{6}{c}{\textbf{$k$}} \\ 
		\cmidrule(lr){2-7} 
		& $20$ & $40$ & $60$ & $80$ & $100$ & $120$ \\ 
		\midrule 
	    \textsc{SSumM} & 2.27 & 2.21 &  1.98 &  2.58 &  2.51 &  3.12 \\
        \textsc{S2L} & 5.18 & 5.49 &  6.58 &  7.09 &  6.99 &  7.26 \\
        \textsc{DeepWalk}-\textsc{Ocsa}-\textsc{KM} & 3.78 & 5.09 &  5.11 &  5.22 &  5.64 &  5.43 \\
        \textsc{SpecSumm} & \underline{6.23} & \underline{8.0} &  \underline{9.65} &  \underline{9.94} & \underline{10.01} & \underline{10.49} \\
        \textsc{SpecSumm}-\textsc{R} & \cellcolor{blue!25}7.38 & \cellcolor{blue!25}9.6 & \cellcolor{blue!25}11.23 & \cellcolor{blue!25}11.92 & \cellcolor{blue!25}12.22 & \cellcolor{blue!25}12.96 \\
        \bottomrule 
    \end{tabular} 
    \caption{\textsc{PPI}}
    \end{subtable}% 
    \begin{subtable}{.485\linewidth}% 
		\centering% 
		\begin{tabular}{lcccccc} 
		\toprule 
		\multirow{2}[2]{*}{\textbf{Algorithm}} & \multicolumn{6}{c}{\textbf{$k$}} \\ 
		\cmidrule(lr){2-7} 
		& $20$ & $40$ & $60$ & $80$ & $100$ & $120$ \\ 
		\midrule 
	    \textsc{SSumM} & 6.03 & 6.52 & 7.01 & 7.42 & 7.79 & 7.86 \\
        \textsc{S2L} & 5.56 & 7.03 & 7.48 & 7.17 & 8.14 & 8.15 \\
        \textsc{DeepWalk}-\textsc{Ocsa}-\textsc{KM} & 5.88 & 6.87 & \underline{7.83} & \underline{7.96} & \cellcolor{blue!25}8.88 & \underline{8.81} \\
        \textsc{SpecSumm} & \underline{6.58} &       \underline{7.33} &  7.7 & 7.65 & 7.85 & 8.41 \\
        \textsc{SpecSumm}-\textsc{R} & \cellcolor{blue!25}6.67 & \cellcolor{blue!25}7.5 & \cellcolor{blue!25}7.98 & \cellcolor{blue!25}8.01 & \underline{8.32} & \cellcolor{blue!25}8.99 \\
        \bottomrule 
    \end{tabular} 
    \caption{\textsc{ca-GrQc}}
    \end{subtable}% 
    \\\vspace{1mm} 
    \begin{subtable}{.485\linewidth}% 
		\centering% 
		\begin{tabular}{lcccccc} 
		\toprule 
		\multirow{2}[2]{*}{\textbf{Algorithm}} & \multicolumn{6}{c}{\textbf{$k$}} \\ 
		\cmidrule(lr){2-7} 
		& $20$ & $40$ & $60$ & $80$ & $100$ & $120$ \\ 
		\midrule 
	    \textsc{SSumM} & 2.33 & 3.05 & 3.08 & 3.81 & 3.83 &  3.9 \\
        \textsc{S2L} & 3.34 & 4.22 & 5.01 & 5.34 & 6.02 & 6.22 \\
        \textsc{DeepWalk}-\textsc{Ocsa}-\textsc{KM} & 3.62 & 4.89 & 5.87 & \underline{6.87} & \underline{7.69} & \underline{8.31} \\
        \textsc{SpecSumm} & \underline{3.91} & \underline{5.28} & \underline{6.25} & 6.76 & 7.54 & 7.98 \\
        \textsc{SpecSumm}-\textsc{R} & \cellcolor{blue!25}3.99 & \cellcolor{blue!25}5.43 & \cellcolor{blue!25}6.48 & \cellcolor{blue!25}7.03 & \cellcolor{blue!25}7.94 & \cellcolor{blue!25}8.41 \\
        \bottomrule 
    \end{tabular} 
    \caption{\textsc{LastFM-Asia}}
    \end{subtable}% 
    \begin{subtable}{.485\linewidth}% 
		\centering% 
		\begin{tabular}{lcccccc} 
		\toprule 
		\multirow{2}[2]{*}{\textbf{Algorithm}} & \multicolumn{6}{c}{\textbf{$k$}} \\ 
		\cmidrule(lr){2-7} 
		& $20$ & $40$ & $60$ & $80$ & $100$ & $120$ \\ 
		\midrule 
	    \textsc{SSumM} & 70.78 &  70.87 &  65.88 &  64.42 &  67.62 &  67.44 \\
        \textsc{S2L} & \cellcolor{blue!25}96.52 & \underline{105.66} &  \underline{108.9} & 112.03 & 113.62 & 112.31 \\
        \textsc{DeepWalk}-\textsc{Ocsa}-\textsc{KM} & 58.14 &  59.56 & 106.68 &  \underline{114.8} & 109.96 & \underline{121.01} \\
        \textsc{SpecSumm} & 86.07 &  99.65 & 100.81 &  112.9 & \underline{116.07} & 116.83 \\
        \textsc{SpecSumm}-\textsc{R} & \underline{91.17} & \cellcolor{blue!25}107.19 & \cellcolor{blue!25}109.75 & \cellcolor{blue!25}121.27 & \cellcolor{blue!25}123.94 & \cellcolor{blue!25}124.88 \\
        \bottomrule 
    \end{tabular} 
    \caption{\textsc{Blogcatalog}}
    \end{subtable}% 
    \\ \vspace{1mm} 
    \begin{subtable}{.485\linewidth}% 
		\centering% 
		\begin{tabular}{lcccccc} 
		\toprule 
		\multirow{2}[2]{*}{\textbf{Algorithm}} & \multicolumn{6}{c}{\textbf{$k$}} \\ 
		\cmidrule(lr){2-7} 
		& $20$ & $40$ & $60$ & $80$ & $100$ & $120$ \\ 
		\midrule 
	    \textsc{SSumM} & 13.15 & 13.23 & 13.63 &  13.1 & \cellcolor{blue!25}60.68 & 61.05 \\
        \textsc{S2L} & 17.62 & 31.64 & 39.35 & 45.66 & 51.56 &  57.2 \\
        \textsc{DeepWalk}-\textsc{Ocsa}-\textsc{KM} & 20.05 & 34.25 & \cellcolor{blue!25}50.16 & 54.24 & 59.77 & \cellcolor{blue!25}64.32 \\
        \textsc{SpecSumm} & \underline{26.96} &  \underline{40.7} & 49.04 & \underline{54.26} & 59.65 & 63.48 \\
        \textsc{SpecSumm}-\textsc{R} & \cellcolor{blue!25}27.16 & \cellcolor{blue!25}40.95 & \underline{49.33} & \cellcolor{blue!25}54.64 &  \underline{60.1} &  \underline{64.0} \\
        \bottomrule 
    \end{tabular} 
    \caption{\textsc{Facebook}}
    \end{subtable}% 
    \begin{subtable}{.485\linewidth}% 
		\centering% 
		\begin{tabular}{lcccccc} 
		\toprule 
		\multirow{2}[2]{*}{\textbf{Algorithm}} & \multicolumn{6}{c}{\textbf{$k$}} \\ 
		\cmidrule(lr){2-7} 
		& $20$ & $40$ & $60$ & $80$ & $100$ & $120$ \\ 
		\midrule 
	    \textsc{SSumM} & 3.11 & 12.23 & 12.11 & 12.24 & 19.44 & 19.61 \\
        \textsc{S2L} & 16.92 & \cellcolor{blue!25}23.36 & \cellcolor{blue!25}25.99 & \cellcolor{blue!25}27.76 & \underline{29.23} & 31.57 \\
        \textsc{DeepWalk}-\textsc{Ocsa}-\textsc{KM} & 14.2 & 15.77 &  19.4 & 23.71 & 24.16 &  26.3 \\
        \textsc{SpecSumm} & \underline{17.25} &  21.0 & 23.81 & 27.09 & 29.02 & \underline{32.44} \\
        \textsc{SpecSumm}-\textsc{R} & \cellcolor{blue!25}17.4 & \underline{21.31} & \underline{24.09} & \underline{27.41} & \cellcolor{blue!25}29.43 & \cellcolor{blue!25}32.98 \\
        \bottomrule 
    \end{tabular} 
    \caption{\textsc{Email-Enron}}
    \end{subtable}% 
\end{table*}

\subsection{Experimental Results}
\label{subsec:results}

\spara{Results for the Relaxed Problem}
Figure~\ref{fig:relaxed_baselines_objval} presents the trace objective value (\FZ{\Z{}}) achieved by \lmeigvecs and \opt as a function of \summarysize.
As expected, \FZ{\Z{}} always increases with \summarysize.
\lmeigvecs attains the highest objective value across \summarysize in each dataset, with a maximum relative improvement of up to 52.12\% over the nearest competitor, \deepwalk-\opt ($\summarysize = 20$ on \PPI).
Also, \lmeigvecs-\opt achieves exactly the same value of \FZ{\Z{}} as \lmeigvecs because \opt always exits immediately after the first iteration (Line 10, Algorithm~\ref{algo:optstiefelgbb}) thereby implying that it cannot find an ascent step that improves the initial solution.
Lastly, we analyze the convergence of \opt on \cora and \PPI by allowing it to run for up to 500 iterations. While \opt significantly improves upon the naive variants, i.e., \random and \deepwalk, given sufficiently many iterations, it converges to the \FZ{\Z{}} value achieved by \lmeigvecs (cf.~Figure~\ref{fig:opt_iterations}).
This provides empirical support for our conjecture that eigenvectors are a stationary point representing at least a local maxima.

% Appendix~\ref{app:t1_t2_relaxed_experiments} reports additional results analyzing the individual sets of terms $T_1$ and $T_2$.
% Also, choosing the DeepWalk embedding as \Zinit results in improved performance across datasets over a random \Zinit.
% This improvement in the objective value ranges between 16031.91 vs. 15257.31 for $\summarysize = 100$ on \lastfm (4.83\% relative improvement) and 69194.66 and 23638.04 for $\summarysize = 60$ on \enron (65.84\% relative improvement).
% That is, for instance, after 500 iterations, \random-\opt attains $\FZ{\Z{}} = 1452.69$ while \lmeigvecs attains $\FZ{\Z{}} = 1482.05$ for $\summarysize = 20$ on \cora.
% This ranges from 38.04 seconds for $\summarysize = 20$ for \cora to 2.83 hours for $\summarysize = 120$ for \enron averaged over 5 random seeds.
% We also analyze the convergence and running time of \opt by allowing it to run for up to 500 iterations on the \cora and \PPI datasets.
% The results are included in the supplementary.
% However, this comes at the cost of lower efficiency.
% As expected, the larger the summary size and/or the graph, the longer it takes.

% # 4.83% relative improvement for DeepWalk-Opt over Random-Opt (16031.91 vs 15257.31 for k=100 on LastFM-Asia)
% # 65.84% relative improvement for DeepWalk-Opt over Random-Opt (69194.66 vs 23638.04 for k=60 on Email-Enron)
% # LM-EigVecs for k=20 on Cora = 1482.05
% # Random-Opt after 500 iterations for k=20 on Cora = 1452.69

\begin{figure}[t!]
  \centering
  \captionsetup{skip=3pt}
  \includegraphics[width=0.475\textwidth]{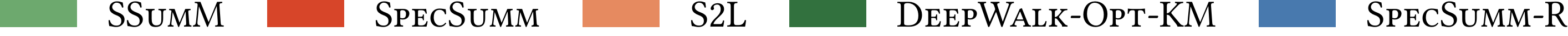}
  \\
  \includegraphics[width=0.42\textwidth]{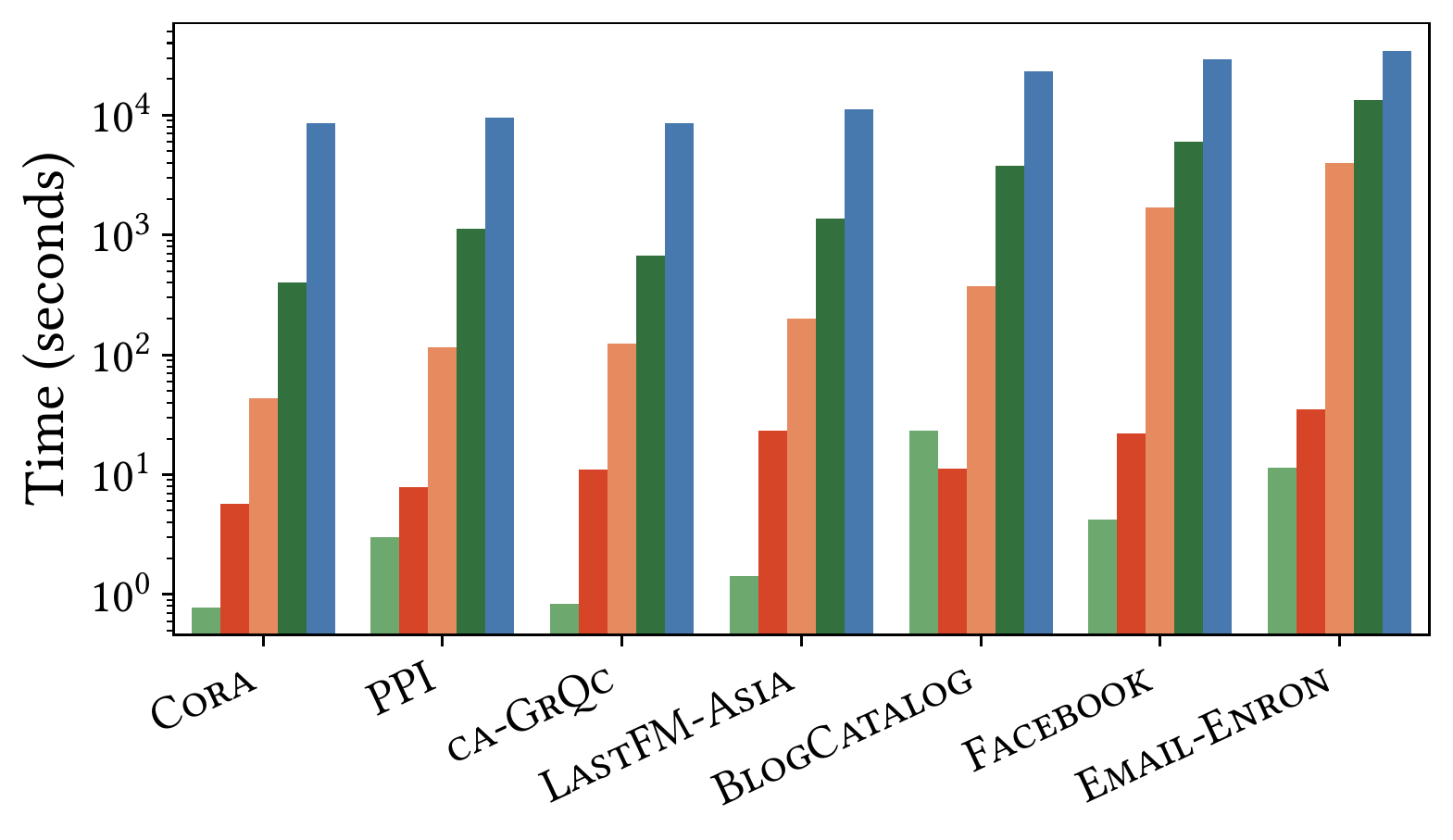}
  \caption{Running time (in log-scale) of different algorithms when the summary size $\summarysize = 120$.}
  \label{fig:total_runtimes}
  \Description{Experiment 7}
\end{figure}

\begin{figure*}[ht]
  \captionsetup{skip=0pt}
  \captionsetup[sub]{skip=0pt}
  \centering
  % \begin{subfigure}{0.163\textwidth}
  %   \centering
  %   \includegraphics[width=\textwidth]{fig/Amazon_Trace_ObjVal.pdf}
  %   \caption{\Amazon}
  %   \label{fig:Amazon_Trace_ObjVal}
  % \end{subfigure}
  % \begin{subfigure}{0.163\textwidth}
  %   \centering
  %   \includegraphics[width=\textwidth]{fig/Youtube_Trace_ObjVal.pdf}
  %   \caption{\Youtube}
  %   \label{fig:Youtube_Trace_ObjVal}
  %   \end{subfigure}
  %   \begin{subfigure}{0.163\textwidth}
  %   \centering
  %   \includegraphics[width=\textwidth]{fig/Wikitalk_Trace_ObjVal.pdf}
  %   \caption{\Wikitalk}
  %   \label{fig:Wikitalk_Trace_ObjVal}
  % \end{subfigure}
  % \begin{subfigure}{0.163\textwidth}
  %   \centering
  %   \includegraphics[width=\textwidth]{fig/Amazon_Total_Runtime.pdf}
  %   \caption{\Amazon}
  %   \label{fig:Amazon_Total_Runtime}
  %   \end{subfigure}
  %   \begin{subfigure}{0.163\textwidth}
  %   \centering
  %   \includegraphics[width=\textwidth]{fig/Youtube_Total_Runtime.pdf}
  %   \caption{\Youtube}
  %   \label{fig:Youtube_Total_Runtime}
  %   \end{subfigure}
  %   \begin{subfigure}{0.163\textwidth}
  %   \centering
  %   \includegraphics[width=\textwidth]{fig/Wikitalk_Total_Runtime.pdf}
  %   \caption{\Wikitalk}
  %   \label{fig:Wikitalk_Total_Runtime}
  %   \end{subfigure}  
    \includegraphics[width=\textwidth]{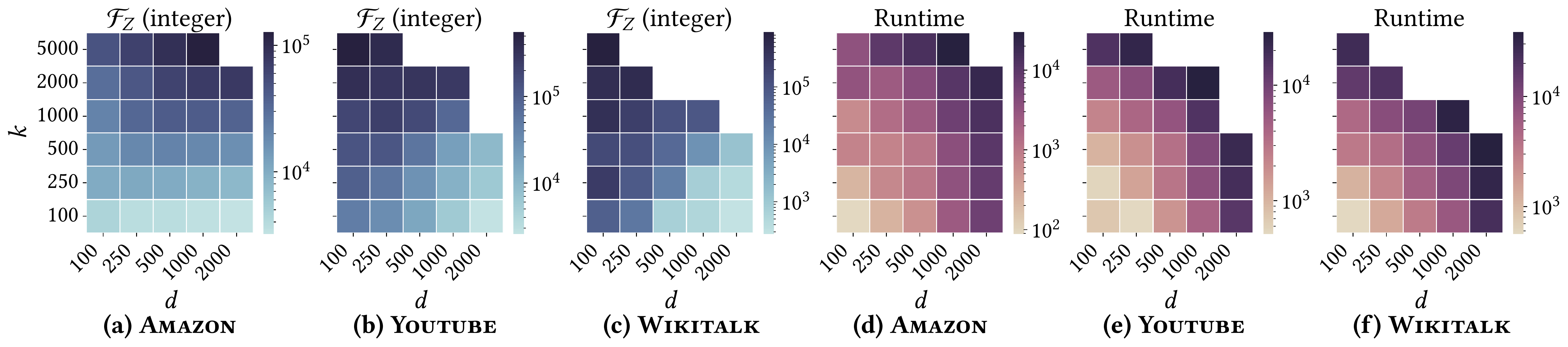}

  \caption{Trade-off between number of eigenvectors ($d$) and summary size ($k$) for the trace objective value ($\mathcal{F}_{Z}$) for \textsc{Amazon}, \textsc{Youtube}, and \textsc{Wikitalk}. Darker shades of blue and red represent higher quality and longer running times, respectively.}%Good quality summaries of larger sizes may be constructed from fewer eigenvectors, but smaller summaries constructed from higher number of eigenvectors are poorer in quality.}
  \label{fig:runtime_objval_big_graph}
  \Description{Experiment 7}
  \end{figure*}

\begin{figure*}[ht]
  \captionsetup{skip=0pt}
  \captionsetup[sub]{skip=0pt}
  \centering
  \begin{subfigure}{0.48\textwidth}
    \centering
    \includegraphics[width=0.4\textwidth]{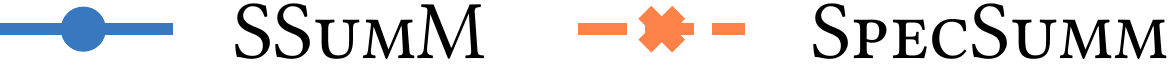}
    \label{fig:legend_scalable_lineplot_compare}
  \end{subfigure}
  \begin{subfigure}{0.48\textwidth}
    \centering
    \includegraphics[width=0.4\textwidth]{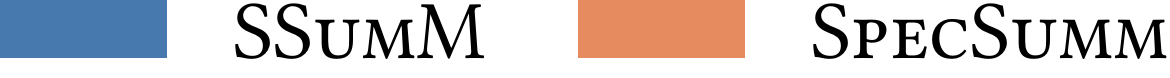}
    \label{fig:legend_scalable_barplot_compare}
  \end{subfigure}
  \\
  \begin{subfigure}{0.163\textwidth}
    \centering
    \includegraphics[width=\textwidth]{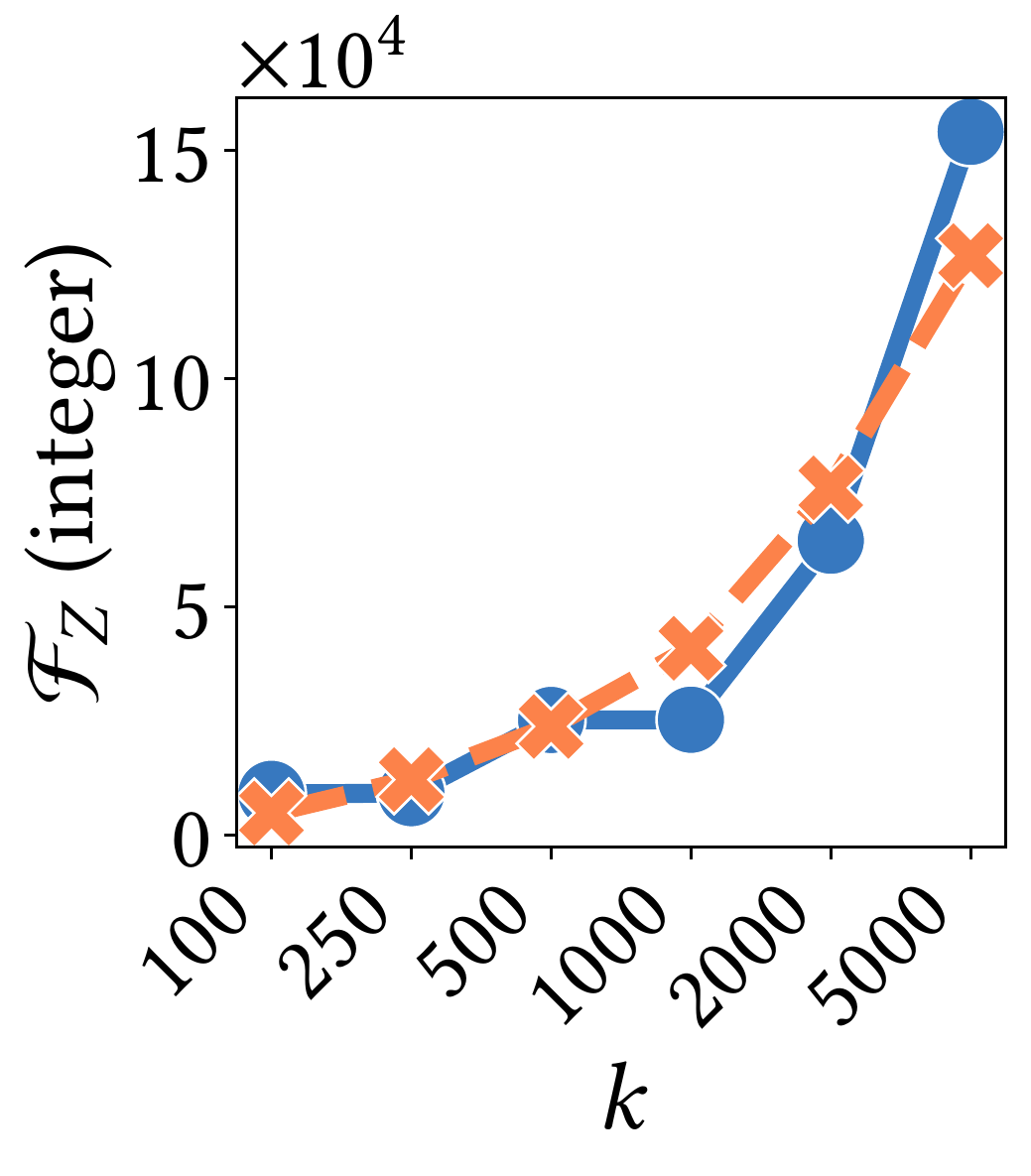}
    \caption{\Amazon}
    \label{fig:compare_Amazon_Trace_ObjVal}
  \end{subfigure}
  \begin{subfigure}{0.163\textwidth}
    \centering
    \includegraphics[width=\textwidth]{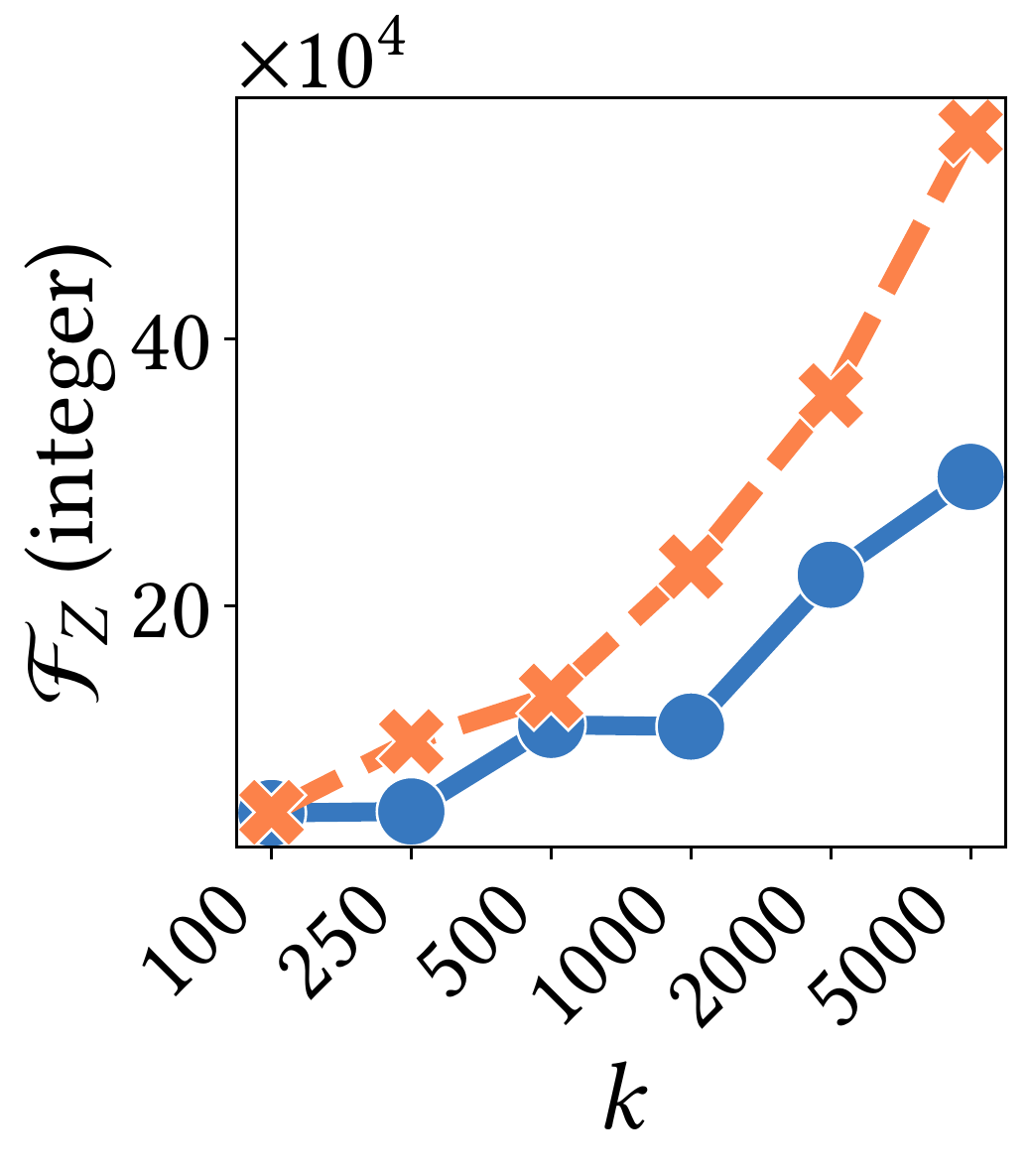}
    \caption{\Youtube}
    \label{fig:compare_Youtube_Trace_ObjVal}
    \end{subfigure}
    \begin{subfigure}{0.163\textwidth}
    \centering
    \includegraphics[width=\textwidth]{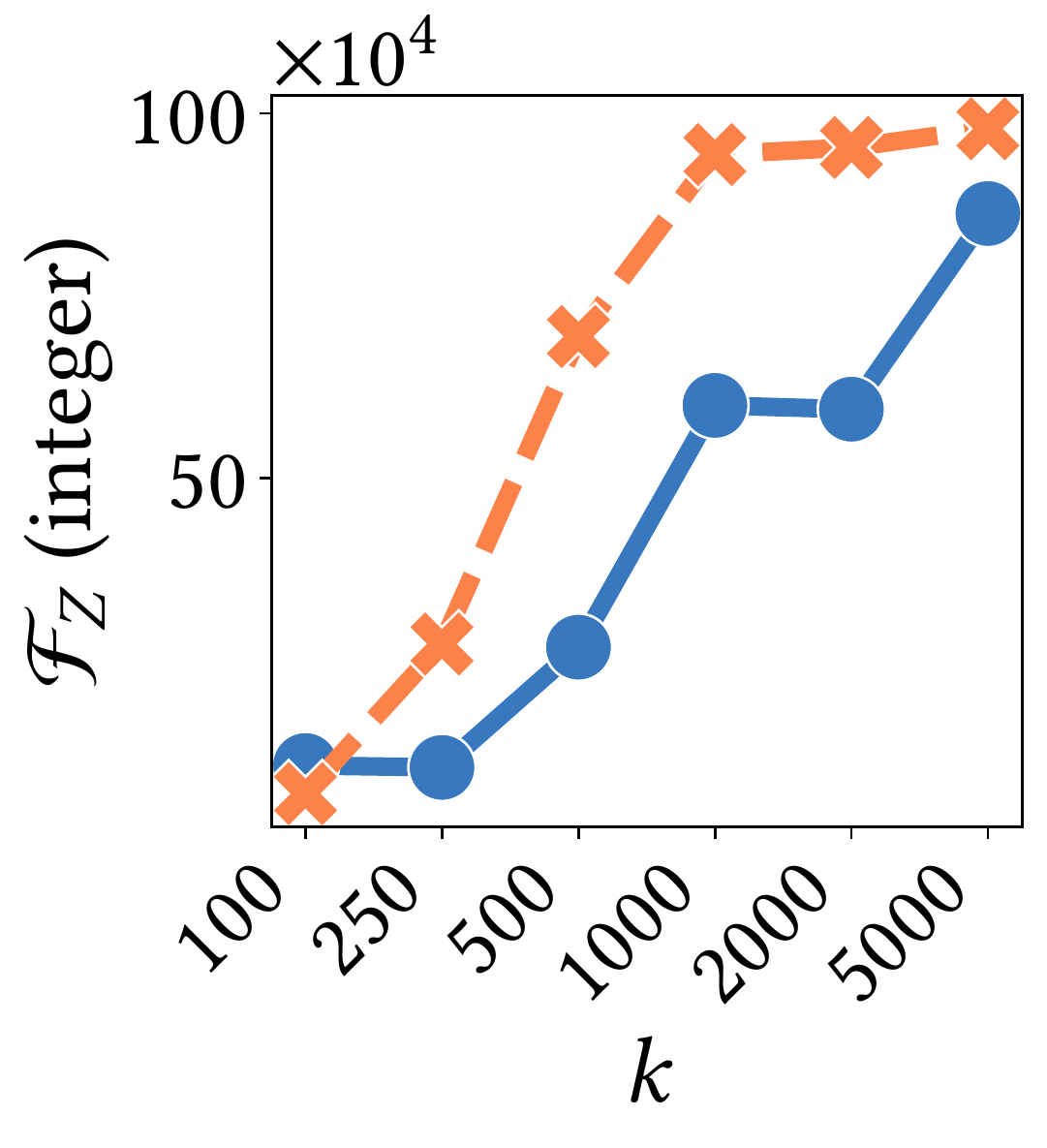}
    \caption{\Wikitalk}
    \label{fig:compare_Wikitalk_Trace_ObjVal}
  \end{subfigure}
  \begin{subfigure}{0.163\textwidth}
    \centering
    \includegraphics[width=\textwidth]{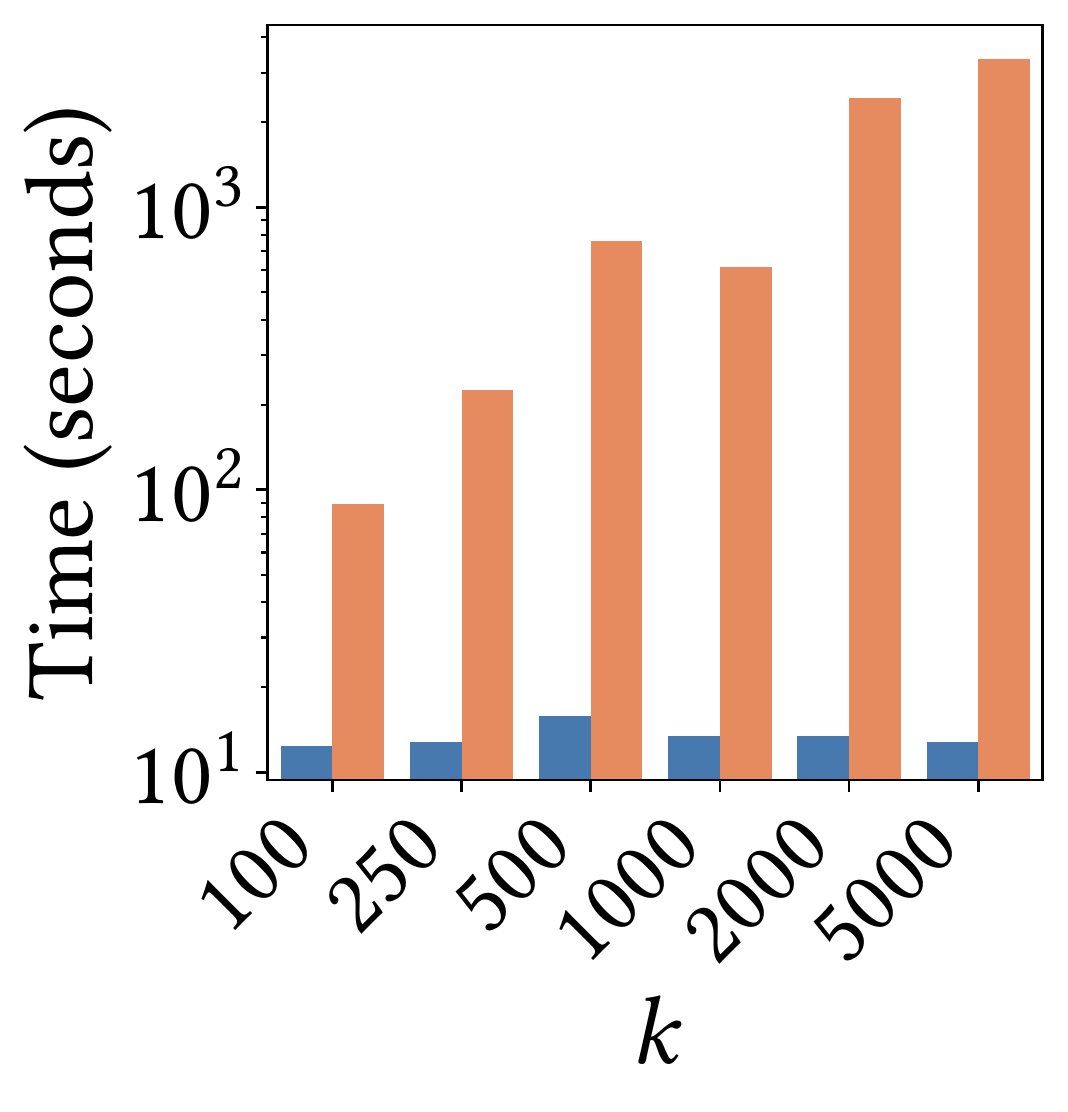}
    \caption{\Amazon}
    \label{fig:compare_Amazon_Total_Runtime}
    \end{subfigure}
    \begin{subfigure}{0.163\textwidth}
    \centering
    \includegraphics[width=\textwidth]{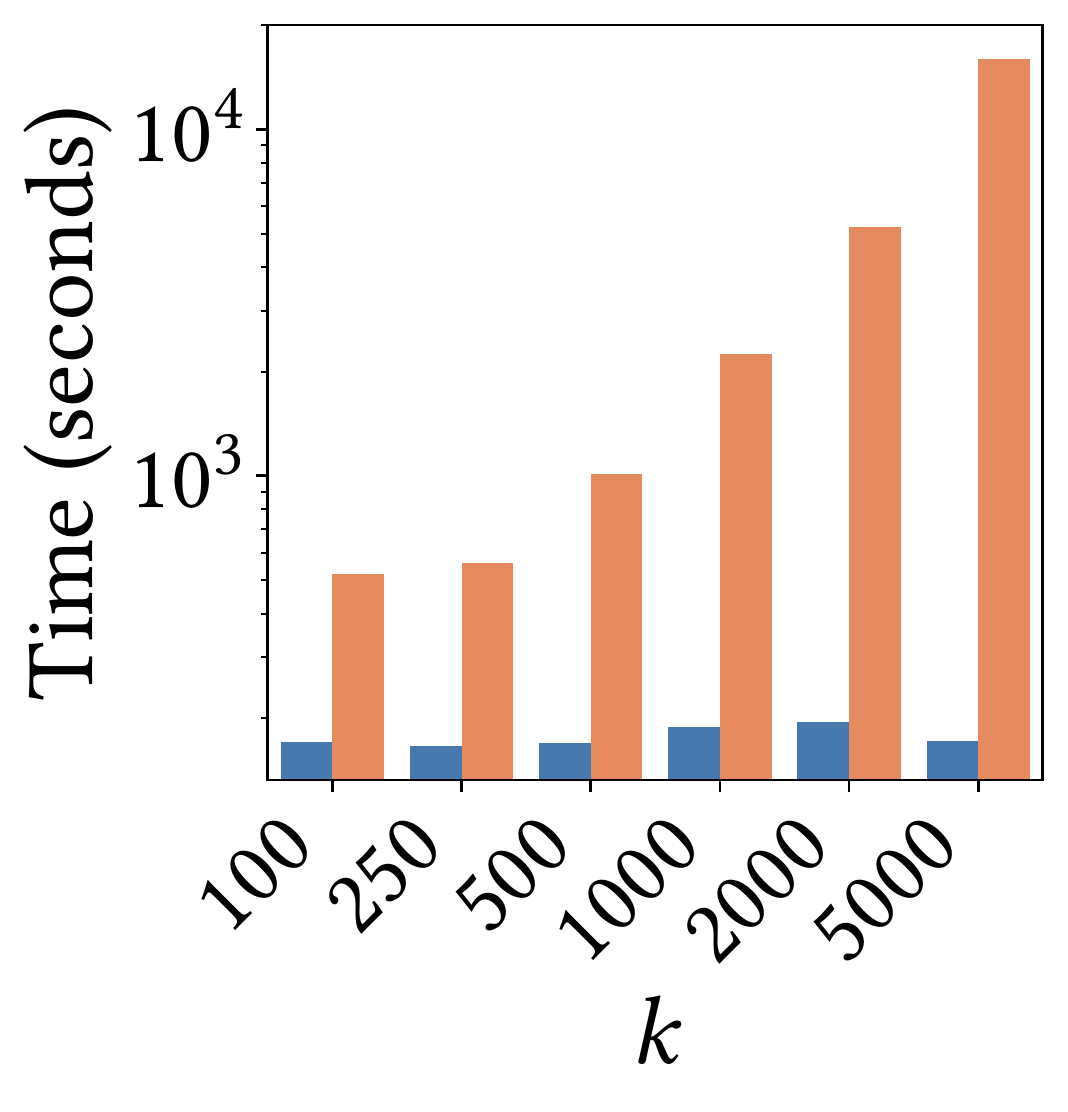}
    \caption{\Youtube}
    \label{fig:compare_Youtube_Total_Runtime}
    \end{subfigure}
    \begin{subfigure}{0.163\textwidth}
    \centering
    \includegraphics[width=\textwidth]{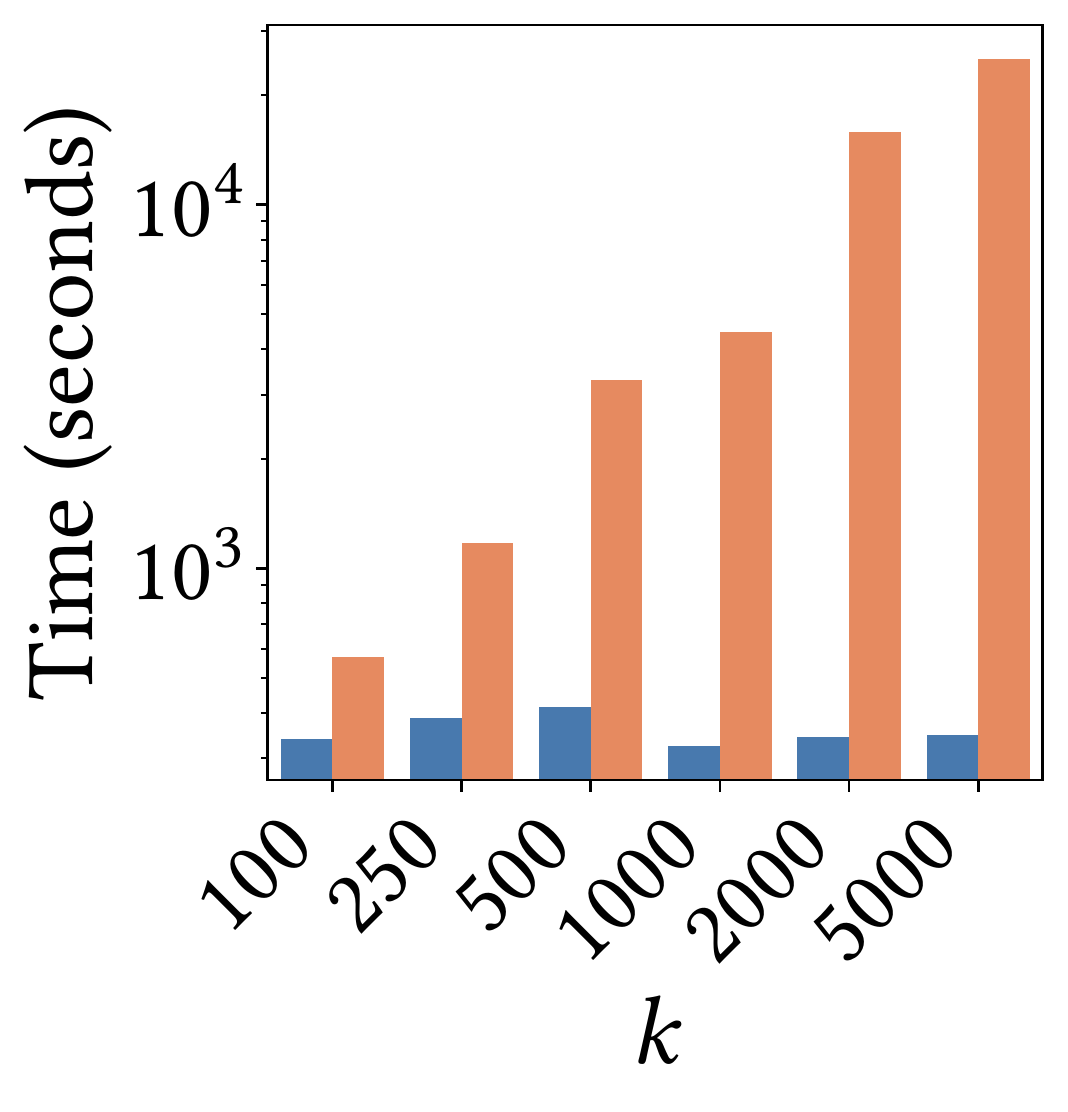}
    \caption{\Wikitalk}
    \label{fig:compare_Wikitalk_Total_Runtime}
    \end{subfigure}

  \caption{Comparison between \specsumm and \ssumm in terms of $\mathcal{F}_{Z}$ and construction time as a function of $k$.}
  \label{fig:compare_ssumm_specsumm_scalable}
  \Description{Experiment 8}
  \end{figure*}

% We evaluate the performance of the algorithms based on two criteria namely, (i) the quality of the summary as measured by the objective in Problem~\ref{prob:trace_maximization_integer}, and (ii) the time required to construct the summary. We do not explicitly consider storage as a criteria because, by definition of the hypergraph summary, for a fixed \summarysize, all algorithms require the same number of bits. 

\spara{Summary Quality}
Table~\ref{tab:fz:small} reports \FZ{\Z{}} values (averaged over 5 random seeds) of Problem~\ref{prob:trace_maximization_integer} attained by each algorithm for varying summary sizes \summarysize on different datasets.
\specsumm-\reassign outperforms other algorithms across datasets while \specsumm mostly achieves the second highest values.
\specsumm-\reassign is particularly effective on \PPI where an improvement of upto 23.5\% over \specsumm is achieved.
\specsumm itself consistently produces higher-quality summaries than \sls -- up to 51.1\% on smaller graphs like \cora for $\summarysize = 120$.
% Also, \specsumm-\reassign consistently improves the quality of summaries upon \specsumm in all cases.
Moreover, the summary quality of \ssumm is (upto 76.1\%) inferior to that of \specsumm as \ssumm over-sparsifies the original graph by minimizing aggregate (over entire $\Ag$) error and destroying topological structure.
The results for $l_2$-reconstruction errors are included in Appendix B.
As shown in Section~\ref{sec:background}, the problems of trace maximization and $l_2$-loss minimization are theoretically equivalent, and thus the results in terms of both objective values are consistent, i.e., any summary attaining a higher \FZ{\Z{}} value than another summary must have a smaller $l_2$-loss as well. 

Beyond aggregate measures such as \FZ{\Z{}}, we also evaluate quality based on estimates for typical graph queries such as the number of triangles (cf.~Appendix B) recovered from the summary. \specsumm provides consistently more accurate estimates than \sls and \ssumm. This further confirms the practical applicability of our approach.

% This ranges from 38.04 seconds for $\summarysize = 20$ for \cora to 2.83 hours for $\summarysize = 120$ for \enron averaged over 5 random seeds.
\spara{Runtime}
Figure~\ref{fig:total_runtimes} presents the average runtime (in seconds) of different algorithms.
\revision{Due to space constraints}, we only provide the results for $\summarysize=120$.
Generally, larger graph and summary sizes indicate longer running times as well.
\specsumm is up to 200\texttimes\ faster than \sls on \enron while still providing a summary of better quality.
On the other hand, \deepwalk-\opt-\km and \specsumm-\reassign are over two orders of magnitude slower than \specsumm, requiring approximately $4$ and $10$ hours on \enron, respectively.
Such high overhead for \deepwalk-\opt-\km comes from the expensive gradient computation of \opt.
And \specsumm-\reassign is slow since it cannot be parallelized and requires recomputing \FZ{\Z{}} during each iteration (Line~10, Algorithm~\ref{algo:specsumm}).
The low efficiencies make both algorithms impractical when the graph sizes are large.
Finally, although \ssumm runs much faster than \specsumm on small graphs (e.g., \cora and \caGrQc), the gaps in time efficiency reduce when the graph size is larger (e.g., \blogcatalog and \enron).
% Also note that the summary quality of \ssumm is inferior to \specsumm in almost all cases.

\spara{Scalability}
% We evaluate the scalability of \specsumm by creating summaries for the three largest graphs, namely \Amazon (\textasciitilde 330K nodes), \Youtube (\textasciitilde 1.1M nodes), and \Wikitalk (\textasciitilde 2.3M nodes).
% We set 12 hours as the time limit in each setting and, as before, use out-of-box implementations to compute eigenvectors and clusters.
We evaluate the scalability of \specsumm by creating summaries for the three largest graphs, namely \Amazon, \Youtube, and \Wikitalk. For these experiments, we set 12 hours as the time limit in each setting and parameter configuration.

Previously, given summary size \summarysize as the only input parameter, we computed \summarysize eigenvectors.
However, it is possible to decouple the number of eigenvectors (say, \neigs) from \summarysize.
We trade off summary quality for efficiency by using fewer than \summarysize eigenvectors.
Figure~\ref{fig:runtime_objval_big_graph} depicts the 
\FZ{\Z{}} and the total running time of \specsumm as a function of \neigs and \summarysize, respectively.
% Figures~\ref{fig:Amazon_Trace_ObjVal}--\ref{fig:Wikitalk_Trace_ObjVal} and Figures~\ref{fig:Amazon_Total_Runtime}--\ref{fig:Wikitalk_Total_Runtime} depict the trace objective (\FZ{\Z{}}) and the total running time of \specsumm as a function of \neigs and \summarysize, respectively.
Darker colors in blue and red indicate higher quality and longer times, accordingly.
The missing regions indicate parameter settings for which \specsumm did not complete within 12 hours.
Results for \sls, \deepwalk-\opt-\km, and \specsumm-\reassign are omitted since they did not finish within 12 hours.

\specsumm builds small summaries of large graphs very quickly.
For $\summarysize=100$ and $\neigs=100$, it only takes 89 and 569 seconds on \Amazon and \Wikitalk, respectively.
As graph size increases, \lmeigvecs scales reasonably while \kmeans is comparatively slower. For \Wikitalk, \lmeigvecs requires up to 5.7 hours to compute 2000 eigenvectors whereas \kmeans taking up to 6.9 hours to create $\summarysize=5000$ clusters when $\neigs=100$.
% \footnote{Due to space constraints, we defer full results to the supplementary material.}
However, choosing appropriate values for \neigs and \summarysize hugely affects quality.
% We make the following observations.
For a fixed \summarysize, increasing \neigs up to \summarysize improves $\FZ{\Z{}}$ values.
Conversely, constructing smaller summaries from larger number of eigenvectors results in even lower values of \FZ{\Z{}}.
However, there exist intermediary settings for \neigs and \summarysize that offer the best trade-off between summary quality and efficiency.
That is, smaller summaries based on higher number of eigenvectors can be constructed up to 17\texttimes\ faster than larger summaries based on fewer eigenvectors while having comparable quality.

Finally, we compare \specsumm with \ssumm on the three largest graphs. Figure~\ref{fig:compare_ssumm_specsumm_scalable} reports \FZ{\Z{}} and runtimes, respectively. While \ssumm runs upto 3 orders of magnitude faster, its summary quality is (upto 2.3\texttimes) worse than that of \specsumm.
Note that we choose the compression ratios such that the size of the summary created by \ssumm is slightly higher than \summarysize (e.g., 5,129 for \summarysize = 5,000 on \Wikitalk) since \ssumm cannot exactly control the summary size.

\section{Conclusion}
\label{sec:conclusion}

In this paper, we propose a novel \specsumm algorithm for graph summarization via node aggregation.
We motivate the use of the top-\summarysize largest in magnitude eigenvectors of the adjacency matrix to reduce the dimensionality of the problem, while also maintaining the relevant objective-specific information.
We additionally provide a greedy reassignment heuristic to further improve the summary quality.
We conduct extensive experiments on 11 real graphs to show that \specsumm yields upto 22.5\% and 76.1\% higher quality summaries compared to \sls and \ssumm and is up to 200\texttimes\ faster than \sls.
Given its efficacy and simplicity, \specsumm can scale to massive graphs and be easily deployed in real-world applications.

\begin{acks}
	\revision{Arpit Merchant would like to thank Ananth Mahadevan and Sachith Pai for useful suggestions regarding \opt and \ssumm. Michael Mathioudakis is supported by University of Helsinki and Academy of Finland Projects MLDB (322046) and HPC-HD (347747).}
\end{acks}  
  
%%%%%%%%%%%%%%%%%%%%%%%%%%%%%%%%%%%%%%%%%%%%%%%%%%%%%%%%%%%%%%%%%%%%%%%%%%%%%%%%
% References
%%%%%%%%%%%%%%%%%%%%%%%%%%%%%%%%%%%%%%%%%%%%%%%%%%%%%%%%%%%%%%%%%%%%%%%%%%%%%%%%

\bibliographystyle{ACM-Reference-Format}
\bibliography{references}


\begin{thebibliography}{50}

%%% the Web address.

\ifx \showCODEN    \undefined \def \showCODEN     #1{\unskip}     \fi
\ifx \showDOI      \undefined \def \showDOI       #1{#1}\fi
\ifx \showISBNx    \undefined \def \showISBNx     #1{\unskip}     \fi
\ifx \showISBNxiii \undefined \def \showISBNxiii  #1{\unskip}     \fi
\ifx \showISSN     \undefined \def \showISSN      #1{\unskip}     \fi
\ifx \showLCCN     \undefined \def \showLCCN      #1{\unskip}     \fi
\ifx \shownote     \undefined \def \shownote      #1{#1}          \fi
\ifx \showarticletitle \undefined \def \showarticletitle #1{#1}   \fi
\ifx \showURL      \undefined \def \showURL       {\relax}        \fi
% invisible to TeX
\providecommand\bibfield[2]{#2}
\providecommand\bibinfo[2]{#2}
\providecommand\natexlab[1]{#1}
\providecommand\showeprint[2][]{arXiv:#2}

\bibitem[Abbe et~al\mbox{.}(2016)]%
        {abbe2015exact}
\bibfield{author}{\bibinfo{person}{Emmanuel Abbe}, \bibinfo{person}{Afonso~S.
  Bandeira}, {and} \bibinfo{person}{Georgina Hall}.}
  \bibinfo{year}{2016}\natexlab{}.
\newblock \showarticletitle{Exact Recovery in the Stochastic Block Model}.
\newblock \bibinfo{journal}{\emph{{IEEE} Trans. Inf. Theory}}
  \bibinfo{volume}{62}, \bibinfo{number}{1} (\bibinfo{year}{2016}),
  \bibinfo{pages}{471--487}.
\newblock


\bibitem[Baglama and Reichel(2005)]%
        {baglama2005lanczos}
\bibfield{author}{\bibinfo{person}{James Baglama} {and} \bibinfo{person}{Lothar
  Reichel}.} \bibinfo{year}{2005}\natexlab{}.
\newblock \showarticletitle{Augmented Implicitly Restarted Lanczos
  Bidiagonalization Methods}.
\newblock \bibinfo{journal}{\emph{SIAM J. Sci. Comput.}} \bibinfo{volume}{27},
  \bibinfo{number}{1} (\bibinfo{year}{2005}), \bibinfo{pages}{19--42}.
\newblock


\bibitem[Beg et~al\mbox{.}(2018)]%
        {beg2018scalable}
\bibfield{author}{\bibinfo{person}{Maham~Anwar Beg}, \bibinfo{person}{Muhammad
  Ahmad}, \bibinfo{person}{Arif Zaman}, {and} \bibinfo{person}{Imdadullah
  Khan}.} \bibinfo{year}{2018}\natexlab{}.
\newblock \showarticletitle{Scalable Approximation Algorithm for Graph
  Summarization}. In \bibinfo{booktitle}{\emph{Advances in Knowledge Discovery
  and Data Mining}}. \bibinfo{publisher}{Springer}, \bibinfo{address}{Cham},
  \bibinfo{pages}{502--514}.
\newblock


\bibitem[Boldi et~al\mbox{.}(2011)]%
        {DBLP:conf/www/BoldiRSV11}
\bibfield{author}{\bibinfo{person}{Paolo Boldi}, \bibinfo{person}{Marco Rosa},
  \bibinfo{person}{Massimo Santini}, {and} \bibinfo{person}{Sebastiano Vigna}.}
  \bibinfo{year}{2011}\natexlab{}.
\newblock \showarticletitle{Layered Label Propagation: A Multiresolution
  Coordinate-Free Ordering for Compressing Social Networks}. In
  \bibinfo{booktitle}{\emph{Proceedings of the 20th International Conference on
  World Wide Web}} \emph{(\bibinfo{series}{WWW '11})}.
  \bibinfo{publisher}{ACM}, \bibinfo{address}{New York, NY, USA},
  \bibinfo{pages}{587--596}.
\newblock


\bibitem[Boldi and Vigna(2004)]%
        {boldi2004webgraph}
\bibfield{author}{\bibinfo{person}{Paolo Boldi} {and}
  \bibinfo{person}{Sebastiano Vigna}.} \bibinfo{year}{2004}\natexlab{}.
\newblock \showarticletitle{The Webgraph Framework I: Compression Techniques}.
  In \bibinfo{booktitle}{\emph{Proceedings of the 13th International Conference
  on World Wide Web}} \emph{(\bibinfo{series}{WWW '04})}.
  \bibinfo{publisher}{ACM}, \bibinfo{address}{New York, NY, USA},
  \bibinfo{pages}{595--602}.
\newblock


\bibitem[Bouritsas et~al\mbox{.}(2021)]%
        {NEURIPS2021_9a4d6e86}
\bibfield{author}{\bibinfo{person}{Giorgos Bouritsas}, \bibinfo{person}{Andreas
  Loukas}, \bibinfo{person}{Nikolaos Karalias}, {and} \bibinfo{person}{Michael
  Bronstein}.} \bibinfo{year}{2021}\natexlab{}.
\newblock \showarticletitle{Partition and Code: learning how to compress
  graphs}.
\newblock \bibinfo{journal}{\emph{Advances in Neural Information Processing
  Systems}}  \bibinfo{volume}{34} (\bibinfo{year}{2021}),
  \bibinfo{pages}{18603--18619}.
\newblock


\bibitem[Buehrer and Chellapilla(2008)]%
        {DBLP:conf/wsdm/BuehrerC08}
\bibfield{author}{\bibinfo{person}{Gregory Buehrer} {and}
  \bibinfo{person}{Kumar Chellapilla}.} \bibinfo{year}{2008}\natexlab{}.
\newblock \showarticletitle{A Scalable Pattern Mining Approach to Web Graph
  Compression with Communities}. In \bibinfo{booktitle}{\emph{Proceedings of
  the 2008 International Conference on Web Search and Data Mining}}
  \emph{(\bibinfo{series}{WSDM '08})}. \bibinfo{publisher}{ACM},
  \bibinfo{address}{New York, NY, USA}, \bibinfo{pages}{95--106}.
\newblock


\bibitem[Chierichetti et~al\mbox{.}(2009)]%
        {chierichetti2009compressing}
\bibfield{author}{\bibinfo{person}{Flavio Chierichetti}, \bibinfo{person}{Ravi
  Kumar}, \bibinfo{person}{Silvio Lattanzi}, \bibinfo{person}{Michael
  Mitzenmacher}, \bibinfo{person}{Alessandro Panconesi}, {and}
  \bibinfo{person}{Prabhakar Raghavan}.} \bibinfo{year}{2009}\natexlab{}.
\newblock \showarticletitle{On Compressing Social Networks}. In
  \bibinfo{booktitle}{\emph{Proceedings of the 15th ACM SIGKDD International
  Conference on Knowledge Discovery and Data Mining}}
  \emph{(\bibinfo{series}{KDD '09})}. \bibinfo{publisher}{ACM},
  \bibinfo{address}{New York, NY, USA}, \bibinfo{pages}{219--228}.
\newblock


\bibitem[Cormode and Muthukrishnan(2005)]%
        {cormode2005improved}
\bibfield{author}{\bibinfo{person}{Graham Cormode} {and} \bibinfo{person}{S.
  Muthukrishnan}.} \bibinfo{year}{2005}\natexlab{}.
\newblock \showarticletitle{An improved data stream summary: the count-min
  sketch and its applications}.
\newblock \bibinfo{journal}{\emph{J. Algorithms}} \bibinfo{volume}{55},
  \bibinfo{number}{1} (\bibinfo{year}{2005}), \bibinfo{pages}{58--75}.
\newblock


\bibitem[Dhulipala et~al\mbox{.}(2016)]%
        {DBLP:conf/kdd/DhulipalaKKOPS16}
\bibfield{author}{\bibinfo{person}{Laxman Dhulipala}, \bibinfo{person}{Igor
  Kabiljo}, \bibinfo{person}{Brian Karrer}, \bibinfo{person}{Giuseppe
  Ottaviano}, \bibinfo{person}{Sergey Pupyrev}, {and} \bibinfo{person}{Alon
  Shalita}.} \bibinfo{year}{2016}\natexlab{}.
\newblock \showarticletitle{Compressing Graphs and Indexes with Recursive Graph
  Bisection}. In \bibinfo{booktitle}{\emph{Proceedings of the 22nd ACM SIGKDD
  International Conference on Knowledge Discovery and Data Mining}}
  \emph{(\bibinfo{series}{KDD '16})}. \bibinfo{publisher}{ACM},
  \bibinfo{address}{New York, NY, USA}, \bibinfo{pages}{1535--1544}.
\newblock


\bibitem[Dunne and Shneiderman(2013)]%
        {dunne2013motif}
\bibfield{author}{\bibinfo{person}{Cody Dunne} {and} \bibinfo{person}{Ben
  Shneiderman}.} \bibinfo{year}{2013}\natexlab{}.
\newblock \showarticletitle{Motif Simplification: Improving Network
  Visualization Readability with Fan, Connector, and Clique Glyphs}. In
  \bibinfo{booktitle}{\emph{Proceedings of the SIGCHI Conference on Human
  Factors in Computing Systems}} \emph{(\bibinfo{series}{CHI '13})}.
  \bibinfo{publisher}{ACM}, \bibinfo{address}{New York, NY, USA},
  \bibinfo{pages}{3247--3256}.
\newblock


\bibitem[Fan et~al\mbox{.}(2012)]%
        {fan2012query}
\bibfield{author}{\bibinfo{person}{Wenfei Fan}, \bibinfo{person}{Jianzhong Li},
  \bibinfo{person}{Xin Wang}, {and} \bibinfo{person}{Yinghui Wu}.}
  \bibinfo{year}{2012}\natexlab{}.
\newblock \showarticletitle{Query Preserving Graph Compression}. In
  \bibinfo{booktitle}{\emph{Proceedings of the 2012 ACM SIGMOD International
  Conference on Management of Data}} \emph{(\bibinfo{series}{SIGMOD '12})}.
  \bibinfo{publisher}{ACM}, \bibinfo{address}{New York, NY, USA},
  \bibinfo{pages}{157--168}.
\newblock


\bibitem[Fan et~al\mbox{.}(2021)]%
        {DBLP:conf/sigmod/FanLLL21}
\bibfield{author}{\bibinfo{person}{Wenfei Fan}, \bibinfo{person}{Yuanhao Li},
  \bibinfo{person}{Muyang Liu}, {and} \bibinfo{person}{Can Lu}.}
  \bibinfo{year}{2021}\natexlab{}.
\newblock \showarticletitle{Making Graphs Compact by Lossless Contraction}. In
  \bibinfo{booktitle}{\emph{Proceedings of the 2021 International Conference on
  Management of Data}} \emph{(\bibinfo{series}{SIGMOD '21})}.
  \bibinfo{publisher}{ACM}, \bibinfo{address}{New York, NY, USA},
  \bibinfo{pages}{472--484}.
\newblock


\bibitem[G{\"o}rke et~al\mbox{.}(2010)]%
        {gorke2010modularity}
\bibfield{author}{\bibinfo{person}{Robert G{\"o}rke}, \bibinfo{person}{Pascal
  Maillard}, \bibinfo{person}{Christian Staudt}, {and}
  \bibinfo{person}{Dorothea Wagner}.} \bibinfo{year}{2010}\natexlab{}.
\newblock \showarticletitle{Modularity-Driven Clustering of Dynamic Graphs}. In
  \bibinfo{booktitle}{\emph{Experimental Algorithms}}.
  \bibinfo{publisher}{Springer}, \bibinfo{address}{Berlin, Heidelberg},
  \bibinfo{pages}{436--448}.
\newblock


\bibitem[Hajiabadi et~al\mbox{.}(2021)]%
        {DBLP:conf/kdd/HajiabadiS0T21}
\bibfield{author}{\bibinfo{person}{Mahdi Hajiabadi}, \bibinfo{person}{Jasbir
  Singh}, \bibinfo{person}{Venkatesh Srinivasan}, {and} \bibinfo{person}{Alex
  Thomo}.} \bibinfo{year}{2021}\natexlab{}.
\newblock \showarticletitle{Graph Summarization with Controlled Utility Loss}.
  In \bibinfo{booktitle}{\emph{Proceedings of the 27th ACM SIGKDD Conference on
  Knowledge Discovery \& Data Mining}} \emph{(\bibinfo{series}{KDD '21})}.
  \bibinfo{publisher}{ACM}, \bibinfo{address}{New York, NY, USA},
  \bibinfo{pages}{536--546}.
\newblock


\bibitem[Khan et~al\mbox{.}(2015)]%
        {DBLP:journals/computing/KhanNL15}
\bibfield{author}{\bibinfo{person}{Kifayat{-}Ullah Khan},
  \bibinfo{person}{Waqas Nawaz}, {and} \bibinfo{person}{Young{-}Koo Lee}.}
  \bibinfo{year}{2015}\natexlab{}.
\newblock \showarticletitle{Set-based approximate approach for lossless graph
  summarization}.
\newblock \bibinfo{journal}{\emph{Computing}} \bibinfo{volume}{97},
  \bibinfo{number}{12} (\bibinfo{year}{2015}), \bibinfo{pages}{1185--1207}.
\newblock


\bibitem[Ko et~al\mbox{.}(2020)]%
        {DBLP:conf/kdd/KoKS20}
\bibfield{author}{\bibinfo{person}{Jihoon Ko}, \bibinfo{person}{Yunbum Kook},
  {and} \bibinfo{person}{Kijung Shin}.} \bibinfo{year}{2020}\natexlab{}.
\newblock \showarticletitle{Incremental Lossless Graph Summarization}. In
  \bibinfo{booktitle}{\emph{Proceedings of the 26th ACM SIGKDD International
  Conference on Knowledge Discovery \& Data Mining}}
  \emph{(\bibinfo{series}{KDD '20})}. \bibinfo{publisher}{ACM},
  \bibinfo{address}{New York, NY, USA}, \bibinfo{pages}{317--327}.
\newblock


\bibitem[Koutra et~al\mbox{.}(2015)]%
        {koutra2014vog}
\bibfield{author}{\bibinfo{person}{Danai Koutra}, \bibinfo{person}{U Kang},
  \bibinfo{person}{Jilles Vreeken}, {and} \bibinfo{person}{Christos
  Faloutsos}.} \bibinfo{year}{2015}\natexlab{}.
\newblock \showarticletitle{Summarizing and Understanding Large Graphs}.
\newblock \bibinfo{journal}{\emph{Stat. Anal. Data Min.}} \bibinfo{volume}{8},
  \bibinfo{number}{3} (\bibinfo{year}{2015}), \bibinfo{pages}{183--202}.
\newblock


\bibitem[Kumar and Efstathopoulos(2018)]%
        {DBLP:journals/pvldb/KumarE18}
\bibfield{author}{\bibinfo{person}{K.~Ashwin Kumar} {and}
  \bibinfo{person}{Petros Efstathopoulos}.} \bibinfo{year}{2018}\natexlab{}.
\newblock \showarticletitle{Utility-Driven Graph Summarization}.
\newblock \bibinfo{journal}{\emph{Proc. {VLDB} Endow.}} \bibinfo{volume}{12},
  \bibinfo{number}{4} (\bibinfo{year}{2018}), \bibinfo{pages}{335--347}.
\newblock


\bibitem[Lee et~al\mbox{.}(2020)]%
        {lee2020ssumm}
\bibfield{author}{\bibinfo{person}{Kyuhan Lee}, \bibinfo{person}{Hyeonsoo Jo},
  \bibinfo{person}{Jihoon Ko}, \bibinfo{person}{Sungsu Lim}, {and}
  \bibinfo{person}{Kijung Shin}.} \bibinfo{year}{2020}\natexlab{}.
\newblock \showarticletitle{SSumM: Sparse Summarization of Massive Graphs}. In
  \bibinfo{booktitle}{\emph{Proceedings of the 26th ACM SIGKDD International
  Conference on Knowledge Discovery \& Data Mining}}
  \emph{(\bibinfo{series}{KDD '20})}. \bibinfo{publisher}{ACM},
  \bibinfo{address}{New York, NY, USA}, \bibinfo{pages}{144--154}.
\newblock


\bibitem[LeFevre and Terzi(2010)]%
        {lefevre2010grass}
\bibfield{author}{\bibinfo{person}{Kristen LeFevre} {and}
  \bibinfo{person}{Evimaria Terzi}.} \bibinfo{year}{2010}\natexlab{}.
\newblock \showarticletitle{GraSS: Graph Structure Summarization}. In
  \bibinfo{booktitle}{\emph{Proceedings of the 2010 SIAM International
  Conference on Data Mining (SDM)}}. \bibinfo{publisher}{{SIAM}},
  \bibinfo{pages}{454--465}.
\newblock


\bibitem[Leskovec et~al\mbox{.}(2010)]%
        {leskovec2010predict}
\bibfield{author}{\bibinfo{person}{Jure Leskovec}, \bibinfo{person}{Daniel
  Huttenlocher}, {and} \bibinfo{person}{Jon Kleinberg}.}
  \bibinfo{year}{2010}\natexlab{}.
\newblock \showarticletitle{Predicting Positive and Negative Links in Online
  Social Networks}. In \bibinfo{booktitle}{\emph{Proceedings of the 19th
  International Conference on World Wide Web}} \emph{(\bibinfo{series}{WWW
  '10})}. \bibinfo{publisher}{ACM}, \bibinfo{address}{New York, NY, USA},
  \bibinfo{pages}{641--650}.
\newblock


\bibitem[Leskovec et~al\mbox{.}(2007)]%
        {leskovec2007graph}
\bibfield{author}{\bibinfo{person}{Jure Leskovec}, \bibinfo{person}{Jon~M.
  Kleinberg}, {and} \bibinfo{person}{Christos Faloutsos}.}
  \bibinfo{year}{2007}\natexlab{}.
\newblock \showarticletitle{Graph evolution: Densification and shrinking
  diameters}.
\newblock \bibinfo{journal}{\emph{{ACM} Trans. Knowl. Discov. Data}}
  \bibinfo{volume}{1}, \bibinfo{number}{1}, Article \bibinfo{articleno}{2}
  (\bibinfo{year}{2007}), \bibinfo{numpages}{41}~pages.
\newblock


\bibitem[Leskovec et~al\mbox{.}(2009)]%
        {leskovec2009community}
\bibfield{author}{\bibinfo{person}{Jure Leskovec}, \bibinfo{person}{Kevin~J.
  Lang}, \bibinfo{person}{Anirban Dasgupta}, {and} \bibinfo{person}{Michael~W.
  Mahoney}.} \bibinfo{year}{2009}\natexlab{}.
\newblock \showarticletitle{Community Structure in Large Networks: Natural
  Cluster Sizes and the Absence of Large Well-Defined Clusters}.
\newblock \bibinfo{journal}{\emph{Internet Math.}} \bibinfo{volume}{6},
  \bibinfo{number}{1} (\bibinfo{year}{2009}), \bibinfo{pages}{29--123}.
\newblock


\bibitem[Liang et~al\mbox{.}(2020)]%
        {DBLP:journals/isci/LiangCWLYL20}
\bibfield{author}{\bibinfo{person}{Yuzhi Liang}, \bibinfo{person}{Chen Chen},
  \bibinfo{person}{Yukun Wang}, \bibinfo{person}{Kai Lei}, \bibinfo{person}{Min
  Yang}, {and} \bibinfo{person}{Ziyu Lyu}.} \bibinfo{year}{2020}\natexlab{}.
\newblock \showarticletitle{Reachability preserving compression for dynamic
  graph}.
\newblock \bibinfo{journal}{\emph{Inf. Sci.}}  \bibinfo{volume}{520}
  (\bibinfo{year}{2020}), \bibinfo{pages}{232--249}.
\newblock


\bibitem[Liu et~al\mbox{.}(2018)]%
        {liu2018graph}
\bibfield{author}{\bibinfo{person}{Yike Liu}, \bibinfo{person}{Tara Safavi},
  \bibinfo{person}{Abhilash Dighe}, {and} \bibinfo{person}{Danai Koutra}.}
  \bibinfo{year}{2018}\natexlab{}.
\newblock \showarticletitle{Graph Summarization Methods and Applications: {A}
  Survey}.
\newblock \bibinfo{journal}{\emph{{ACM} Comput. Surv.}} \bibinfo{volume}{51},
  \bibinfo{number}{3}, Article \bibinfo{articleno}{62} (\bibinfo{year}{2018}),
  \bibinfo{numpages}{34}~pages.
\newblock


\bibitem[Maserrat and Pei(2010)]%
        {maserrat2010neighbor}
\bibfield{author}{\bibinfo{person}{Hossein Maserrat} {and}
  \bibinfo{person}{Jian Pei}.} \bibinfo{year}{2010}\natexlab{}.
\newblock \showarticletitle{Neighbor Query Friendly Compression of Social
  Networks}. In \bibinfo{booktitle}{\emph{Proceedings of the 16th ACM SIGKDD
  International Conference on Knowledge Discovery and Data Mining}}
  \emph{(\bibinfo{series}{KDD '10})}. \bibinfo{publisher}{ACM},
  \bibinfo{address}{New York, NY, USA}, \bibinfo{pages}{533--542}.
\newblock


\bibitem[Navlakha et~al\mbox{.}(2008)]%
        {navlakha2008graph}
\bibfield{author}{\bibinfo{person}{Saket Navlakha}, \bibinfo{person}{Rajeev
  Rastogi}, {and} \bibinfo{person}{Nisheeth Shrivastava}.}
  \bibinfo{year}{2008}\natexlab{}.
\newblock \showarticletitle{Graph Summarization with Bounded Error}. In
  \bibinfo{booktitle}{\emph{Proceedings of the 2008 ACM SIGMOD International
  Conference on Management of Data}} \emph{(\bibinfo{series}{SIGMOD '08})}.
  \bibinfo{publisher}{ACM}, \bibinfo{address}{New York, NY, USA},
  \bibinfo{pages}{419--432}.
\newblock


\bibitem[Nejad et~al\mbox{.}(2021)]%
        {DBLP:journals/isci/NejadJT21}
\bibfield{author}{\bibinfo{person}{Amin Emamzadeh~Esmaeili Nejad},
  \bibinfo{person}{Mansoor~Zolghadri Jahromi}, {and} \bibinfo{person}{Mohammad
  Taheri}.} \bibinfo{year}{2021}\natexlab{}.
\newblock \showarticletitle{Graph compression based on transitivity for
  neighborhood query}.
\newblock \bibinfo{journal}{\emph{Inf. Sci.}}  \bibinfo{volume}{576}
  (\bibinfo{year}{2021}), \bibinfo{pages}{312--328}.
\newblock


\bibitem[Nocedal and Wright(1999)]%
        {wright1999numerical}
\bibfield{author}{\bibinfo{person}{Jorge Nocedal} {and}
  \bibinfo{person}{Stephen~J. Wright}.} \bibinfo{year}{1999}\natexlab{}.
\newblock \bibinfo{booktitle}{\emph{Numerical Optimization}}.
\newblock \bibinfo{publisher}{Springer}, \bibinfo{address}{Berlin, Heidelberg}.
\newblock


\bibitem[Perozzi et~al\mbox{.}(2014)]%
        {perozzi2014deepwalk}
\bibfield{author}{\bibinfo{person}{Bryan Perozzi}, \bibinfo{person}{Rami
  Al-Rfou}, {and} \bibinfo{person}{Steven Skiena}.}
  \bibinfo{year}{2014}\natexlab{}.
\newblock \showarticletitle{DeepWalk: Online Learning of Social
  Representations}. In \bibinfo{booktitle}{\emph{Proceedings of the 20th ACM
  SIGKDD International Conference on Knowledge Discovery and Data Mining}}
  \emph{(\bibinfo{series}{KDD '14})}. \bibinfo{publisher}{ACM},
  \bibinfo{address}{New York, NY, USA}, \bibinfo{pages}{701--710}.
\newblock


\bibitem[Petersen and Pedersen(2012)]%
        {petersen2008matrix}
\bibfield{author}{\bibinfo{person}{Kaare~Brandt Petersen} {and}
  \bibinfo{person}{Michael~Syskind Pedersen}.} \bibinfo{year}{2012}\natexlab{}.
\newblock \bibinfo{title}{The Matrix Cookbook}.
\newblock
\newblock
\newblock
\shownote{Version 20121115}.


\bibitem[Qiu et~al\mbox{.}(2018)]%
        {qiu2018network}
\bibfield{author}{\bibinfo{person}{Jiezhong Qiu}, \bibinfo{person}{Yuxiao
  Dong}, \bibinfo{person}{Hao Ma}, \bibinfo{person}{Jian Li},
  \bibinfo{person}{Kuansan Wang}, {and} \bibinfo{person}{Jie Tang}.}
  \bibinfo{year}{2018}\natexlab{}.
\newblock \showarticletitle{Network Embedding as Matrix Factorization: Unifying
  DeepWalk, LINE, PTE, and Node2vec}. In \bibinfo{booktitle}{\emph{Proceedings
  of the Eleventh ACM International Conference on Web Search and Data Mining}}
  \emph{(\bibinfo{series}{WSDM '18})}. \bibinfo{publisher}{ACM},
  \bibinfo{address}{New York, NY, USA}, \bibinfo{pages}{459--467}.
\newblock


\bibitem[Riondato et~al\mbox{.}(2017)]%
        {riondato2017graph}
\bibfield{author}{\bibinfo{person}{Matteo Riondato}, \bibinfo{person}{David
  Garc{\'{\i}}a{-}Soriano}, {and} \bibinfo{person}{Francesco Bonchi}.}
  \bibinfo{year}{2017}\natexlab{}.
\newblock \showarticletitle{Graph summarization with quality guarantees}.
\newblock \bibinfo{journal}{\emph{Data Min. Knowl. Discov.}}
  \bibinfo{volume}{31}, \bibinfo{number}{2} (\bibinfo{year}{2017}),
  \bibinfo{pages}{314--349}.
\newblock


\bibitem[Rozemberczki et~al\mbox{.}(2021)]%
        {rozemberczki2019multiscale}
\bibfield{author}{\bibinfo{person}{Benedek Rozemberczki}, \bibinfo{person}{Carl
  Allen}, {and} \bibinfo{person}{Rik Sarkar}.} \bibinfo{year}{2021}\natexlab{}.
\newblock \showarticletitle{Multi-Scale attributed node embedding}.
\newblock \bibinfo{journal}{\emph{J. Complex Networks}} \bibinfo{volume}{9},
  \bibinfo{number}{2} (\bibinfo{year}{2021}), \bibinfo{numpages}{22}~pages.
\newblock
\newblock
\shownote{cnab014}.


\bibitem[Rozemberczki and Sarkar(2020)]%
        {feather}
\bibfield{author}{\bibinfo{person}{Benedek Rozemberczki} {and}
  \bibinfo{person}{Rik Sarkar}.} \bibinfo{year}{2020}\natexlab{}.
\newblock \showarticletitle{Characteristic Functions on Graphs: Birds of a
  Feather, from Statistical Descriptors to Parametric Models}. In
  \bibinfo{booktitle}{\emph{Proceedings of the 29th ACM International
  Conference on Information \& Knowledge Management}}
  \emph{(\bibinfo{series}{CIKM '20})}. \bibinfo{publisher}{ACM},
  \bibinfo{address}{New York, NY, USA}, \bibinfo{pages}{1325--1334}.
\newblock


\bibitem[Sadri et~al\mbox{.}(2017)]%
        {DBLP:journals/is/SadriSRZCS17}
\bibfield{author}{\bibinfo{person}{Amin Sadri}, \bibinfo{person}{Flora~D.
  Salim}, \bibinfo{person}{Yongli Ren}, \bibinfo{person}{Masoomeh Zameni},
  \bibinfo{person}{Jeffrey Chan}, {and} \bibinfo{person}{Timos Sellis}.}
  \bibinfo{year}{2017}\natexlab{}.
\newblock \showarticletitle{Shrink: Distance preserving graph compression}.
\newblock \bibinfo{journal}{\emph{Inf. Syst.}}  \bibinfo{volume}{69}
  (\bibinfo{year}{2017}), \bibinfo{pages}{180--193}.
\newblock


\bibitem[Sculley(2010)]%
        {sculley2010web}
\bibfield{author}{\bibinfo{person}{David Sculley}.}
  \bibinfo{year}{2010}\natexlab{}.
\newblock \showarticletitle{Web-Scale k-Means Clustering}. In
  \bibinfo{booktitle}{\emph{Proceedings of the 19th International Conference on
  World Wide Web}} \emph{(\bibinfo{series}{WWW '10})}.
  \bibinfo{publisher}{ACM}, \bibinfo{address}{New York, NY, USA},
  \bibinfo{pages}{1177--1178}.
\newblock


\bibitem[Sen et~al\mbox{.}(2008)]%
        {sen2008collective}
\bibfield{author}{\bibinfo{person}{Prithviraj Sen}, \bibinfo{person}{Galileo
  Namata}, \bibinfo{person}{Mustafa Bilgic}, \bibinfo{person}{Lise Getoor},
  \bibinfo{person}{Brian Gallagher}, {and} \bibinfo{person}{Tina
  Eliassi{-}Rad}.} \bibinfo{year}{2008}\natexlab{}.
\newblock \showarticletitle{Collective Classification in Network Data}.
\newblock \bibinfo{journal}{\emph{{AI} Mag.}} \bibinfo{volume}{29},
  \bibinfo{number}{3} (\bibinfo{year}{2008}), \bibinfo{pages}{93--106}.
\newblock


\bibitem[Shi and Malik(2000)]%
        {shi2000normalized}
\bibfield{author}{\bibinfo{person}{Jianbo Shi} {and} \bibinfo{person}{Jitendra
  Malik}.} \bibinfo{year}{2000}\natexlab{}.
\newblock \showarticletitle{Normalized Cuts and Image Segmentation}.
\newblock \bibinfo{journal}{\emph{{IEEE} Trans. Pattern Anal. Mach. Intell.}}
  \bibinfo{volume}{22}, \bibinfo{number}{8} (\bibinfo{year}{2000}),
  \bibinfo{pages}{888--905}.
\newblock


\bibitem[Shin et~al\mbox{.}(2019)]%
        {DBLP:conf/www/ShinG0R19}
\bibfield{author}{\bibinfo{person}{Kijung Shin}, \bibinfo{person}{Amol
  Ghoting}, \bibinfo{person}{Myunghwan Kim}, {and} \bibinfo{person}{Hema
  Raghavan}.} \bibinfo{year}{2019}\natexlab{}.
\newblock \showarticletitle{SWeG: Lossless and Lossy Summarization of Web-Scale
  Graphs}. In \bibinfo{booktitle}{\emph{The World Wide Web Conference}}
  \emph{(\bibinfo{series}{WWW '19})}. \bibinfo{publisher}{ACM},
  \bibinfo{address}{New York, NY, USA}, \bibinfo{pages}{1679--1690}.
\newblock


\bibitem[Spielman(2007)]%
        {spielman2007spectral}
\bibfield{author}{\bibinfo{person}{Daniel~A. Spielman}.}
  \bibinfo{year}{2007}\natexlab{}.
\newblock \showarticletitle{Spectral Graph Theory and its Applications}. In
  \bibinfo{booktitle}{\emph{{FOCS}}}. \bibinfo{publisher}{IEEE},
  \bibinfo{pages}{29--38}.
\newblock


\bibitem[Stella and Shi(2003)]%
        {stella2003multiclass}
\bibfield{author}{\bibinfo{person}{X~Yu Stella} {and} \bibinfo{person}{Jianbo
  Shi}.} \bibinfo{year}{2003}\natexlab{}.
\newblock \showarticletitle{Multiclass spectral clustering}. In
  \bibinfo{booktitle}{\emph{ICCV}}. \bibinfo{pages}{313--319}.
\newblock


\bibitem[Tang et~al\mbox{.}(2016)]%
        {tang2016graph}
\bibfield{author}{\bibinfo{person}{Nan Tang}, \bibinfo{person}{Qing Chen},
  {and} \bibinfo{person}{Prasenjit Mitra}.} \bibinfo{year}{2016}\natexlab{}.
\newblock \showarticletitle{Graph Stream Summarization: From Big Bang to Big
  Crunch}. In \bibinfo{booktitle}{\emph{Proceedings of the 2016 International
  Conference on Management of Data}} \emph{(\bibinfo{series}{SIGMOD '16})}.
  \bibinfo{publisher}{ACM}, \bibinfo{address}{New York, NY, USA},
  \bibinfo{pages}{1481--1496}.
\newblock


\bibitem[Tian et~al\mbox{.}(2008)]%
        {DBLP:conf/sigmod/TianHP08}
\bibfield{author}{\bibinfo{person}{Yuanyuan Tian}, \bibinfo{person}{Richard~A.
  Hankins}, {and} \bibinfo{person}{Jignesh~M. Patel}.}
  \bibinfo{year}{2008}\natexlab{}.
\newblock \showarticletitle{Efficient Aggregation for Graph Summarization}. In
  \bibinfo{booktitle}{\emph{Proceedings of the 2008 ACM SIGMOD International
  Conference on Management of Data}} \emph{(\bibinfo{series}{SIGMOD '08})}.
  \bibinfo{publisher}{ACM}, \bibinfo{address}{New York, NY, USA},
  \bibinfo{pages}{567--580}.
\newblock


\bibitem[Toivonen et~al\mbox{.}(2011)]%
        {toivonen2011compression}
\bibfield{author}{\bibinfo{person}{Hannu Toivonen}, \bibinfo{person}{Fang
  Zhou}, \bibinfo{person}{Aleksi Hartikainen}, {and} \bibinfo{person}{Atte
  Hinkka}.} \bibinfo{year}{2011}\natexlab{}.
\newblock \showarticletitle{Compression of Weighted Graphs}. In
  \bibinfo{booktitle}{\emph{Proceedings of the 17th ACM SIGKDD International
  Conference on Knowledge Discovery and Data Mining}}
  \emph{(\bibinfo{series}{KDD '11})}. \bibinfo{publisher}{ACM},
  \bibinfo{address}{New York, NY, USA}, \bibinfo{pages}{965--973}.
\newblock


\bibitem[Wen and Yin(2013)]%
        {wen2013feasible}
\bibfield{author}{\bibinfo{person}{Zaiwen Wen} {and} \bibinfo{person}{Wotao
  Yin}.} \bibinfo{year}{2013}\natexlab{}.
\newblock \showarticletitle{A feasible method for optimization with
  orthogonality constraints}.
\newblock \bibinfo{journal}{\emph{Math. Program.}} \bibinfo{volume}{142},
  \bibinfo{number}{1-2} (\bibinfo{year}{2013}), \bibinfo{pages}{397--434}.
\newblock


\bibitem[Yan et~al\mbox{.}(2009)]%
        {yan2009fast}
\bibfield{author}{\bibinfo{person}{Donghui Yan}, \bibinfo{person}{Ling Huang},
  {and} \bibinfo{person}{Michael~I. Jordan}.} \bibinfo{year}{2009}\natexlab{}.
\newblock \showarticletitle{Fast Approximate Spectral Clustering}. In
  \bibinfo{booktitle}{\emph{Proceedings of the 15th ACM SIGKDD International
  Conference on Knowledge Discovery and Data Mining}}
  \emph{(\bibinfo{series}{KDD '09})}. \bibinfo{publisher}{ACM},
  \bibinfo{address}{New York, NY, USA}, \bibinfo{pages}{907--916}.
\newblock


\bibitem[Yang and Leskovec(2012)]%
        {leskovec2012define}
\bibfield{author}{\bibinfo{person}{Jaewon Yang} {and} \bibinfo{person}{Jure
  Leskovec}.} \bibinfo{year}{2012}\natexlab{}.
\newblock \showarticletitle{Defining and Evaluating Network Communities Based
  on Ground-Truth}. In \bibinfo{booktitle}{\emph{Proceedings of the ACM SIGKDD
  Workshop on Mining Data Semantics}} \emph{(\bibinfo{series}{MDS '12})}.
  \bibinfo{publisher}{ACM}, \bibinfo{address}{New York, NY, USA}, Article
  \bibinfo{articleno}{3}, \bibinfo{numpages}{8}~pages.
\newblock


\bibitem[Yong et~al\mbox{.}(2021)]%
        {DBLP:conf/sigmod/YongH0T21}
\bibfield{author}{\bibinfo{person}{Quinton Yong}, \bibinfo{person}{Mahdi
  Hajiabadi}, \bibinfo{person}{Venkatesh Srinivasan}, {and}
  \bibinfo{person}{Alex Thomo}.} \bibinfo{year}{2021}\natexlab{}.
\newblock \showarticletitle{Efficient Graph Summarization Using Weighted LSH at
  Billion-Scale}. In \bibinfo{booktitle}{\emph{Proceedings of the 2021
  International Conference on Management of Data}}
  \emph{(\bibinfo{series}{SIGMOD '21})}. \bibinfo{publisher}{ACM},
  \bibinfo{address}{New York, NY, USA}, \bibinfo{pages}{2357--2365}.
\newblock


\end{thebibliography}

\clearpage
\appendix

\section{Omitted Proofs}
\label{app:proofs}

In this section, we provide the proofs of lemmas and theorems in the paper that are omitted due to space limitations.

\subsection{Connection between k-Summary and Normalized Cut}

The $\summarysize$-way normalized cut problem amounts to finding $\summarysize$ disjoint subsets of \vertexset such that the total weights of edges that cross different partitions is minimized while the sizes of the subsets are roughly balanced~\cite{shi2000normalized}. This results in an optimization problem similar to Problem~\ref{prob:trace_maximization_integer} with one notable difference: the objective function is $ \tr{\round{Z^{\top} \Ag Z}} $ and not $\tr{\round{Z^{\top} \Ag Z}^2}$. The $\summarysize$-way normalized cut is NP-Hard~\cite{stella2003multiclass}. And its relaxed version is optimized by the eigenvectors corresponding to the \dimension (\emph{algebraically}) largest eigenvalues.

\subsection{Proof of Lemma~\ref{lemma:traceloss}}
\label{app:traceloss_lemma_proof}

\textsc{Lemma 3.1.}\quad$\normloss\round{\Ag, \Aslift} = \tr{\Ag^2} - \underbrace{\tr{\round{\ZAZ}^2}}_{\FZ{\Z{\S}}}$

\begin{proof}
    Using the basic matrix identities $\|L\|_2^2 = \tr{L^\top L}$, $\tr{L + M} = \tr{L} + \tr{M}$, $\tr{c L} = c \cdot \tr{L}$, and from trace invariance under cyclic permutation, $\tr{L M N} = \tr{M N L} = \tr{N L M}$:
    \begin{align*}
        \norm{\Ag - \Aslift}_2^2
        & = \tr{\round{\Ag - \Aslift}^\top \round{\Ag - \Aslift}} \\
        & = \tr{\Ag^2 - \Ag\PAP} - \tr{\PAP \Ag} + \tr{\PAP \PAP}\\
        & = \tr{\Ag^2 - \Ag\PAP} - \tr{\PAP \Ag} + \tr{P_{\S{}}\PAP \Ag }\\
        & = \tr{\Ag^2} - \tr{\Ag\PAP}\\
        & = \tr{\Ag^2} - \tr{\round{\ZAZ}^2}
    \end{align*}
    where the fourth equation follows from substituting $P^2 = P$ and the last equation follows from substituting $P = Z Z^\top$ and again using the invariance under cyclic permutation property.
\end{proof}

\subsection{Proof of Lemma~\ref{lemma:k_equals_n}}\label{app:k_equals_n}

\kequalsn*
\begin{proof}
    By definition, \normloss is non-negative and so we have an upper bound on the objective function, i.e. $\tr{\round{\ZAZ}^2} \leq \tr{\Ag^2}$.
    Let $\B = \squares{\e_1, \ldots, \e_{\summarysize}}$. Substituting into $\tr{\round{\Z{}^\top \Ag \Z{}}^2}$:
    \begin{equation}
        \begin{aligned}
        \tr{\round{\Z{}^\top \Ag \Z{}}^2} 
            & = \tr{\round{B^\top \B\D\B^\top \B}^2} \\
            & = \tr{(\B^\top \B)\D\B^\top \B (\B^\top \B)\D\B^\top \B} \\
            & = \tr{\D\B^\top \B \D\B^\top \B} \\ % rotate
            & \stackrel{\text{rotate}}{=} \tr{\B\D\B^\top \B \D\B^\top} \\
            & = \tr{\Ag^2}
        \end{aligned}
    \end{equation}
    \noindent
    Therefore, the upper bound of the objective function is achieved when $\Z{} = \B$ and this is a feasible solution for Problem~\ref{prob:trace_maximization_relaxed} because \B is orthonormal.
\end{proof}

\subsection{Proof of Lemma~\ref{lemma:k_equals_one}} \label{app:k_equals_one}

\kequalsone*

\begin{proof}
    Let the Rayleigh quotient with respect to a fixed $\Ag$ be:
    \begin{equation*}
      R\round{\v} = \frac{\round{\v^\top \Ag \v}^2}{\round{\v^\top \v}^2}
    \end{equation*}
    Since the Rayleigh quotient is homogeneous\footnote{A function is called homogeneous with degree $k$, if it satisfies the condition $f\round{\alpha x, \alpha y} = \alpha^k f\round{x, y}$.}, the square of the Rayleigh quotient is also homogeneous~\cite{spielman2007spectral}. And so, it suffices to consider unit vectors \v. Since the set of unit vectors is closed and compact, the function has a maximum value. The partial differential of the quotient with respect to \Ag and \v is given by: 
    \begin{equation*} \label{eq:nabla_RA}
        \nabla \frac{\round{\v^\top \Ag \v}^2}{\round{\v^\top \v}^2}
          = \frac{4 \round{\v^\top \A \v} \round{\A \v} \round{\v^\top \v}^2 - 4 \round{\v^\top \A \v}^2 \cdot \round{\v^\top \v} \cdot \round{\v} }{\round{\round{\v^\top \v}^2}^2}
    \end{equation*}
    \noindent
    Let $\v^*$ be a non-zero vector that maximizes $R\round{\v}$. The gradient of a function at it's maximum value must equal the zero vector. Therefore:
    \begin{equation*} \label{eq:nabla_RA_zero}
      \begin{split}
        & \nabla \frac{\round{\round{\v^*}^\top \A \v^*}^2}{\round{\round{\v^*}^\top \v^*}^2} = 0 \\
        & \A \v^* = \round{\frac{\round{\v^*}^\top \A \v^*}{\round{\v^*}^\top \v^*}} \cdot \v^* \\
      \end{split}
    \end{equation*}
    This implies, that $\v^*$ maximizes $R\round{\v}$ if and only if $\v^*$ is an eigenvector of \Ag with eigenvalue equal to the Rayleigh quotient. And therefore, the maximum value of $R\round{\v} = \lambda_{\text{max}}^2$ where $\lambda_{\text{max}}$ is the largest (in magnitude) eigenvalue of \Ag and $\v^*$ is the corresponding eigenvector.
\end{proof}

\subsection{Proof of Lemma~\ref{lemma:higher_k}} \label{app:higher_k}

\kbetweenonen*

\begin{proof}
    Consider the subspace orthogonal to the subspace defined by the first (say) $m$ largest (in magnitude) eigenvectors of $\Ag$. This lemma shows that the unit vector $\v$ from this orthogonal subspace that maximizes $\round{\v^\top \Ag \v}^2$ is the $\round{m+1}$-th largest eigenvector of $\Ag$. Subsequently, the maximum value of $\sum\limits_{j=1}^{k} \round{\v^\top_j \Ag \v_j}^2$ is achieved by eigenvectors corresponding to the $k$ largest (in magnitude) eigenvalues of $\Ag$.

    Let $\lambda_{\text{min}}^2$ be the minimum value of $R\round{\v}$ for some vector $\v_{\text{min}}$. Matrices $\Ag$ and $\tilde{\Ag} = \Ag + \round{1 - \mu_{\text{min}}^2}\identitymat$ have the same eigenvectors. For all unit norm vectors $\v$, $\tilde{\Ag}$ is positive definite because $\v^\top \tilde{\Ag} \v = \v^\top \Ag \v + 1 - \mu_{\text{min}}^2 \geq 1$. So it suffices to prove the following result for positive definite matrices.
    \begin{equation} \label{eq:argmax_vtAv}
        \psi_i \in \argmax_{\substack{\norm{\v}{} = 1 \\  \v^\top \psi_j=0, \text{for } j < i}} \round{\v^\top \Ag \v}^2.
    \end{equation}
    The base case is true for $\psi_1$ due to Lemma~\ref{lemma:k_equals_one}. Assume that Equation~\ref{eq:argmax_vtAv} holds for the first $m$ eigenvectors $\psi_1, \ldots, \psi_m$. We now show that the result is valid for $i = m+1$ and $\psi_{m+1}$. Define,
    \begin{equation*}
    \Ag_m = \Ag - \sum\limits_{i=1}^{m} \mu_i \psi_i \psi_i^\top.
    \end{equation*}

    For all $j \leq m$, due to the orthogonality of eigenvectors, we have
    \begin{equation} \label{eq:Al_psij_ortho}
    \begin{split}
        \Ag_m \psi_j
        & = \Ag  \psi_j - \sum\limits_{i=1}^{m} \mu_i \psi_i \psi_i^\top \psi_j \\
        & = \Ag  \psi_j - \mu_j \psi_j \\
        & = 0
    \end{split}
    \end{equation}
    
    For all vectors \v orthogonal to $\psi_1, \ldots, \psi_m$, we have
    \begin{equation} \label{eq:vtAkv_vtAv}
    \begin{split}
        \Ag_m \v & = \Ag \v \\
        \round{\v^\top \Ag_m \v}^2 & = \round{\v^\top \Ag \v}^2 \\
        \argmax_{\substack{\norm{\v}{} = 1 \\  \v^\top \psi_j=0, j \leq m}} \round{\v^\top \Ag \v}^2 & = \argmax_{\substack{\norm{\v}{} = 1 \\  \v^\top \psi_j=0, j \leq m}} \round{\v^\top \Ag_m \v}^2
        \subseteq \argmax_{\norm{\v}{} = 1} \round{\v^\top \Ag_m \v}^2
    \end{split}
    \end{equation}

    Consider a unit vector $\u$ that maximizes $\round{\v^\top \Ag_m \v}^2$. Since $\Ag_m$ is a symmetric matrix, according to Lemma~\ref{lemma:k_equals_one}, \u must be an eigenvector of $\Ag_m$. If we show that \u is orthogonal to $\psi_1, \ldots, \psi_m$, then from Equation \ref{eq:vtAkv_vtAv}, we know that \u is also an eigenvector of \Ag. Define the projection of \u orthogonal to $\psi_1, \ldots, \psi_m$.
    \begin{equation*}
    \tilde{\u} = \u - \sum\limits_{j=1}^{m} \psi_j \round{\psi_j^\top \u}
    \end{equation*}
    If $\tilde{\u} = \u$, then we are done. We show this by contradiction. Say that there exists some $\round{\psi_i^\top \u} \neq 0$. This implies, $\norm{\tilde{\u}}{} < \norm{\u}{}$. We have
    \begin{equation} \label{eq:utAlut_uAlu}
    \begin{split}
        \tilde{\u}^\top \Ag_m \tilde{\u}
        & = \tilde{\u}^\top \Ag_m \round{\u - \sum\limits_{j=1}^{m} \psi_j \round{\psi_j^\top \u}} \\
        & = \tilde{\u}^\top \Ag_m \u - \tilde{\u}^\top \round{\sum\limits_{j=1}^{m} \cancelto{0}{\round{\Ag_m \psi_j}} \round{\psi_j^\top \u}} \\
        & = \tilde{\u}^\top \Ag_m \u \\
        & = \round{\u - \sum\limits_{j=1}^{m} \psi_j \round{\psi_j^\top \u}}^\top \Ag_m \u \\
        & = \u^\top \Ag_m \u
    \end{split}
    \end{equation}
    So $\round{\tilde{\u}^\top \Ag_m \tilde{\u}}^2 = \round{\u^\top \Ag_m \u}^2$.

    Define $\hat{\u} = \tilde{\u}/\norm{\tilde{\u}}{}$. Substituting into Equation~\ref{eq:utAlut_uAlu}, we get
    \begin{equation}
    \begin{split}
        \round{\tilde{\u}^\top \Ag_m \tilde{\u}}^2 & = \round{\u^\top \Ag_m \u}^2 \\
        \round{\round{\norm{\tilde{\u}}{}\hat{\u}}^\top \Ag_m \round{\norm{\tilde{\u}}{}\hat{\u}}}^2 & = \round{\norm{\u}{} \u^\top \Ag_m \u \norm{\u}{}}^2 \\
        \round{\frac{\norm{\tilde{\u}}{}^2}{\norm{\u}{}^2}}^2 \round{\hat{\u}^\top \Ag_m \hat{\u}}^2 & =  \round{\u^\top \Ag_m \u}^2
    \end{split}
    \end{equation}
    where the equality holds because \u is a unit vector. But $\norm{\tilde{\u}}{}^2 / \norm{\u}{}^2 < 1$ and therefore $\round{\hat{\u}^\top \Ag_m \hat{\u}}^2 > \round{\u^\top \Ag_m \u}^2 $. This is a contradiction because by definition, \u maximizes $\round{\v^\top \Ag_m \v}^2$ for all unit vectors \v. Therefore $\tilde{\u} = \u$ and \u is orthogonal to $\psi_1, \ldots, \psi_m$. We can thus set $\u = \psi_{m+1}$ and this completes the proof.
\end{proof}

\subsection{Proof of Lemma~\ref{lemma:nonzero_cross_terms}} \label{app:nonzero_cross_terms}

\nonzerocrossterms*

\begin{proof}
    We shall prove this result using the theory of Lagrange multipliers on an individual term of $T_2$. Given \Ag, $\v_1^\top \Ag \v_2 = \v_2^\top \Ag \v_1$. The individual term to optimize is thus:
    \begin{equation}
      \text{max} \round{\v_1^\top \Ag \v_2}^2 \, \text{ s.t. } \, \v_1^\top \v_1 = 1, \, \v_2^\top \v_2 = 1, \, \v_1^\top \v_2 = 0
    \end{equation}
    
    Since $\round{\v_1^\top \Ag \v_2}$ is a scalar quantity, the optimization is equivalent to:
    \begin{equation}
        \begin{aligned}
            \max & \quad \round{\braces{\text{max} \round{\v_1^\top \Ag \v_2}, \text{min} \round{\v_1^\top \Ag \v_2}}}^2 \\
            \text{ s.t. } & \quad \v_1^\top \v_1 = 1, \v_2^\top \v_2 = 1 \\
            & \quad \v_1^\top \v_2 = 0
        \end{aligned}
    \end{equation}
    
    As \Ag is a real, symmetric, square matrix, we rewrite $\Ag = \B^\top \D \B$ where $\B^\top \B = \B \B^\top = \identitymat$. Note $\v_1^\top \v_2 = \v_1^\top \round{\B^\top \B} \v_2 = \round{\B \v_1}^\top \round{\B \v_2}$. Similarly, $\v_1^\top \v_1 = \round{\B \v_1}^\top \round{\B \v_1}$, $\v_2^\top \v_2 = \round{\B \v_2}^\top \round{\B \v_2}$. For ease of notation, we substitute $\x = \round{\B \v_1}$ and $\y = \round{\B \v_2}$. Substituting into the equation above, we get:
    \begin{equation}
        \begin{aligned}
            \underset{\x, \y}\max & \quad \braces{\underset{\x, \y}{\text{max}} \, \x^\top \D \y, \, \underset{\x, \y}{\text{min}} \, \x^\top \D \y} \\
            \text{ s.t. } & \x^\top \x = 1, \, \y^\top \y = 1 \\
            & \quad \x^\top \y = 0
        \end{aligned}
    \end{equation}

    Using Lagrange multipliers with scalars $-\alpha/2, \, -\beta/2, \, \text{and} \, \, -\gamma$, we have
    \begin{equation}\label{eq:opt_final_lemma3}
    \begin{aligned}
      \L \round{\x, \y, -\alpha/2, -\beta/2, -\gamma} & = \x^\top \D \y - \alpha/2 \round{\x^\top \x - 1} \\ & \quad \quad \quad -\beta/2 \round{\y^\top \y - 1} - \gamma \round{\x^\top \y - 0}
    \end{aligned}
    \end{equation}
    
    \begin{equation} \label{eq:partial_Lx}
      \begin{aligned}
        \frac{\partial \L}{\partial \x} = \D \y - \alpha \x - \gamma \y
          & = 0 \\
          \round{\D - \gamma \identitymat} \y & = \alpha \x
      \end{aligned}
    \end{equation}
    
    \begin{equation} \label{eq:partial_Ly}
      \begin{aligned}
        \frac{\partial \L}{\partial \y} = \D \x - \beta \y - \gamma \x
          & = 0 \\
          \round{\D - \gamma \identitymat}\x \ & = \beta \y
      \end{aligned}
    \end{equation}
    
    \begin{equation} \label{eq:partial_Lalpha}
      \begin{aligned}
        \frac{\partial \L}{\partial \round{-\alpha/2}} = \x^\top \x - 1
          & = 0 \\
          \x^\top \x & = 1
      \end{aligned}
    \end{equation}
    
    \begin{equation} \label{eq:partial_Lbeta}
      \begin{aligned}
        \frac{\partial \L}{\partial \round{-\beta/2}} = \y^\top \y - 1
          & = 0 \\
          \y^\top \y & = 1
      \end{aligned}
    \end{equation}
    
    \begin{equation} \label{eq:partial_Lgamma}
      \begin{aligned}
        \frac{\partial \L}{\partial \round{-\gamma}} = \x^\top \y - 0
          & = 0 \\
          \x^\top \y & = 0
      \end{aligned}
    \end{equation}

    Lets multiply Equation~\ref{eq:partial_Lx} with $\beta$ and Equation~\ref{eq:partial_Ly} with $\alpha$ on both sides. Substituting Equation~\ref{eq:partial_Lx} into Equation~\ref{eq:partial_Ly} and Equation~\ref{eq:partial_Ly} into Equation~\ref{eq:partial_Lx}, we get
    \begin{equation} \label{eq:Lxy}
      \round{\D - \gamma \identitymat}^2 \x = \alpha \beta \x
    \end{equation}
    
    \begin{equation} \label{eq:Lyx}
      \round{\D - \gamma \identitymat}^2 \y = \alpha \beta \y
    \end{equation}
    
    This implies that $\x$ and $\y$ are eigenvectors of $\round{\D - \gamma \identitymat}^2$ corresponding to the same eigenvalue, i.e. $\alpha \beta$. Since we know that $\x$ and $\y$ are distinct, $\alpha \beta$ is an eigenvalue of $\round{\D - \gamma \identitymat}^2$ with multiplicity $m \geq 2$. Define $\alpha \beta = l = \round{\lambda_{i_1} - \gamma}^2 = \round{\lambda_{i_2} - \gamma}^2 = \ldots = \round{\lambda_{i_m} - \gamma}^2$, where $\lambda_{i_j}, \, j \in \squares{m}$ represents the $i_j$-th diagonal entry of $\D$.
    
    There are two solutions to the above equation.
    \begin{itemize}
      \item \textbf{Case 1:} $\round{\lambda_{i_1} - \gamma} = \round{\lambda_{i_2} - \gamma} = \ldots = \round{\lambda_{i_m} - \gamma}$. This implies $\lambda = \lambda_{i_1} = \lambda_{i_2} = \ldots = \lambda_{i_m}$.
      \item \textbf{Case 2:} Without loss of generality, assume $\round{\lambda_{i_1} - \gamma} = \ldots = \round{\lambda_{i_{m'}} - \gamma}$ and $\round{\lambda_{i_{m'+1}} - \gamma} = \ldots = \round{\lambda_{i_m} - \gamma}$ and $\round{\lambda_{i_1} - \gamma} = - \round{\lambda_{i_m} - \gamma}$. This implies $\lambda_1 = \lambda_{i_1} = \ldots = \lambda_{i_{m'}}$, $\lambda_2 = \lambda_{i_{m'+1}} = \ldots = \lambda_{i_m}$, and $\gamma = \round{\lambda_1 + \lambda_2}/2$.
    \end{itemize}
    
    Note that $\round{\D - \gamma \identitymat}^2$ is a diagonal matrix. Therefore, one set of solutions to this system of linear equations is given by $E = \braces{\e_{i_1}, \ldots, \e_{i_m}}$ where $\e{i_j}, \, j \in \squares{m}$ are the respective vectors of the canonical basis of the space $\R^n$ corresponding to eigenvalue $l = \alpha \beta$. Define $\z = c_1 \e_{i_1} + \ldots + c_m \e_{i_m}$ where $c_1, \ldots, c_m$ are scalars. Then
    \begin{equation}
      \begin{aligned}
        \round{\D - \gamma \identitymat}^2 \z
          & =\round{\D - \gamma \identitymat}^2 \round{c_1 \e_{i_1} + \ldots + c_m \e_{i_m}} \\
          & = c_1 \round{\D - \gamma \identitymat}^2 \e_{i_1} + \ldots + c_m \round{\D - \gamma \identitymat}^2 \e_{i_m}\\
          & = c_1 l_{i_1} \e_{i_1} + \ldots + c_m l_{i_m} \e_{i_m} \\
          & = c_1 l \e_{i_1} + \ldots + c_m l \e_{i_m} \\
          & = l \round{c_1 \e_{i_1} + \ldots + c_m \e_{i_m}} \\
          & = l \, \z
      \end{aligned}
    \end{equation}
    
    If $\z$ is a linear combination of the support $E' \supset E$, then $\round{\D - \gamma \identitymat}^2 \z \neq l \, \z$ and $\z$ would not be an eigenvector of $\round{\D - \gamma \identitymat}^2$. This implies that $\x$ and $\y$ must be a linear combination of $E = \braces{\e_{i_1}, \ldots, \e_{i_m}}$. Let's write $\x = a_1 \e_{i_1} + \ldots + a_m \e_{i_m}$ and $\y = b_1 \e_{i_1} + \ldots + b_m \e_{i_m}$.
    
    \textbf{Case 1:} $\lambda = \lambda_{i_1} = \lambda_{i_2} = \ldots = \lambda_{i_m}$.
    
    Substituting into $\x^\top \D \y$, we get
    \begin{equation}
      \begin{aligned}
        \x^\top \D \y
          & = \x^\top \D \round{b_1 \e_{i_1} + \ldots + b_m \e_{i_m}} \\
        %   & = \x^\top \round{b_1 \D \e_{i_1} + \ldots + b_m \D \e_{i_m}}\\
          & = \x^\top \round{b_1 \lambda_{i_1} \e_{i_1} + \ldots + b_m \lambda_{i_m} \e_{i_m}} \\
          & = \x^\top \round{b_1 \lambda \e_{i_1} + \ldots + b_m \lambda \e_{i_m}}\\
          & = \x^\top \, \lambda \round{b_1 \e_{i_1} + \ldots + b_m \e_{i_m}} \\
          & = \x^\top \, \lambda \, \y \\
        %   & = \lambda \x^\top \y\\
          & = 0
      \end{aligned}
    \end{equation}
    Hence, the maximum value of the objective function is 0 in this case.
    
    \textbf{Case 2:} $\lambda_1 = \lambda_{i_1} = \ldots = \lambda_{i_{m'}}$, $\lambda_2 = \lambda_{i_{m'+1}} = \ldots = \lambda_{i_m}$, and $\gamma = \round{\lambda_1 + \lambda_2}/2$.
    
    Substituting into $\x^\top \D \y$, we get
    \begin{align*}
        \x^\top \D \y
        & = \x^\top \D \round{b_1 \e_{i_1} + \ldots + b_m \e_{i_m}} \\
        & = \x^\top \round{b_1 \D \e_{i_1} + \ldots + b_m \D \e_{i_m}}\\
        & = \x^\top \round{b_1 \lambda_{i_1} \e_{i_1} + \ldots + b_m \lambda_{i_m} \e_{i_m}} \\
        & = \x^\top \round{b_1 \lambda_{i_1} \e_{i_1} + \ldots + b_{m'}\lambda_{i_{m'}} \e_{i_{m'}}} \\ 
        & \quad \quad + \x^\top \round{b_{m'+1} \lambda_{i_{m'+1}} \e_{i_{m'+1}} + \ldots + b_m \lambda_{i_m} \e_{i_m}} \\
        & = \x^\top \round{b_1 \lambda_{1} \e_{i_1} + \ldots + b_{m'}\lambda_{1} \e_{i_{m'}}} \\ 
        & \quad \quad + \x^\top \round{b_{m'+1} \lambda_{2} \e_{i_{m'+1}} + \ldots + b_m \lambda_{2} \e_{i_m}} \\
        & = \x^\top \lambda_1 \round{b_1 \e_{i_1} + \ldots + b_{m'} \e_{i_{m'}}} \\
        & \quad \quad + \x^\top\lambda_2 \round{b_{m'+1} \e_{i_{m'+1}} + \ldots + b_m \e_{i_m}} \\
        & = \x^\top \round{\lambda_1 \round{b_1 \e_{i_1} + \ldots + b_{m'} \e_{i_{m'}}} + \lambda_2 \round{b_{m'+1} \e_{i_{m'+1}} + \ldots + b_m \e_{i_m}}} \\
        & \quad \quad + \x^\top \round{\round{\lambda_1 - \lambda_1} \round{b_{m'+1} \e_{i_{m'+1}} + \ldots + b_m \e_{i_m}}} \\
        & = \x^\top \lambda_1 \round{b_1 \e_{i_1} + \ldots + b_{m'} \e_{i_{m'}} + b_{m'+1} \e_{i_{m'+1}} + \ldots + b_m \e_{i_m}} \\
        & \quad \quad + \x^\top\round{\lambda_2 - \lambda_1} \round{b_{m'+1} \e_{i_{m'+1}} + \ldots + b_m \e_{i_m}} \\
        & = \x^\top \round{\lambda_1 \y + \round{\lambda_2 - \lambda_1} \round{b_{m'+1} \e_{i_{m'+1}} + \ldots + b_m \e_{i_m}}} \\
        & = \x^\top \lambda_1 \y + \x^\top \round{\round{\lambda_2 - \lambda_1} \round{b_{m'+1} \e_{i_{m'+1}} + \ldots + b_m \e_{i_m}}} \\
        & = \lambda_1 \cancelto{0}{\x^\top \y} + \round{\lambda_2 - \lambda_1} \x^\top \round{b_{m'+1} \e_{i_{m'+1}} + \ldots + b_m \e_{i_m}} \\
        & = \round{\lambda_2 - \lambda_1} \round{a_1 \e_{i_1} + \ldots + a_m \e_{i_m}}^\top \round{b_{m'+1} \e_{i_{m'+1}} + \ldots + b_m \e_{i_m}} \\
        & = \round{\lambda_2 - \lambda_1} \round{a_{m'+1} b_{m'+1} \e_{i_{m'+1}}^\top \e_{i_{m'+1}} + \ldots + a_m b_m \e_{i_m}^\top \e_{i_m}} \\
        & = \round{\lambda_2 - \lambda_1} \round{a_{m'+1} b_{m'+1} + \ldots + a_m b_m} \\
    \end{align*}
    And since $\lambda_2 \neq \lambda_1$ and $\sum_i a_i b_i = 0$, the above quantity can be non-zero.
\end{proof}

\subsection{Proof of Lemma~\ref{lemma:gradienttrace}} \label{app:proof_traceloss_gradient}

\gradienttrace*

\begin{proof}
Let $\U{} = \Z{}^{\top} \Ag \Z{}$ and $\FZ{\U{}} = \U{}^2$. \FZ{\U{}} is a differentiable function of each of the elements of \Z{}. Then, $\g{\FZ{\U{}}} = \tr{\U{}^2}$. Since the trace function is linear, the differential of \g{\cdot} at \Z{} is the composition of trace and the differential of \FZ{\U{}}. Applying the general rule for differentiating a scalar function of a matrix and the chain rule of differentiation for each element \Z{ij}: 
\begin{equation}\label{eq:differential_rule}
\begin{aligned}
    \frac{\partial ~}{\partial ~\Z{ij}} ~\g{\FZ{\U{}}}
    & = \tr{\round{\frac{\partial ~\g{\U{}^2}}{\partial ~\U{}}}^\top \frac{\partial ~\U{}}{\partial ~\Z{ij}}} \\
    & = \tr{\round{\frac{\partial ~\tr{\U{}^2}}{\partial ~\U{}}}^\top \frac{\partial ~\Z{}^{\top} \Ag \Z{}}{\partial ~\Z{ij}}} \\
    & = \tr{\round{~2\U{}^\top}^\top \times \round{\Z{}^\top\Ag \J{ij} + \J{ji}\Ag\Z{}}} \\
    & = \tr{2 \round{\Z{}^\top \Ag \Z{}} \times \round{\Z{}^\top\Ag \J{ij} + \J{ji}\Ag\Z{}}}
\end{aligned}
\end{equation}
where \J{ij} is a single-entry matrix and the derivatives follow from Equation 106 and Equation 80 of the Matrix Cookbook, respectively~\cite{petersen2008matrix}.
\end{proof}

\section{Additional Experiments}\label{app:extra_experiments}

In this section, we present the additional experimental results omitted from the paper.

% \subsection{Analyzing $T_1$ and $T_2$ of the Relaxed Trace Objective}\label{app:t1_t2_relaxed_experiments}

% \input{plot_relaxed_baselines_first_term_compare.tex}

% \input{plot_relaxed_baselines_second_term_compare.tex}

\subsection{Runtime of \opt for the Relaxed Problem}

\begin{figure}[ht]
	\centering
	\includegraphics[width=0.46\textwidth]{fig/legend_opt_iterations.pdf}
	\\
	\centering
	\includegraphics[width=0.46\textwidth]{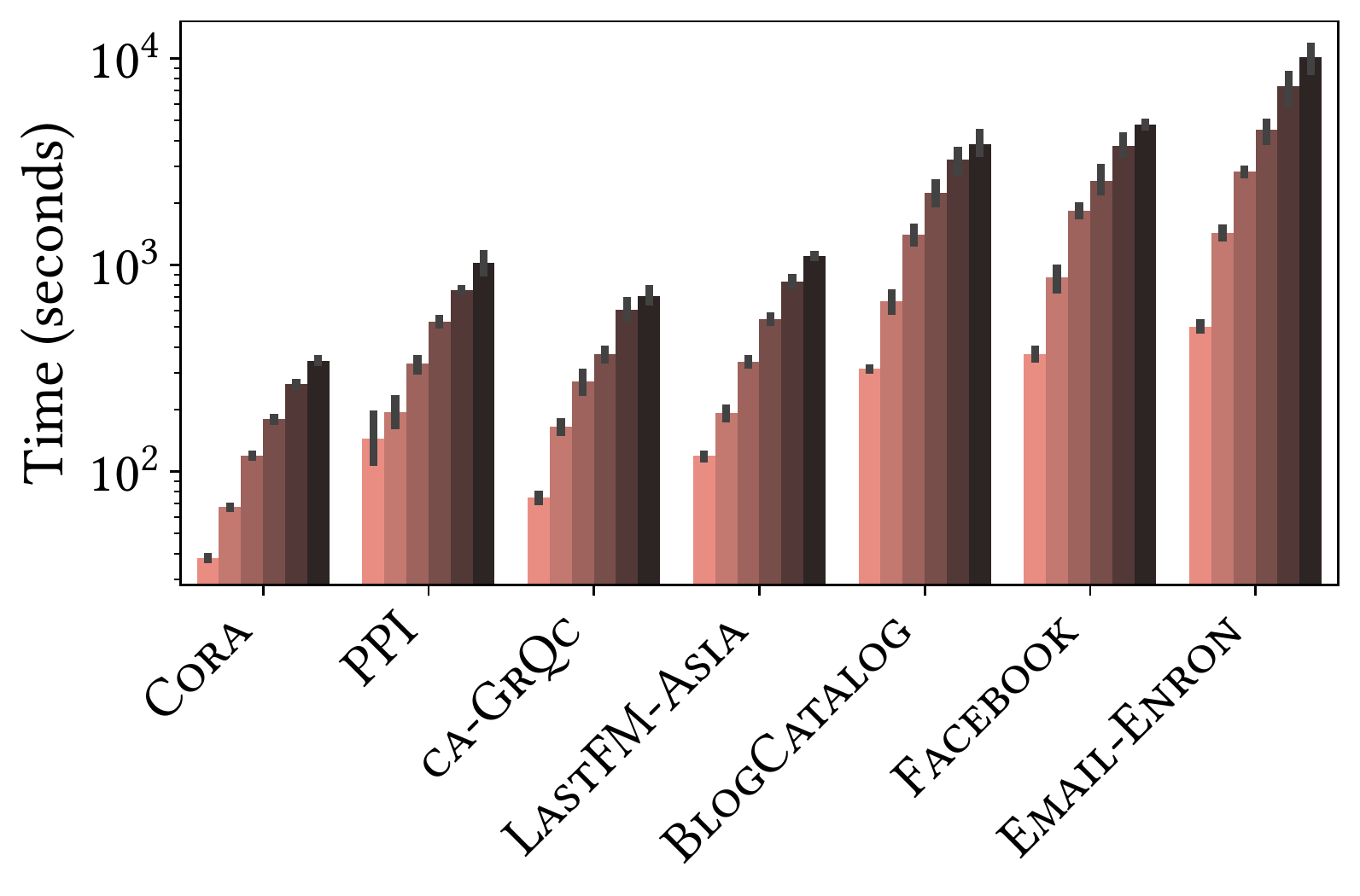}
	\caption{Total runtime (in seconds) of \random-\opt for various summary sizes and datasets averaged over 5 runs.}
	\label{fig:opt_runtime}
	\Description{Experiment 3}
  \end{figure}

Figure~\ref{fig:opt_runtime} shows the time in seconds required by our implementation of \random-\opt for each $\summarysize \in \braces{20, 40, 60, 80, 100, 120}$ on all datasets for 100 iterations. As \summarysize and size of the graph increases, so does the runtime of \opt. The primary computation bottleneck in each iteration here is the gradient computation.

\subsection{Reconstruction Errors}\label{app:normloss_results}

In this subsection, we include the results for summary quality in terms of $l_2$-reconstruction errors in Table~\ref{tab:main_integer_baselines}.
We note that the results are always consistent with those in Table~\ref{tab:fz:small} because the trace maximization objective and the $l_2$-loss minimization objective are theoretically equivalent.

\begin{table}[t!]

		\setlength\tabcolsep{12pt}

		% \fontsize{8}{6}

		\fontsize{8.5}{8.4}\selectfont 

		\centering 

		\begin{tabular}{cccc} 

		\toprule 

		\multirow{2}[2]{*}{\textbf{$k$}} & \multicolumn{3}{c}{\textbf{Algorithm}} \\ 

		\cmidrule(lr){2-4} 

		& \textsc{S2L} & \textsc{SSumM} & \textsc{SpecSumm} \\ 

		\midrule 

	 20 &                     368.08 & 518.57 &  \cellcolor{blue!25}585.92 \\
 40 &                     870.56 &   4.40 & \cellcolor{blue!25}1140.33 \\
 60 & \cellcolor{blue!25}1452.51 & 534.23 &                    1864.09 \\
 80 & \cellcolor{blue!25}1607.07 &  71.63 &                    2828.58 \\
100 &  \cellcolor{blue!25}2424.8 & 681.24 &                     2647.5 \\
120 &                     2988.9 & 304.78 & \cellcolor{blue!25}2054.96 \\
\bottomrule 
 \end{tabular} 
\caption{Expected number of triangles in \textsc{Cora}. Blue denotes summary with the closest estimate to exact value (1558).}
\label{tab:Cora_estimate_num_triangles}
\end{table}
\begin{table}[t!]

		\setlength\tabcolsep{12pt}

		% \fontsize{8}{6}

		\fontsize{8.5}{8.4}\selectfont 

		\centering 

		\begin{tabular}{cccc} 

		\toprule 

		\multirow{2}[2]{*}{\textbf{$k$}} & \multicolumn{3}{c}{\textbf{Algorithm}} \\ 

		\cmidrule(lr){2-4} 

		& \textsc{S2L} & \textsc{SSumM} & \textsc{SpecSumm} \\ 

		\midrule 

	 20 &                     75900.87 & 11872.76 &  \cellcolor{blue!25}87202.76 \\
 40 &                    105716.55 & 22992.23 & \cellcolor{blue!25}104782.71 \\
 60 & \cellcolor{blue!25}121510.11 & 21340.82 &                    148717.02 \\
 80 & \cellcolor{blue!25}158771.29 & 15492.11 &                     162440.6 \\
100 & \cellcolor{blue!25}144888.47 & 12541.86 &                     179208.1 \\
120 & \cellcolor{blue!25}131449.58 & 35264.23 &                    166575.54 \\
\bottomrule 
 \end{tabular} 
\caption{Expected number of triangles in \textsc{PPI}. Blue denotes summary with the closest estimate to exact value (91461).}
\label{tab:PPI_estimate_num_triangles}
\end{table}

\begin{table*}[t]
    \captionsetup{skip=3pt}
	\captionsetup[sub]{skip=0pt} 
	\centering  
	\footnotesize 
	\caption{The $l_2$-reconstruction errors of the summaries computed by each algorithm across different datasets. The values highlighted in blue denote the best quality and the underlined values denote the second-best quality.}
	\label{tab:main_integer_baselines} 
	\begin{subtable}{.485\linewidth}% 
	\centering% 
	\begin{tabular}{lcccccc} 
	\toprule 
	\multirow{2}[2]{*}{\textbf{Algorithm}} & \multicolumn{6}{c}{\textbf{$k$}} \\ 
	\cmidrule(lr){2-7} 
	& $5$ & $10$ & $15$ & $20$ & $25$ & $30$ \\ 
	\midrule
	\textsc{SSumM} &237.53 &237.77 &237.39 &237.67 &237.05 &237.03 \\
    \textsc{S2L} &237.01 &236.98 &236.92 &236.9 &236.88 &236.87 \\
    \textsc{DeepWalk}-\textsc{Ocsa}-\textsc{KM} &236.66 &236.55 & 236.4 &236.32 &236.24 &235.85 \\
    \textsc{SpecSumm} &\underline{236.35} &\underline{235.62} &\underline{234.91} &\underline{234.2} & \underline{233.8} &\underline{233.73} \\
    \textsc{SpecSumm}-\textsc{R} & \cellcolor{blue!25}236.17 & \cellcolor{blue!25}235.27 & \cellcolor{blue!25}234.28 & \cellcolor{blue!25}233.5 & \cellcolor{blue!25}233.07 & \cellcolor{blue!25}232.95 \\
    \bottomrule 
\end{tabular} 
\caption{\textsc{SBM}}
\end{subtable}% 
\begin{subtable}{.485\linewidth}% 
	\centering% 
	\begin{tabular}{lcccccc} 
	\toprule 
	\multirow{2}[2]{*}{\textbf{Algorithm}} & \multicolumn{6}{c}{\textbf{$k$}} \\ 
	\cmidrule(lr){2-7} 
	& $20$ & $40$ & $60$ & $80$ & $100$ & $120$ \\ 
	\midrule 
	\textsc{SSumM} &99.02 &98.11 &96.94 &96.19 &96.31 &95.79 \\
    \textsc{S2L} &99.59 &98.56 &97.16 &96.16 &95.89 &95.31 \\
    \textsc{DeepWalk}-\textsc{Ocsa}-\textsc{KM} &99.07 &96.36 &95.12 & 94.0 &93.61 &\underline{92.42} \\
    \textsc{SpecSumm} &\underline{98.24} &\underline{96.33} &\underline{94.27} &\underline{93.02} &\underline{93.32} &92.47 \\
    \textsc{SpecSumm}-\textsc{R} & \cellcolor{blue!25}97.77 & \cellcolor{blue!25}95.45 & \cellcolor{blue!25}93.47 & \cellcolor{blue!25}91.85 & \cellcolor{blue!25}91.75 & \cellcolor{blue!25}90.74 \\
    \bottomrule 
\end{tabular} 
\caption{\textsc{Cora}}
\end{subtable}% 
\\\vspace{1mm} 
\begin{subtable}{.485\linewidth}% 
	\centering% 
	\begin{tabular}{lcccccc} 
	\toprule 
	\multirow{2}[2]{*}{\textbf{Algorithm}} & \multicolumn{6}{c}{\textbf{$k$}} \\ 
	\cmidrule(lr){2-7} 
	& $20$ & $40$ & $60$ & $80$ & $100$ & $120$ \\ 
	\midrule 
	\textsc{SSumM} &270.94 &271.06 &271.49 &270.36 &270.49 &269.36 \\
    \textsc{S2L} &265.52 &264.93 &262.88 & 261.9 &262.09 &261.58 \\
    \textsc{DeepWalk}-\textsc{Ocsa}-\textsc{KM} &268.14 &265.69 &265.66 &265.46 &264.65 &265.06 \\
    \textsc{SpecSumm} &\underline{263.53} &\underline{260.17} &\underline{256.97} & \underline{256.4} &\underline{256.26} &\underline{255.32} \\
    \textsc{SpecSumm}-\textsc{R} & \cellcolor{blue!25}261.36 & \cellcolor{blue!25}257.06 & \cellcolor{blue!25}253.88 & \cellcolor{blue!25}252.51 & \cellcolor{blue!25}251.92 & \cellcolor{blue!25}250.44 \\
    \bottomrule 
\end{tabular} 
\caption{\textsc{PPI}}
\end{subtable}% 
\begin{subtable}{.485\linewidth}% 
	\centering% 
	\begin{tabular}{lcccccc} 
	\toprule 
	\multirow{2}[2]{*}{\textbf{Algorithm}} & \multicolumn{6}{c}{\textbf{$k$}} \\ 
	\cmidrule(lr){2-7} 
	& $20$ & $40$ & $60$ & $80$ & $100$ & $120$ \\ 
	\midrule 
	\textsc{SSumM} & 144.3 &142.6 &140.86 & 139.4 &138.09 &137.84 \\
    \textsc{S2L} & 145.9 &140.78 &139.18 &140.27 &136.76 &136.73 \\
    \textsc{DeepWalk}-\textsc{Ocsa}-\textsc{KM} &144.79 &141.34 & \underline{137.9} &\underline{137.44} & \cellcolor{blue!25}134.06 &\underline{134.28} \\
    \textsc{SpecSumm} &\underline{142.37} &       \underline{139.69} &138.38 &138.56 &137.84 &135.76 \\
    \textsc{SpecSumm}-\textsc{R} & \cellcolor{blue!25}142.06 & \cellcolor{blue!25}139.1 & \cellcolor{blue!25}137.35 & \cellcolor{blue!25}137.26 &\underline{136.12} & \cellcolor{blue!25}133.63 \\
    \bottomrule 
\end{tabular} 
\caption{\textsc{ca-GrQc}}
\end{subtable}% 
\\\vspace{1mm}
\begin{subtable}{.485\linewidth}% 
	\centering% 
	\begin{tabular}{lcccccc} 
	\toprule 
	\multirow{2}[2]{*}{\textbf{Algorithm}} & \multicolumn{6}{c}{\textbf{$k$}} \\ 
	\cmidrule(lr){2-7} 
	& $20$ & $40$ & $60$ & $80$ & $100$ & $120$ \\ 
	\midrule 
	\textsc{SSumM} &230.83 &229.27 & 229.2 & 227.6 &227.56 & 227.4 \\
    \textsc{S2L} &228.62 &226.69 &224.94 &224.22 &222.68 &222.23 \\
    \textsc{DeepWalk}-\textsc{Ocsa}-\textsc{KM} &228.01 &225.21 &223.02 &\underline{220.79} &\underline{218.92} &\underline{217.48} \\
    \textsc{SpecSumm} &       \underline{227.37} &\underline{224.36} &\underline{222.18} &221.02 &219.24 &218.25 \\
    \textsc{SpecSumm}-\textsc{R} & \cellcolor{blue!25}227.2 & \cellcolor{blue!25}224.01 & \cellcolor{blue!25}221.65 & \cellcolor{blue!25}220.42 & \cellcolor{blue!25}218.34 & \cellcolor{blue!25}217.26 \\
    \bottomrule 
\end{tabular} 
\caption{\textsc{LastFM-Asia}}
\end{subtable}% 
\begin{subtable}{.485\linewidth}% 
	\centering% 
	\begin{tabular}{lcccccc} 
	\toprule 
	\multirow{2}[2]{*}{\textbf{Algorithm}} & \multicolumn{6}{c}{\textbf{$k$}} \\ 
	\cmidrule(lr){2-7} 
	& $20$ & $40$ & $60$ & $80$ & $100$ & $120$ \\ 
	\midrule 
	\textsc{SSumM} &772.77 &772.72 &775.93 &776.87 & 774.8 &774.93 \\
    \textsc{S2L} & \cellcolor{blue!25}755.9 &\underline{749.87} &\underline{747.71} &745.61 &744.54 &745.41 \\
    \textsc{DeepWalk}-\textsc{Ocsa}-\textsc{KM} &780.9 &779.99 &749.19 &\underline{743.75} & 747.0 &\underline{739.56} \\
    \textsc{SpecSumm} &762.81 &753.87 &753.09 &745.02 & \underline{742.9} &742.38 \\
    \textsc{SpecSumm}-\textsc{R} &       \underline{759.46} & \cellcolor{blue!25}748.85 & \cellcolor{blue!25}747.13 & \cellcolor{blue!25}739.38 & \cellcolor{blue!25}737.58 & \cellcolor{blue!25}736.94 \\
    \bottomrule 
\end{tabular} 
\caption{\textsc{Blogcatalog}}
\end{subtable}% 
\\ \vspace{1mm} 
\begin{subtable}{.485\linewidth}% 
	\centering% 
	\begin{tabular}{lcccccc} 
	\toprule 
	\multirow{2}[2]{*}{\textbf{Algorithm}} & \multicolumn{6}{c}{\textbf{$k$}} \\ 
	\cmidrule(lr){2-7} 
	& $20$ & $40$ & $60$ & $80$ & $100$ & $120$ \\ 
	\midrule 
	\textsc{SSumM} &573.46 &573.39 &573.04 & 573.5 & \cellcolor{blue!25}530.4 &530.05 \\
    \textsc{S2L} &569.39 &556.94 &549.97 & 544.2 &538.76 & 533.5 \\
    \textsc{DeepWalk}-\textsc{Ocsa}-\textsc{KM} &567.25 &554.59 & \cellcolor{blue!25}540.06 &536.27 &531.09 & \cellcolor{blue!25}526.78 \\
    \textsc{SpecSumm} &\underline{561.12} &\underline{548.75} & 541.1 &\underline{536.25} &531.19 &527.58 \\
    \textsc{SpecSumm}-\textsc{R} & \cellcolor{blue!25}560.95 & \cellcolor{blue!25}548.51 &\underline{540.82} & \cellcolor{blue!25}535.89 &       \underline{530.78} &\underline{527.09} \\
    \bottomrule 
\end{tabular} 
\caption{\textsc{Facebook}}
\end{subtable}% 
\begin{subtable}{.485\linewidth}% 
	\centering% 
	\begin{tabular}{lcccccc} 
	\toprule 
	\multirow{2}[2]{*}{\textbf{Algorithm}} & \multicolumn{6}{c}{\textbf{$k$}} \\ 
	\cmidrule(lr){2-7} 
	& $20$ & $40$ & $60$ & $80$ & $100$ & $120$ \\ 
	\midrule 
	\textsc{SSumM} &598.76 &591.09 & 591.2 &591.09 &584.96 &584.82 \\
    \textsc{S2L} &587.11 & \cellcolor{blue!25}581.6 & \cellcolor{blue!25}579.34 & \cellcolor{blue!25}577.8 &\underline{576.53} & 574.5 \\
    \textsc{DeepWalk}-\textsc{Ocsa}-\textsc{KM} &589.42 &588.09 &584.99 &581.3 &580.91 &579.07 \\
    \textsc{SpecSumm} &       \underline{586.83} &583.63 &581.21 &578.38 &576.71 &\underline{573.74} \\
    \textsc{SpecSumm}-\textsc{R} & \cellcolor{blue!25}586.7 &       \underline{583.36} &\underline{580.98} &\underline{578.1} & \cellcolor{blue!25}576.36 & \cellcolor{blue!25}573.27 \\
    \bottomrule 
\end{tabular} 
\caption{\textsc{Email-Enron}}
\end{subtable}
\end{table*}

\subsection{Estimating Number of Triangles}

In this subsection, we present results on using the \ksummary to estimate the number of triangles in the original graph. Let $n_i = |V_i|$ be the size of supernode $i$ and let $d_{ij} = A_{\S}(V_i, V_j)$ be the $\round{i,j}$-th entry of the density matrix of summary \S (c.f. Equation~\ref{eq:summary_edge_weight}). Further, define $\forall 1 \leq i,~j, \leq \summarysize:~ \pi_{ij} = d_{ij}$ if $i \neq j$ or if $i = 1$ or if $j = 1$, and $\pi_{ij} = \frac{d_{ij} n_i}{n_i-1}$ if $i = j$. Then, as per \citet{riondato2017graph} (Lemma 7), the expected number of triangles is:
\begin{align*}
	\E\squares{\Delta} 
		= \sum\limits_{i=1}^{\summarysize} & \left( {n_i \choose 3}\pi_{ii}^{3} + \sum\limits_{j=i+1}^{\summarysize} \left( \pi_{ij}^2 \round{{n_i \choose 2} n_j \pi_{ii} + {n_j \choose 2} n_i \pi_{jj}} + \right. \right. \\
		& \left. \left. + \sum\limits_{w = j + 1}^{\summarysize} n_i n_j n_w \pi_{ij} \pi_{jw} \pi_{wj} \right) \right)
\end{align*}

Tables~\ref{tab:Cora_estimate_num_triangles} and ~\ref{tab:PPI_estimate_num_triangles} report the expected number of triangles in the graph estimated by the summary for different \summarysize for \sls, \ssumm, and \specsumm. The estimates from \specsumm and \sls are significantly close to the exact values as compared to the estimates from \ssumm. In fact, after taking standard deviation into account, the estimates by \sls and \specsumm are similar. These results indicate that although \ssumm minimizes aggregate reconstruction error, it does not preserve graph structure information, making practical applications very limited. 
% \subsection{Detailed Results on Large Graphs}\label{app:large_graphs}

% \input{tables/large_graphs_objvals.tex}

\end{document}